\documentclass[structabstract]{aa}  

\usepackage{tabularx}
\usepackage{graphicx}
\usepackage{natbib}
\usepackage{txfonts}
\usepackage{longtable}
\usepackage{multirow}
\usepackage{lscape}
\usepackage{amssymb}

\newcommand{\hersc}{{\it  Herschel}}

\newcommand{\spitz}{{\it   Spitzer}}

\newcommand{\iras}{{\it  IRAS}}

\newcommand{\wise}{{\it WISE}}

\newcommand{\msun}{$M_\odot$}
\newcommand{\ysun}{$Y_\odot$}
\newcommand{\zsun}{$Z_\odot$}
\newcommand{\mic}{$\mu$m}

\newcommand{\xco}{X$_{{\rm CO}}$}
\newcommand{\xcogal}{X$_{{\rm CO, MW}}$}
\newcommand{\xcoz}{X$_{{\rm CO}, {\rm Z}}$}
\newcommand{\mmol}{{\it M$_{{\rm H_2}}$}}
\newcommand{\mmolgal}{{\it M$_{{\rm H_2, MW}}$}}
\newcommand{\mmolz}{{\it M$_{{\rm H_2, Z}}$}}
\newcommand{\matom}{{\it M$_{{\rm H{\sc I}}}$}}

\newcommand{\mgas}{{\it M$_{{\rm gas}}$}}

\newcommand{\mdust}{{\it M$_{{\rm dust}}$}}
\newcommand{\HI}{H{\sc i}}
\newcommand{\mion}{{\it M$_{{\rm H{\sc II}}}$}}

\renewcommand*{\figurename}{ Fig.}
\renewcommand*{\tablename}{Table}
\newlength{\pointwidth}
\DeclareGraphicsExtensions{.pdf,.png,.jpg, .eps}

\def\revised{}

\begin{document}

  \title{Gas-to-Dust mass ratios in {\revised local} galaxies over a 2 dex metallicity range}
  
  \author{A. R\'emy-Ruyer\inst{1} 
  	\and S. C. Madden\inst{1}
  	\and F. Galliano\inst{1}
	\and M. Galametz\inst{2}
	\and T. T. Takeuchi\inst{3}
	\and R. S. Asano\inst{3}
	\and S. Zhukovska\inst{4,5}
	\and V. Lebouteiller\inst{1}
	\and D. Cormier\inst{5}
	\and A. Jones\inst{6}
	\and M. Bocchio\inst{6}
	\and M. Baes\inst{7}
	\and G. J. Bendo\inst{8}
	\and M. Boquien\inst{9}
	\and A. Boselli\inst{9}
	\and I. DeLooze\inst{7}
	\and V. Doublier-Pritchard\inst{10}
	\and T. Hughes\inst{7}
	\and O. \L. Karczewski\inst{11}
	\and L. Spinoglio\inst{12}}	      
	      
	      \institute{Laboratoire AIM, CEA/IRFU/Service d'Astrophysique, Universit\'{e} Paris Diderot, Bat. 709, 91191 Gif-sur-Yvette, France, \\
	      \email{aurelie.remy@cea.fr}
	       \and Institute of Astronomy, University of Cambridge, Madingley Road, Cambridge CB3 0HA, UK
                \and Division of Particle and Astrophysical Science, Nagoya University, Furo-cho, Chikusa-ku, Nagoya 464-8602, Japan
                \and Max-Planck-Institut f¬ur Astronomie, K¬onigstuhl 17, 69117, Heidelberg, Germany
                \and Zentrum f\"ur Astronomie der Universit\"at Heidelberg, Institut f\"ur Theoretische Astrophysik, Albert-Ueberle-Str. 2, 69120 Heidelberg, Germany
               \and Institut d'Astrophysique Spatiale, CNRS, UMR8617, 91405, Orsay, France
               \and Sterrenkundig Observatorium, Universiteit Gent, Krijgslaan 281 S9, B-9000 Gent, Belgium               
                \and UK ALMA Regional Centre Node, Jodrell Bank Centre for Astrophysics, School of Physics \& Astronomy, University of Manchester, Oxford Road, Manchester M13 9PL, UK
                \and Laboratoire dÕAstrophysique de Marseille - LAM, Universit\'e d'Aix-Marseille \& CNRS, UMR7326, 38 rue F. Joliot-Curie, 13388 Marseille Cedex 13, France
               \and Max Planck fŸr Extraterrestrische Physik, Giessenbachstr. 1, 85748 Garching
               \and Department of Physics and Astronomy, University of Sussex, Brighton, BN1 9QH, UK
                \and Instituto di Astrofisica e Planetologia Spaziali, INAF-IAPS, Via Fosso del Cavaliere 100, I-00133 Roma, Italy
        }

\date{Received date/Accepted date}


 \abstract
{}
{This paper analyses the behaviour of the gas-to-dust mass ratio (G/D) of local Universe galaxies over a large metallicity range. We especially focus on the low-metallicity part of the G/D vs metallicity relation and investigate explanations for the observed relation and scatter.}
{We combine three samples: the Dwarf Galaxy Survey, the KINGFISH survey and a subsample from Galametz et al. (2011) totalling 126 galaxies, covering a 2 dex metallicity range, with 30\% of the sample with 12+log(O/H) $\leq$ 8.0. The dust masses are homogeneously determined with a semi-empirical dust model, including submm constraints. The atomic and molecular gas masses are compiled from the literature. Two \xco\ are used to estimate molecular gas masses: the Galactic \xco\, and a \xco\ depending on the metallicity ($\propto Z^{-2}$). Correlations with morphological types, stellar masses, star formation rates and specific star formation rates are discussed. The trend between G/D and metallicity is empirically modelled using power-laws (slope of -1 and free) and a broken power-law. We then compare the evolution of the G/D with predictions from chemical evolution models.}
{We find that out of the five tested galactic parameters, metallicity is the galactic property driving the observed G/D. The observed G/D versus metallicity relation cannot be represented by a power-law with a slope of -1 over the whole metallicity range. The observed trend is steeper for metallicities lower than $\sim$ 8.0. A large scatter is observed in the G/D for a given metallicity, with a dispersion of 0.37 dex on average in metallicity bins of $\sim$0.1 dex, for both \xco\ values. The broken power-law reproduces best the observed G/D compared to the single power laws and provides estimates of the G/D that are accurate to a factor of 1.6. The good agreement of the G/D and its scatter with respect to metallicity with the predictions from the three tested chemical evolution models allows us to infer that the scatter is intrinsic to galactic properties, reflecting the different star formation histories, dust destruction efficiencies, dust grain size distributions and chemical compositions across the sample.}
{Our results show that the chemical evolution of low-metallicity galaxies, traced by their G/D, depends strongly on their local internal conditions and individual histories. The large scatter in the observed G/D at a given metallicity reflects the impact of various processes occurring during the evolution of a galaxy. Disentangling between these various processes, despite the numerous degeneracies affecting them, is now the next step.
}


     \authorrunning{R\'emy-Ruyer et al.}
     \titlerunning{Gas-to-Dust mass ratios in {\revised local} galaxies over a 2 dex metallicity range}

 \maketitle


\section{Introduction}\label{intro}

Metallicity is a key parameter in studying the evolution of galaxies as it traces metal enrichment. Elements are injected by stars in the interstellar medium (ISM) via stellar winds and/or supernovae (SN) explosions \citep{DwekScalo1980}, and become available for the next generation of stars. The metallicity {\it a priori} traces the history of the stellar activity of a galaxy, i.e. the number of stellar generations already produced. Metallicity is thus expected to increase with age as the galaxy undergoes chemical enrichment through successive star formation events. However, this metal enrichment is in fact a more complex process and depends on external and internal processes occurring during galaxy evolution. Indeed, the gas phase abundance can be affected by metal-poor gas inflows that will dilute the ISM \citep{Montuori2010,DiMatteo2011} and decrease the metallicity of the galaxy; or by outflows driven by stellar feedback \citep[stellar winds or SN shocks; e.g.][]{Dahlem1998,Frye2002} that will eject metal-rich gas into the intergalactic medium, resulting again in a galaxy whose metallicity does not simply increase with time. Elements are also processed by gas and dust in the ISM.  
Dust grains form from the available metals in the ISM and those metals can thus be depleted from the gas phase \citep{SavageSembach1996, Whittet2003}. Metals are returned to the gas phase when dust is destroyed \citep[e.g. by SN blast waves][]{Jones1994,Jones1996}. The gas-to-dust mass ratio (G/D) links the amount of metals locked up in dust and in the gas phase and is thus a powerful tracer of the evolutionary stage of a galaxy. Investigations of the relation between the observed G/D and metallicity can thus place important constraints on the physical processes governing galaxy evolution and more specifically on chemical evolution models. Because of their low-metallicity, dwarf galaxies can be considered as chemically young objects that are at an early stage of their evolution. In this picture, they can be seen as our closest analogues of the primordial environments present in the early Universe, from which the present-day galaxies formed. G/D derived from dwarf galaxies are thus crucial for constraining chemical evolution models at low metallicities.

The observed G/D of integrated galaxies as a function of metallicity has been intensively studied over the past decades \citep[e.g.][]{Issa1990, LisenfeldFerrara1998, Hirashita2002, James2002, Hunt2005, Draine2007, Engelbracht2008, Galliano2008, MunozMateos2009, Bendo2010b, Galametz2011,Magrini2011}. In the disk of our Galaxy the proportion of heavy elements in the gas and in the dust has been shown to scale with the metallicity \citep{Dwek1998} if one assumes that the time dependence of the dust formation timescale is the same as that of the dust destruction timescale. This results in a constant dust-to-metal mass ratio and gives a dependence of the G/D on metallicity as: G/D~$\propto$~Z$^{-1}$ (that we call hereafter the ``reference'' trend). This reference trend between G/D and metallicity seems consistent with the observations of galaxies with near-solar metallicities \citep[e.g.][]{James2002, Draine2007, Bendo2010b, Magrini2011}. However some studies also show that the G/D of some low-metallicity dwarf galaxies deviate from this reference trend, with a higher G/D than expected for their metallicity \citep{LisenfeldFerrara1998, Galliano2003,Hunt2005, Galliano2005, Galliano2008, Bernard2008, Engelbracht2008, Galametz2011, Galliano2011}. 

The G/D is often used to empirically estimate the ``CO-free'' molecular gas. This molecular gas, not directly traced by CO measurements, was first proposed in low-metallicity galaxies using the 158 \mic\ [C{\sc ii}] line \citep{Poglitsch1995, Israel1996, Madden1997, Madden2012}. 
Alternatively, using the dust mass determined from far-infrared (FIR) measurements and assuming a G/D, given the metallicity of the galaxy, a gas mass can be estimated. 
This gas mass is then compared to the observed \HI\ and CO gas masses to estimate the ``CO-free" gas mass. This method is also used to estimate the CO-to-H2 conversion factor in local \citep[e.g.][]{Guelin1993,Guelin1995,Neininger1996,Boselli2002, Sandstrom2012} and high-{\it z} galaxies \citep[e.g.][]{Magdis2012,Magnelli2012}. However using these methods requires an accurate estimation of the G/D at a given metallicity. 

A certain number of instrumental limitations and/or model caveats have limited former studies of the G/D. 
First, limits in wavelength coverage in the FIR have hampered the precise determination of the dust masses. For the earliest studies, the dust masses were derived from \iras\ or \spitz\ measurements, not extending further than 100-160 \mic, therefore not tracing the cooler dust. As the bulk of the dust mass in galaxies often resides in the cold dust component, this has important consequences for the dust mass and the G/D determination. Before \hersc, using \spitz\ and ground-based data, \cite{Galametz2011} indeed showed that a broad wavelength coverage of the FIR-to-submillimeter (submm) part of the spectral energy distribution (SED) was critical to obtain accurate estimates of the dust masses. 
Second, some of the studies presented previously used modified blackbody models to derive dust masses. Using \hersc\ data and a semi-empirical dust model, \cite{Dale2012} showed that the dust mass modelled with a modified blackbody could be underestimated by a factor of $\sim$ 2. \cite{Bianchi2013} showed on the contrary that a modified blackbody modelling, using a fixed emissivity index, provides a good estimate of the dust mass. However the case of low-metallicity dwarf galaxies has not been investigated yet and the two dust mass estimates may not be consistent with each other at low metallicities. 
And finally, the limited sensitivities of the pre-\hersc\ era instruments only allowed the detection of dust for the brightest and highest-metallicity dwarf galaxies, limiting G/D studies to metallicities~$\geq$~1/5~\zsun\footnote{Throughout we assume (O/H)$_\odot$ = 4.90 $\times$ 10$^{-4}$, i.e., 12+log(O/H)$_\odot$~=~8.69, and a total solar mass fraction of metals \zsun\ =0.014 \citep{Asplund2009}.} (12+log(O/H) = 8.0).

In this work we investigate the relation between the G/D and metallicity avoiding the limitations and caveats mentioned in the previous paragraph. Using new \hersc\ data to constrain a semi-empirical dust model, we have a more accurate determination of the dust masses comparing to \spitz-only dust masses and/or modified blackbody dust masses. Our sample also covers a wide range in metallicity (2 dex, from 12+log(O/H) = 7.1 to 9.1), with a significant fraction of the sample below 12+log(O/H) = 8.0 ($\sim$ 30\%, see Section \ref{sec:data}), thanks to the increased sensitivity of \hersc\ which enables us to access the dust in the lowest metallicity galaxies. 
We are thus able to provide better constraints on the G/D at low-metallicities.

In Section \ref{sec:data}, we describe the sample and the method used to estimate dust and total gas masses. Then we investigate the relation of the observed G/D with metallicity {\revised and other galactic parameters} in Section \ref{sec:analysis} and fit several empirical relations to the data. We then interpret our results with the aid of several chemical evolution models in Section \ref{sec:Chemvol}. Throughout we consider 
(G/D)$_\odot$ = 162\footnote{This value is from Table 6 from \cite{Zubko2004} for the BARE-GR-S model, which corresponds to the dust composition used for our modelling (see Section \ref{ssec:dust}).} \citep{Zubko2004}.


\section{Sample and Masses}\label{sec:data}

\subsection{Sample}

We combine 3 different samples for our study of the G/D: the Dwarf Galaxy Survey \citep[DGS,][]{Madden2013}, the KINGFISH survey \citep{Kennicutt2011} and a subsample of the sample presented in \cite{Galametz2011} (called the ``G11 sample'' hereafter). The basic parameters for all of the galaxies such as positions and distances can be found in \cite{Madden2013} for the DGS, \cite{Kennicutt2011} for KINGFISH and \cite{Galametz2011} for the G11 sample.

\begin{figure}

\begin{center}
\includegraphics[width=8.8cm]{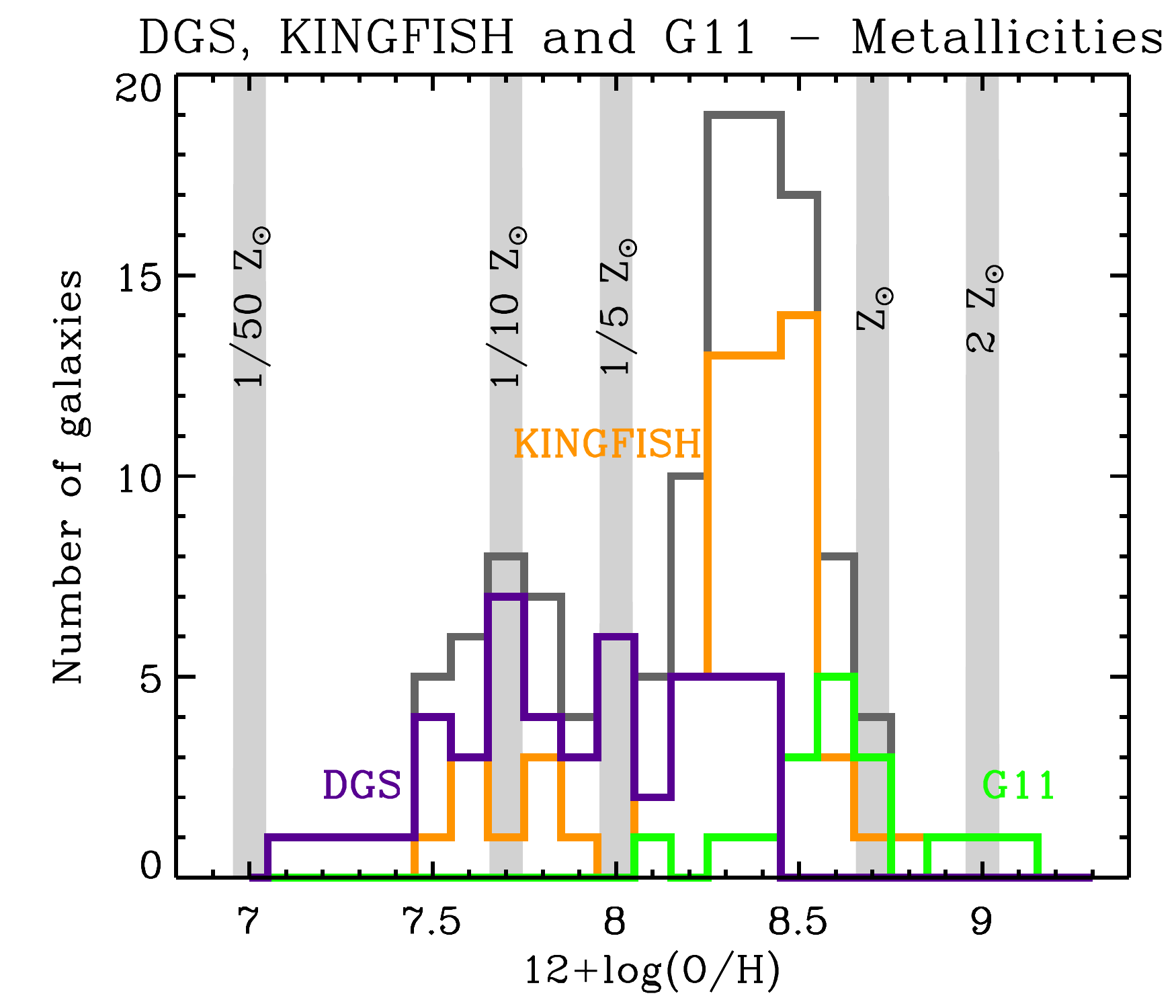}
\caption{Metallicity distribution of the DGS (purple), KINGFISH (orange) and G11 (green) samples from 12+log(O/H) = 7.14 to 9.1. The total distribution is indicated in grey. Solar metallicity is indicated here as a guide to the eye, as well as the 1/50, 1/10, 1/5 and 2 \zsun values.}
\label{met}
\end{center}
\end{figure}

The DGS sample consists of 48 star-forming dwarf galaxies {\revised (mostly dwarf irregulars and blue compact dwarfs (BCDs))} covering metallicities from 12+log(O/H) = 7.14 to 8.43, whereas the KINGFISH sample probes more metal-rich environments (61 galaxies including spiral, early-type and a few irregular galaxies), from 12+log(O/H) = 7.54 to 8.77. The G11 sample consists of all of the galaxies in \cite{Galametz2011} that are neither already in the DGS nor in the KINGFISH samples, and except those galaxies which show a submm excess (see Section \ref{ssec:dust}). This gives 17 additional galaxies, mostly solar or super solar environments (mostly spiral galaxies), with metallicities from 12+log(O/H) = 8.14 to 9.1. The metallicity distribution for each of the 3 samples is presented in Fig. \ref{met}. 

All of these metallicities have been derived using empirical strong emission line methods \citep[see][for the DGS, Kennicutt et al. 2011 for KINGFISH and Galametz et al. 2011 for G11 metallicity determination]{Madden2013}. The DGS and KINGFISH metallicities have been obtained through the R$_{23}$ ratio\footnote{R$_{23}$=([OII]$\lambda$3727+[OIII]$\lambda\lambda$4959,5007)/H$\beta$} with the \cite{PilyuginThuan2005} calibration.
\cite{Galametz2011} do not indicate precisely which calibration they use to convert R$_{23}$ into metallicity, thus several metallicities for the G11 galaxies were re-estimated from the original line intensities, {\revised available in the literature}, with the \cite{PilyuginThuan2005} calibration. We also assume a conservative 0.1 dex uncertainty for the G11 metallicities. On average for the total sample, the uncertainty on the metallicity measurements is $\sim$ 0.1 dex. The metallicities for the whole sample are listed in Table \ref{t:Gas}.
Other methods exist to determine metallicities and can lead to very different values, but this will only introduce a systematic offset in the adopted values here \citep{KewleyEllison2008}. Note that these metallicity values correspond to global estimates. On smaller scales within galaxies, differences can occur due to inhomogeneous mixing of metals: metallicity gradients have been observed in large spiral galaxies \citep{Garnett2004, Bendo2010b, Moustakas2010}. Dwarf galaxies, however, are smaller in size than metal-rich galaxies and we can presume, for this study, that metallicity is more homogeneous within these environments \citep{RevazJablonka2012,Valcke2008} . 

This gives us a total of 126 galaxies spanning a 2 dex range in metallicity (Fig. \ref{met}). We see that the low-metallicity end of the distribution is fairly well sampled as we have $\sim$ 30 \% of the total sample with metallicities below 1/5 \zsun.

\subsection{Dust masses}\label{ssec:dust}

To ensure a consistent determination of the dust masses throughout our sample, all of the galaxies are modelled with the dust SED model presented in \cite{Galliano2011}.
It is a phenomenological model based on a two steps approach: first the modelling of one mass element of the ISM with a uniform illumination, and second the synthesis of several mass elements to account for the different illumination conditions. In the first step we assume that the interstellar radiation field (ISRF) has the spectral shape of the solar neighbourhood ISRF \citep{Mathis1983} and use the \cite{Zubko2004} grain properties with updated PAH optical properties from \cite{DraineLi2007}. In the second step, we assume that the total SEDs of the galaxies can be well represented by the combination of the emission from regions with different properties. In order to do so, we assume that the dust properties are uniform and that only the illumination conditions vary in the different regions. The various regions are then combined using the \cite{Dale2001} prescription: the distribution of the starlight intensities per unit dust mass can be represented by a power-law:

\begin{equation}
\frac{dM_{dust}}{dU} \propto U^{-\alpha_{U}} \ with \  U_{min} \leq U \leq U_{min} +  \Delta U
\end{equation}
\noindent where {\it M$_{dust}$} is the dust mass, {\it U} is the starlight intensity and $\alpha_U$ is the index of the power-law describing the starlight intensity distribution. 

We use the ``standard'' grain composition in our model, i.e., PAHs, silicate dust and carbon dust in form of graphite\footnote{This corresponds to the BARE-GR-S model of \cite{Zubko2004}.}. The submm emissivity index, $\beta$, for this grain composition is $\sim$ 2.0 \citep{Galliano2011}. We discuss the impact of having another dust composition on  our results (e.g. the use amorphous carbons instead of graphite grains) in Section \ref{ssec:discussion}.

The free parameters of the model are:  the dust mass, {\it M$_{dust}$}, the minimum starlight intensity, {\it U$_{min}$}, the difference between the maximum and minimum starlight intensities, {\it $\Delta$U}, the starlight intensity distribution power-law index,  {\it $\alpha_U$}, the PAH-to-total dust mass ratio, {\it f$_{PAH}$}, the mass fraction of ionised PAHs compared to the total PAH mass,  {\it f$_{ion}$}, the mass fraction of very small grains (i.e. non-PAH grains with sizes $\leq$ 10 nm), {\it f$_{vsg}$}, and the contribution of the near-IR stellar continuum, {\it M$_{star}$} \citep[see][for details and a full description of the model]{Galliano2011}. This model has previously been used to model dwarf galaxies, notably by \cite{Galametz2009,Galametz2011, OHalloran2010, Cormier2010, Hony2010, Meixner2010}. 

For the DGS sample, we collect IR-to-submm photometrical data from various instruments (i.e., 2MASS, \spitz, \wise, \iras\ and \hersc) for the largest possible number of galaxies. IRS spectra are also used to constrain the MIR slope of the SEDs. {\revised For 12 DGS galaxies, the MIR continuum shape outlined by the IRS spectrum cannot be well fitted by our model. In these cases, we add an extra modified blackbody component in the MIR, with a fixed $\beta$ = 2.0 and a temperature varying between 80 and 300 K. This affect our dust masses by $\sim$ 5\%, which is well within the error bars for the dust masses ($\sim$ 26\%), and thus this does not affect our following results. For one DGS galaxy, SBS1533+574, the impact on the dust mass is important, but the addition of this warm modified blackbody is necessary to have a MIR to submm shape of the SED consistent with the observations (R\'emy-Ruyer et al., in prep).} Out of the 48 DGS galaxies, only five do not have enough constraints (i.e. no observations are available or there are too many non-detections) to obtain a dust mass. \hersc\ data for the DGS is presented in \cite{RemyRuyer2013a}. 2MASS, \wise\ and \iras\ flux densities for the DGS are compiled from the NASA/IPAC IRSA databases and the literature. \spitz-MIPS measurements are taken from \cite{Bendo2012}. \spitz-IRAC and IRS data together with a complete description of the dust modelling and the presentation of the final SEDs and dust masses for the DGS galaxies are presented in R\'emy-Ruyer et al., (in prep.). 


For the KINGFISH sample, we use IR-to-submm fluxes from \cite{Dale2007, Dale2012} to build the observed SEDs (i.e. observational constraints from 2MASS, \spitz, \iras\ and \hersc). The \cite{Dale2012} fluxes have been updated to the new values of the SPIRE beam areas\footnote{SPIRE photometer reference spectrum values: http://herschel.esac. \\ esa.int/twiki/bin/view/Public/SpirePhotometerBeamProfileAnalysis: 465, 822 and 1768 square arcseconds at 250, 350, 500 \mic\ (September 2012 values).}. The dust masses for the KINGFISH galaxies are presented in Table \ref{t:Dust}.
A submm excess is observed in some DGS and KINGFISH galaxies at 500 \mic\ \citep[][R\'emy-Ruyer et al., in prep.]{Dale2012}. If present, the excess starts at 500 \mic\ and can increase as we go to longer wavelengths. However, because of the unknown origin of this excess and because of the uncertainties it can bring in the dust mass estimation, we do not attempt to model this excess with additional modifications to the model, {\revised and we thus leave aside the 500 \mic\ point in our model. Including the 500 \mic\ point results in a median difference of $\sim$3\% for the dust masses in the DGS and KINGFISH samples}. We discuss {\revised the influence of the presence of a submm excess} in Section \ref{ssec:discussion}.

We also model galaxies from the G11 sample, as some model assumptions were different in \cite{Galametz2011}. \hersc\ constraints are not present but other submm constraints are taken into account such as JCMT/SCUBA at 850 \mic\ and/or APEX/LABOCA at 870 \mic, allowing a precise determination of the dust masses (given in Table \ref{t:Dust}). \cite{Galametz2011} also observed a submm excess in nine galaxies of their original sample and modelled it with a Very Cold Dust (VCD) component. However the submm excess is not fully understood yet and this extra VCD component may lead to an overestimation of the dust mass. Because we do not have constraints between 160 \mic\ and the available ground-based submm fluxes to see where the submm excess starts, we do not consider nor model these galaxies here.

\begin{figure}
\begin{center}
\includegraphics[width=8.8cm]{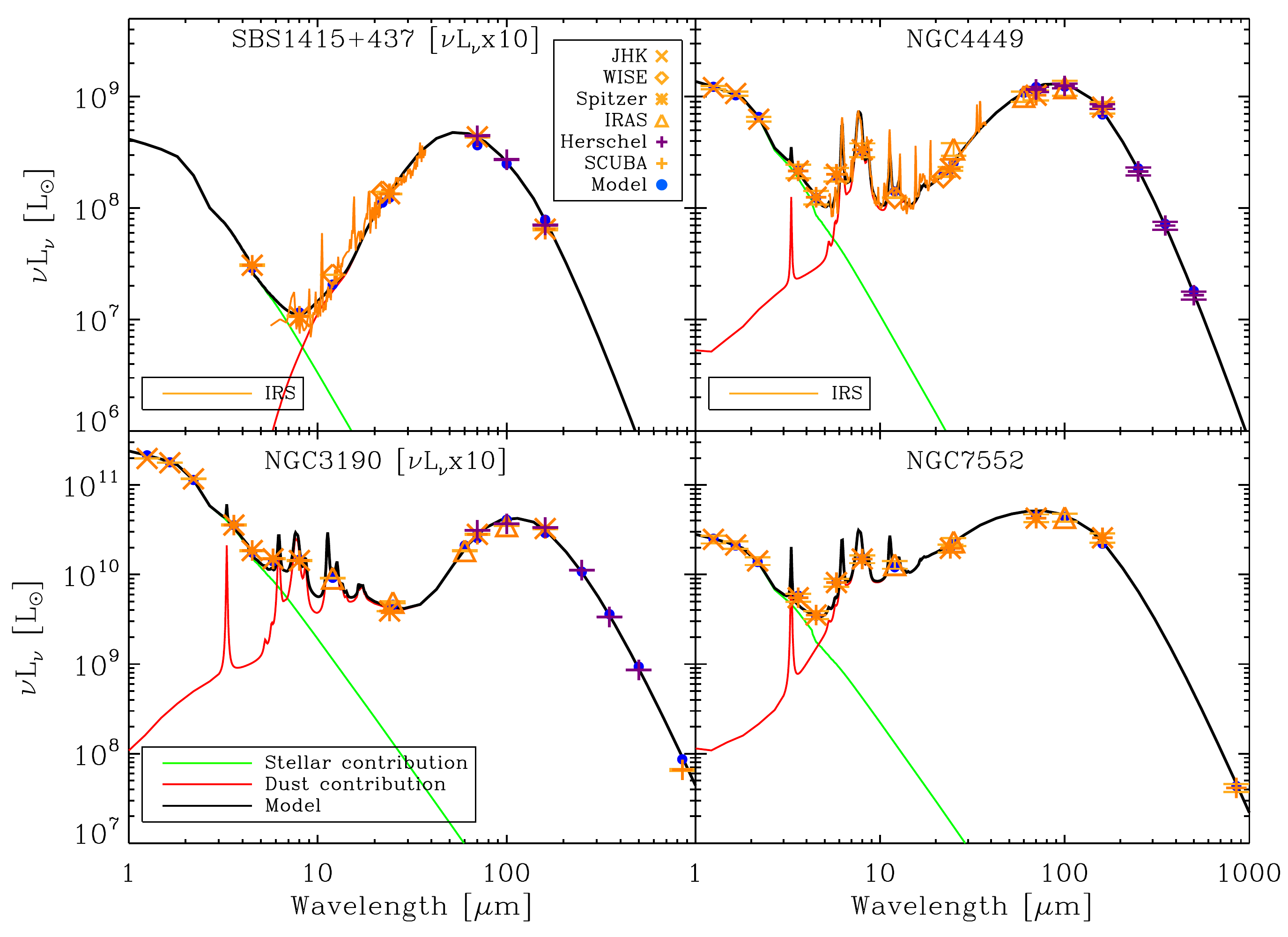}
\caption{{\revised Examples of SEDs modelled with the \cite{Galliano2011} model: {\it (top left)} SBS1415+437, {\it (top right)} NGC4449, {\it (bottom left)} NGC3190 and {\it (bottom right)} NGC7552. The SEDs have been multiplied by 10 for SBS1415+437 and NGC3190 for display purposes. The observed SED includes the \hersc\ data (purple crosses) as well as ancillary data (in orange). The different symbols code for the different instruments: Xs for 2MASS bands, diamonds for WISE, stars for \spitz\ IRAC and MIPS, triangles for IRAS and orange crosses for SCUBA. The IRS spectrum is also displayed in orange for the two DGS galaxies. The total modelled SED in black is the sum of the stellar (green) and dust (red) contributions. The modelled points in the different bands are the filled blue circles.}}
\label{f:exSEDs}
\end{center}
\end{figure}

The wavelength coverage is not exactly the same from galaxy to galaxy. The most important constraints for the determination of the dust mass are constraints sampling the peak of the dust SED. \hersc\ provides such constraints for all of the DGS and KINGFISH galaxies. Some dwarf galaxies are not detected with \hersc\ at submm wavelengths (beyond 250 \mic) and are noted in Fig. \ref{G/Dline} (see Section \ref{ssec:gas}). These galaxies harbour particularly warm dust \citep{RemyRuyer2013a}; the peak of their SED is thus shifted towards shorter wavelengths and is then well sampled by constraints until 160 \mic\ where the galaxy is still detected. For galaxies in the G11 sample, the peak of the dust SED is probed by \spitz\ observations and the Rayleigh Jeans slope of the SED by longer submm wavelength observations. Thus we are confident in the dust masses we derive with these sets of constraints. {\revised As an illustration, four SEDs are presented in Fig. \ref{f:exSEDs}: one very low metallicity DGS galaxy (SBS1415+437, 12+log(O/H)=7.55) not observed with SPIRE, one low metallicity DGS galaxy (NGC4449, 12+log(O/H) = 8.20), and two spiral galaxies from the KINGFISH sample (NGC3190, 12+log(O/H)= 8.49) and from the G11 sample (NGC7552, 12+log(O/H)=8.35).}

The errors on the dust masses are estimated by generating 300 random realisations of the SED, perturbed according to the random and systematic noise, in order to get a distribution for the dust mass. The error bars on the dust mass are taken to be the 66.67 \% confidence interval of the distribution (i.e., the range of the parameter values between 0.1667 and 0.8333 of the repartition function). The detailed procedure of the error estimation is presented in R\'emy-Ruyer et al., (in prep.).

\subsection{Gas masses}\label{ssec:gas}

\paragraph{\HI\ masses -}

The \HI\ masses and their errors are compiled from the literature, and rescaled to the distances used here. Most of the atomic gas masses are given in \cite{Galametz2011} for the G11 sample, and in \cite{DraineLi2007} for the KINGFISH survey. They are presented in \cite{Madden2013} for the DGS. The errors were not available for all of the \HI\ measurements. When no error was available for the \HI\ mass, we adopted the mean value of all of the relative errors on the \HI\ masses compiled from the literature: $\sim$ 16\%.

However the \HI\ extent of a galaxy is not necessarily the same as the aperture used to probe the dust SED, 
as the \HI\ often extends beyond the optical radius of a galaxy \citep{Hunter1997}. This can be particularly true for dwarf galaxies where the \HI\ halo can be very extended: some irregular galaxies present unusually extended \HI\ gas \citep[up to seven times the optical radius,][]{Huchtmeier1979, Huchtmeier1981, Carignan1989, Carignan1990, Thuan2004}. We also note that in some galaxies (e.g. NGC 4449), gas morphology may be highly perturbed due to past interactions or mergers \citep[e.g.][]{Hunter1999}. This may also lead to significant uncertainties in the \HI\ mass and thus on the derived G/D \citep[e.g.][]{Karczewski2013}.
Thus we check the literature for the DGS sample for the size of the \HI\ halo to compare it to the dust aperture. It was not possible to find this information for $\sim$ 38\% of the sample (\HI\ not detected or no map available). For the rest, 25\% of the DGS galaxies have a \HI\ extent that corresponds to the dust IR aperture, which has been chosen to be 1.5 times the optical radius \citep[for most cases, see][]{RemyRuyer2013a}; and 35\% have a \HI\ halo that is more extended. 
{\revised If we assume that the \HI\ mass distribution follows a gaussian profile \citep[based on the observed high central gas concentration seen in BCDs, e.g.,][]{VanZee1998, VanZee2001, Simpson2000}, we can correct the total \HI\ mass for these galaxies to find the \HI\ mass corresponding to the dust aperture.}
In reality, the \HI\ profile can show a complicated structure with clumps and shells, {\revised rendering the profile more assymetric.}

Our correction corresponds to a factor of $\sim$ 1.55 on average, for these galaxies. Several studies \citep{Thuan1981, Swaters2002, Lee2002, Begum2005, Pustilnik2007} have tried to quantify the extent of the \HI\ halo for dwarf galaxies and found that the ratio of \HI\ size to the optical size is typically 2, which gives a {\revised correction of $\sim$ 1.4. }
This is similar to what we find here with our assumptions. 
The atomic gas masses for the sample, after correction if needed, are presented in Table \ref{t:Gas}.

\paragraph{H$_2$ masses -}

The H$_2$ masses have been compiled from the literature. They have been rescaled, when necessary, to the distance adopted here to derive the dust masses.

\begin{figure}
\begin{center}
\includegraphics[width=8.8cm]{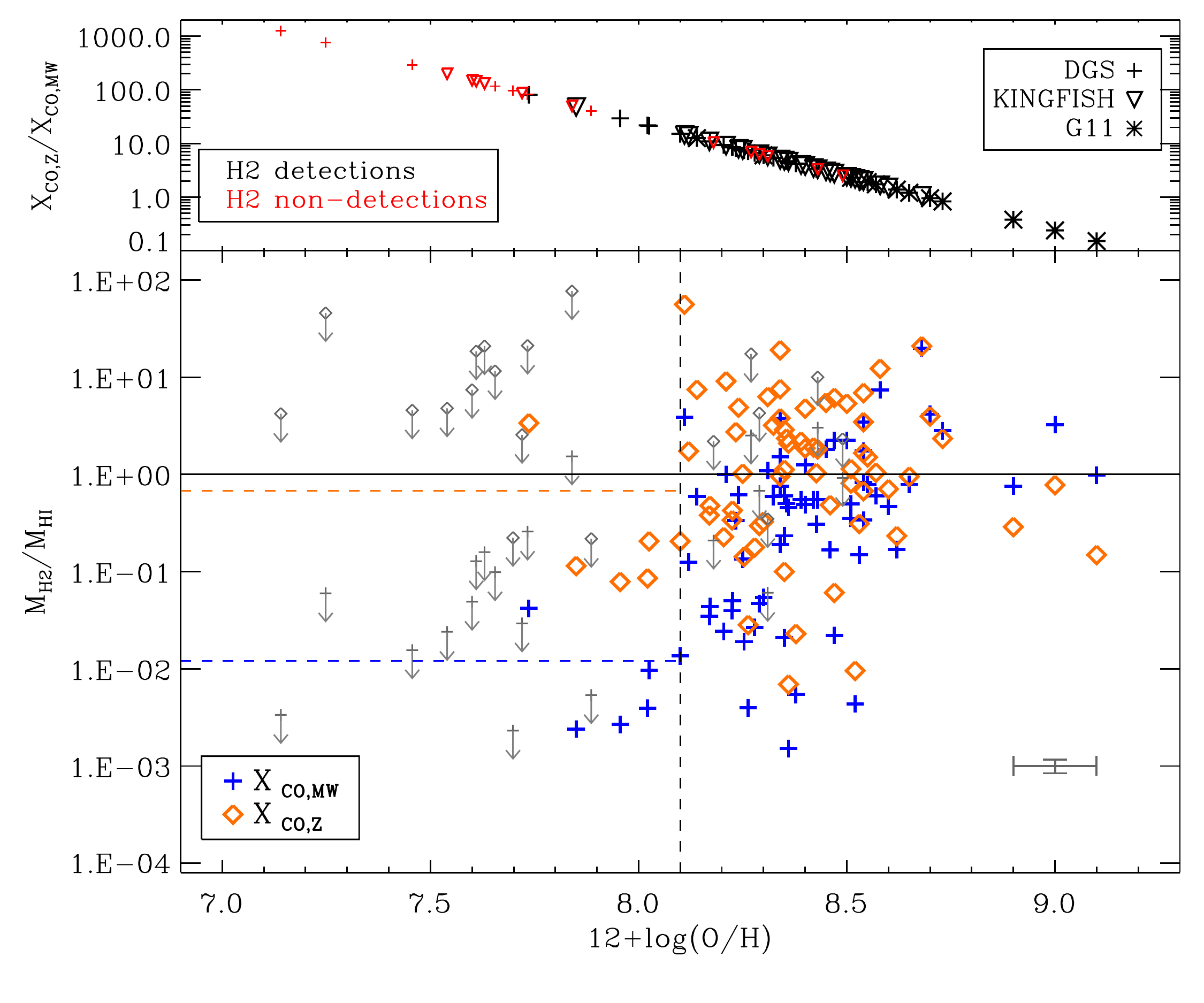}
\caption{{\it (bottom)} \mmol/\matom\ as a function of metallicity for the whole sample. The blue crosses are for molecular gas masses computed with \xcogal\ and the orange diamonds are for molecular gas masses computed with \xcoz. Upper limits in the molecular gas mass are indicated with grey arrows and smaller grey symbols. The mean error for the data points is shown in grey on the bottom right of the plot. The plain line shows the unity line. The dashed blue and orange lines show the 1.2\% and 68\% molecular-to-atomic gas mass fractions respectively and represent the mean H$_2$-to-\HI\ ratio of the detected galaxies with 12+log(O/H) $<$ 8.1 (see text). {\revised The horizontal dashed black line shows the metallicity threshold 12+log(O/H) = 8.1 to guide the eye. {\it (top)} \xcoz/\xcogal\ illustrating the (O/H)$^{-2}$ dependence adopted to compute \xcoz. The symbols delineate between the three samples: crosses, downward triangles and stars for the DGS, KINGFISH and G11 samples respectively. The colours differentiate between H$_2$ detections (in black, corresponding to the coloured points on the bottom pannel) and the H$_2$ non-detections (in red, corresponding to the grey points in the bottom pannel).}}
\label{f:gasratio}
\end{center}
\end{figure}

The molecular gas mass is usually derived through CO measurements as H$_2$ is not directly observable. 
There are two main issues in determining the molecular gas masses. First, detection of CO in low-metallicity galaxies is challenging: sensitivity has limited CO detections to galaxies with 12+log(O/H)$\gtrsim$ 8.0 \citep[e.g.,][]{Leroy2009, Schruba2012}. The other issue in the H$_2$ mass determination is the choice of the conversion factor between CO intensities and molecular gas masses, \xco. 
Indeed the variation of this factor with metallicity is poorly constrained, and a number of studies have been dedicated to quantifying the dependence of \xco\ on metallicity \citep{Wilson1995, Israel1997,Boselli2002, Israel2003,Strong2004, Leroy2011, Schruba2012, Bolatto2013}. From a sample of 16 dwarf galaxies, and assuming a constant H$_2$ depletion timescale, \cite{Schruba2012} found a \xco\ scaling with (O/H)$^{-2}$. This relation takes into account possible ``CO-free'' gas as the \xco\ conversion factor is estimated from the total reservoir of molecular gas needed for star formation \citep{Schruba2012}.
Following Cormier et al. (A\&A in press.), we estimate the molecular gas masses from a constant \xco\ factor using the Galactic value: \xcogal\ = 2.0 $\times$ 10$^{20}$ cm$^{-2}$ (K km s$^{-1}$)$^{-1}$ \citep{Ackermann2011}, giving us \mmolgal, and from a \xco\ factor depending on (O/H)$^{-2}$: \xcoz, giving us \mmolz. This provides a conservative range of molecular gas mass estimates that reflects how uncertain the molecular gas mass determination is. For this reason, we do not give any error bar on our molecular gas mass. 

In order to go beyond the CO upper limits and to constrain the G/D behaviour at low metallicities we find a way to estimate the amount of molecular gas for the lowest metallicity galaxies. Figure \ref{f:gasratio} shows the ratio of \mmol-to-\matom\ as a function of metallicity for our sample and for both cases of \xco. We note that around 12+log(O/H) $\sim$ 8.1 the ratio \mmol/\matom\ drops suddenly for the detected galaxies for both \xco. For these very low-metallicity galaxies with 12+log(O/H) $\leq$ 8.1, 
the mean ratio between the detected \mmol\ and \matom\ is 1.2\%, for \xcogal. Using \xcoz, this ratio goes up to 68\%.
Thus for galaxies with non-detections in CO, or without any CO observations, and with 12+log(O/H) $\leq$ 8.1, we replace the upper limit values by 0.012$\times$\matom\ for \mmolgal\ and 0.68$\times$\matom\ for \mmolz. Given the low molecular gas fraction we find, this will not greatly affect our interpretation of G/D nor the conclusions. From now on, the galaxies for which we apply this correction will be treated as detections. This 12+log(O/H) value of $\sim$ 8.1 has already been noted as being special for dwarf galaxies \citep[e.g. for the strength of the PAH features:][]{Engelbracht2005, Madden2006, Draine2007, Engelbracht2008, Galliano2008}. 
The molecular gas masses we use in the following analysis are presented in Table \ref{t:Gas}.

\paragraph{Total gas masses -}

We get the total gas mass, \mgas, by adding all of the different gas contributions: the atomic gas mass, the molecular gas mass, the helium gas mass and the gaseous metal mass:

\begin{equation}
M_{gas} = M_{HI} + M_{H_2}+ M_{He} + Z_{gal}\times M_{gas},
\end{equation}

\noindent where $M_{{\rm He}}$ is the helium mass and {\it Z$_{{\rm gal}}$} the mass fraction of metals in the galaxy. Assuming $M_{{\rm He}}$= \ysun\ $M_{{\rm gas}}$, where \ysun\ is the Galactic mass fraction of Helium, \ysun\ = 0.270 \citep{Asplund2009}, we have:

\begin{equation}
M_{gas} = \mu_{gal} (M_{HI} + M_{H_2}),
\label{eq:totgas}
\end{equation}

\noindent with $\mu_{gal}$ = 1/(1-\ysun - Z$_{{\rm gal}}$) the mean atomic weight. $\mu_{gal}$ has been computed for each galaxy and the mean value for our sample is 1.38$\pm$0.01 (see Table \ref{t:Gas}). We get {\it Z$_{{\rm gal}}$} assuming (Z$_{{\rm gal}}$/\zsun)~=(O/H$_{{\rm gal}}$)/(O/H$_\odot$) and \zsun = 0.014 \citep{Asplund2009}.

We assume here that the ionised gas mass (\mion) is negligible compared to the \HI\ mass. We perform the test for 67 galaxies of the sample, with \mion\ derived from H$\alpha$ measurements of \cite{dePaz2003, Kennicutt2009, Skibba2011} and found \mion/\matom $\sim$ 0.2\%.  However, we found two dwarf galaxies for which the ionised gas mass should be taken into account as it contributes equally or more than the atomic gas mass: Haro11 \citep[\mion~$\sim$~1.2~$\times$~\matom,][]{Cormier2012} and Pox186 \citep[\mion~$\sim$~\matom,][]{dePaz2003}. For these two galaxies, the total gas mass also includes \mion.

\begin{figure*}
\begin{center}
\begin{tabular}{ p{8cm}p{8cm}}
    {\LARGE a)} & {\LARGE b)} \\
    \includegraphics[width=8.5cm]{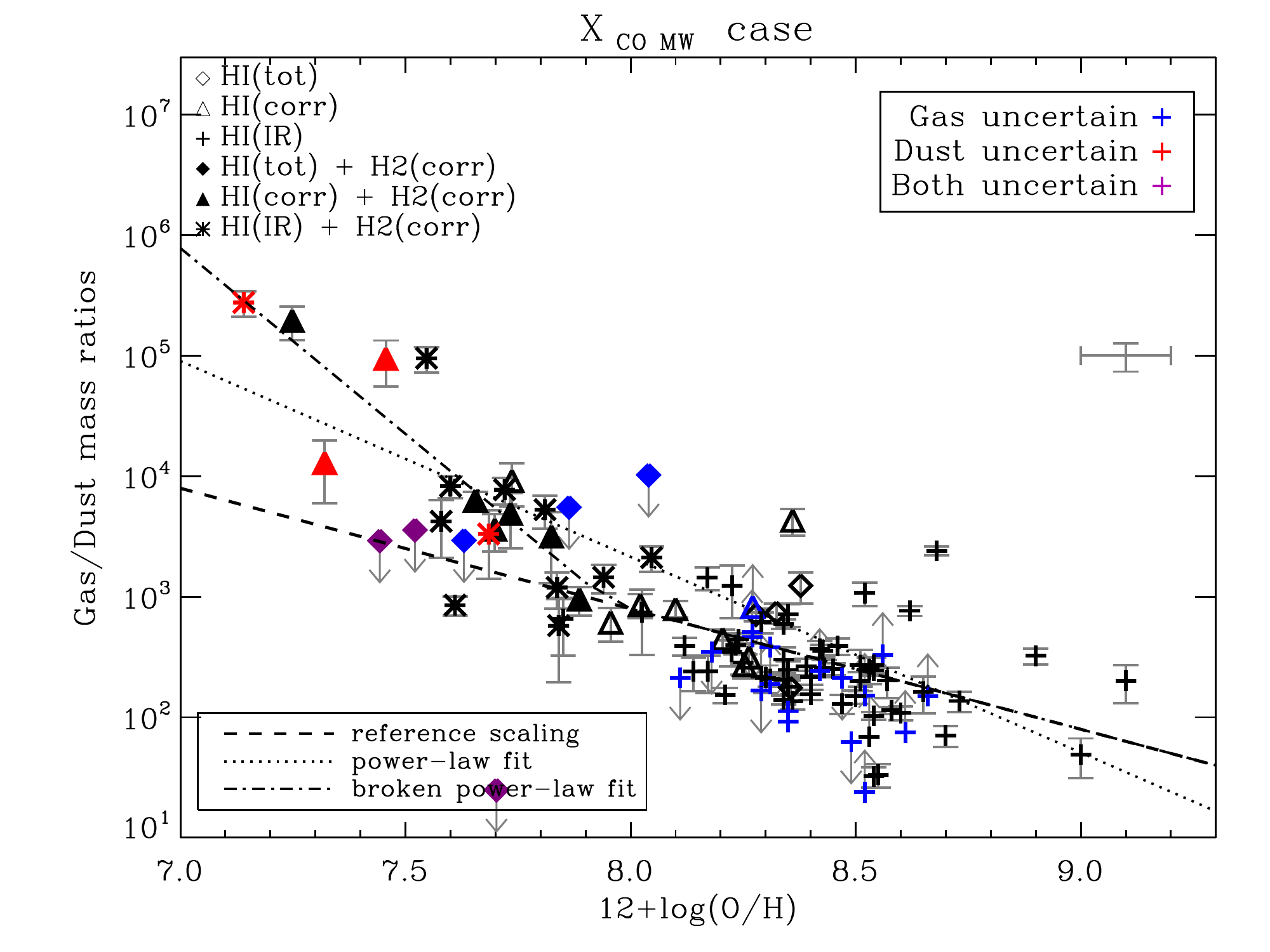} &
   \includegraphics[width=8.5cm]{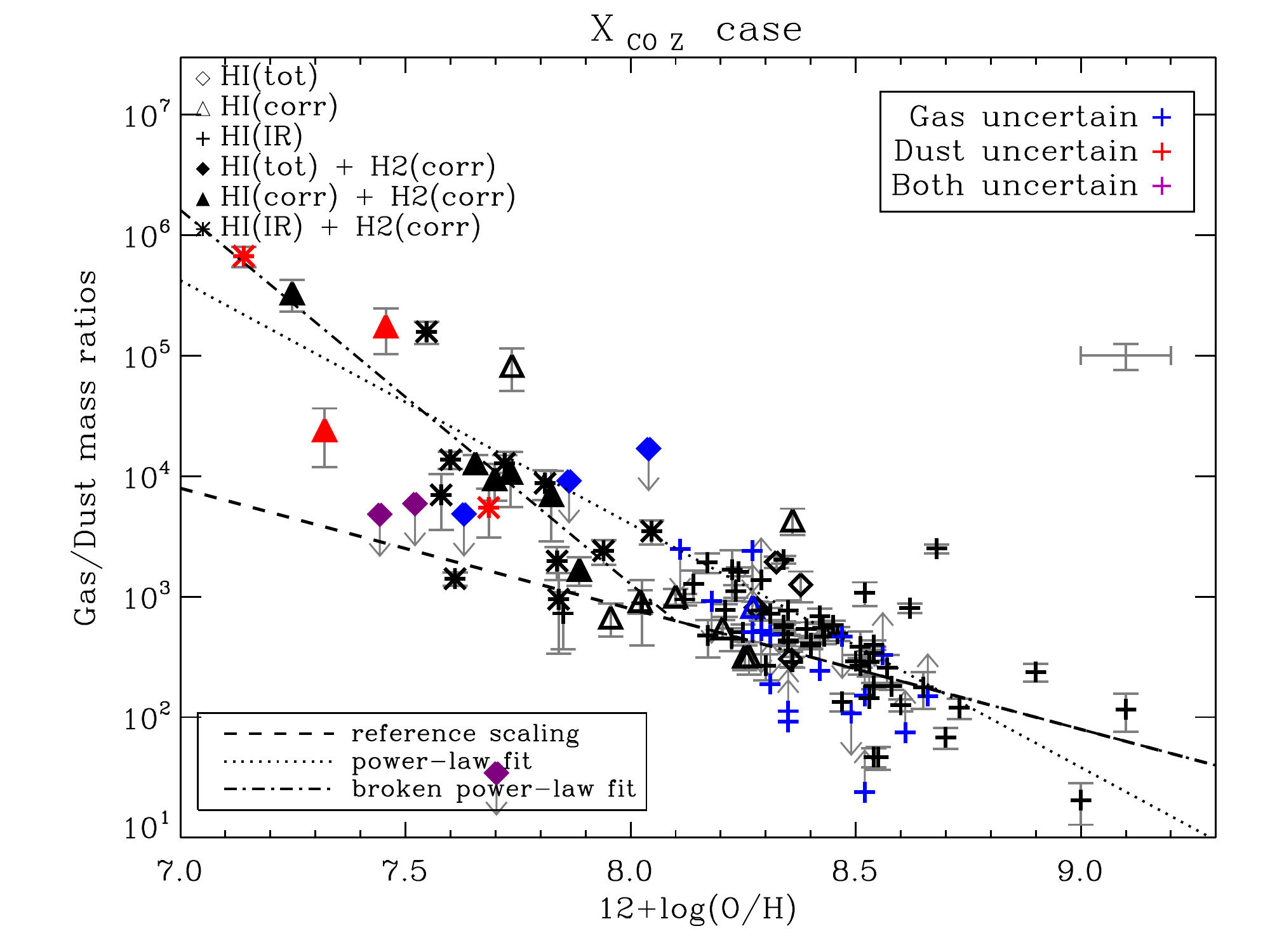} \\
    {\LARGE c)} & {\LARGE d)} \\ 
    \includegraphics[width=8.5cm]{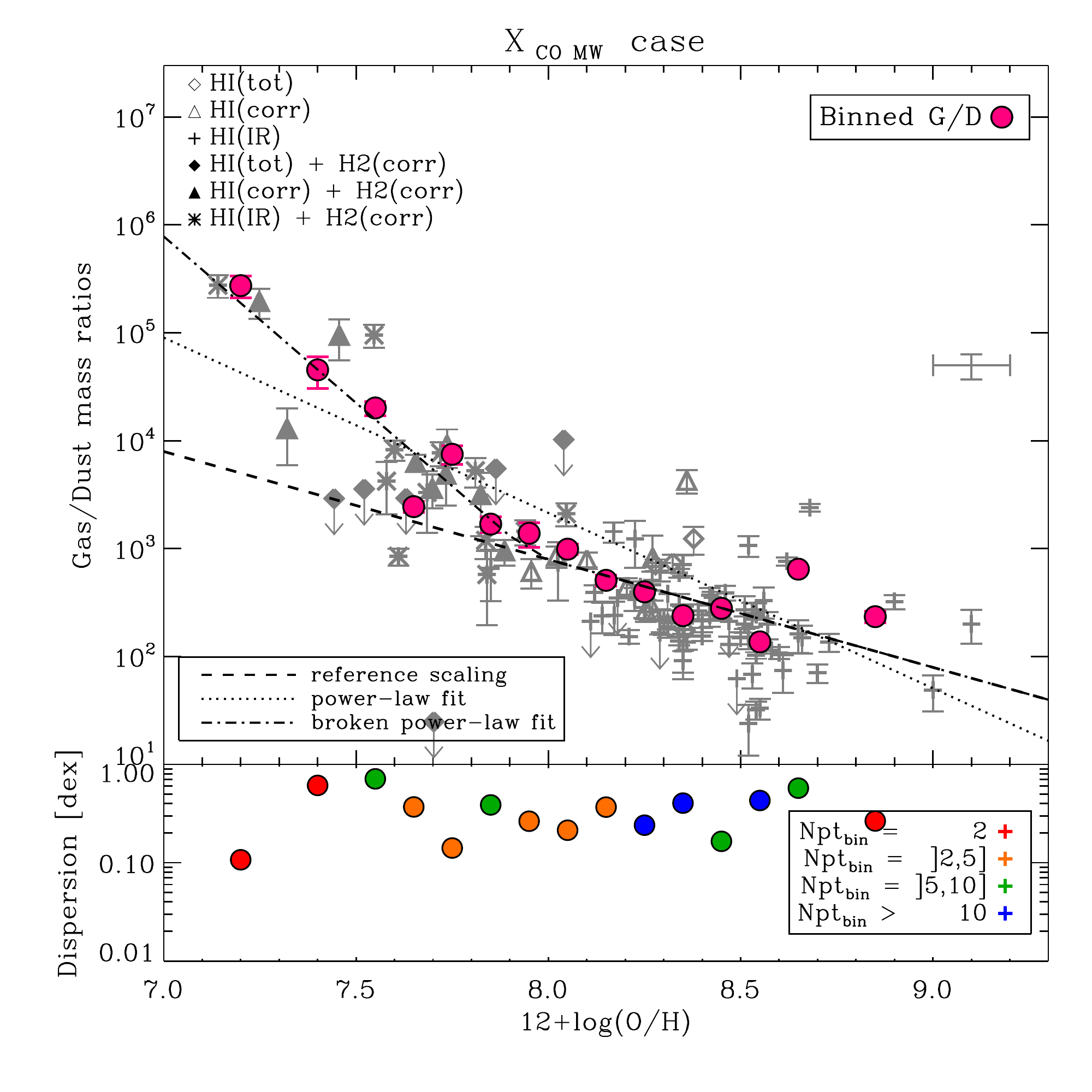} &
    \includegraphics[width=8.5cm]{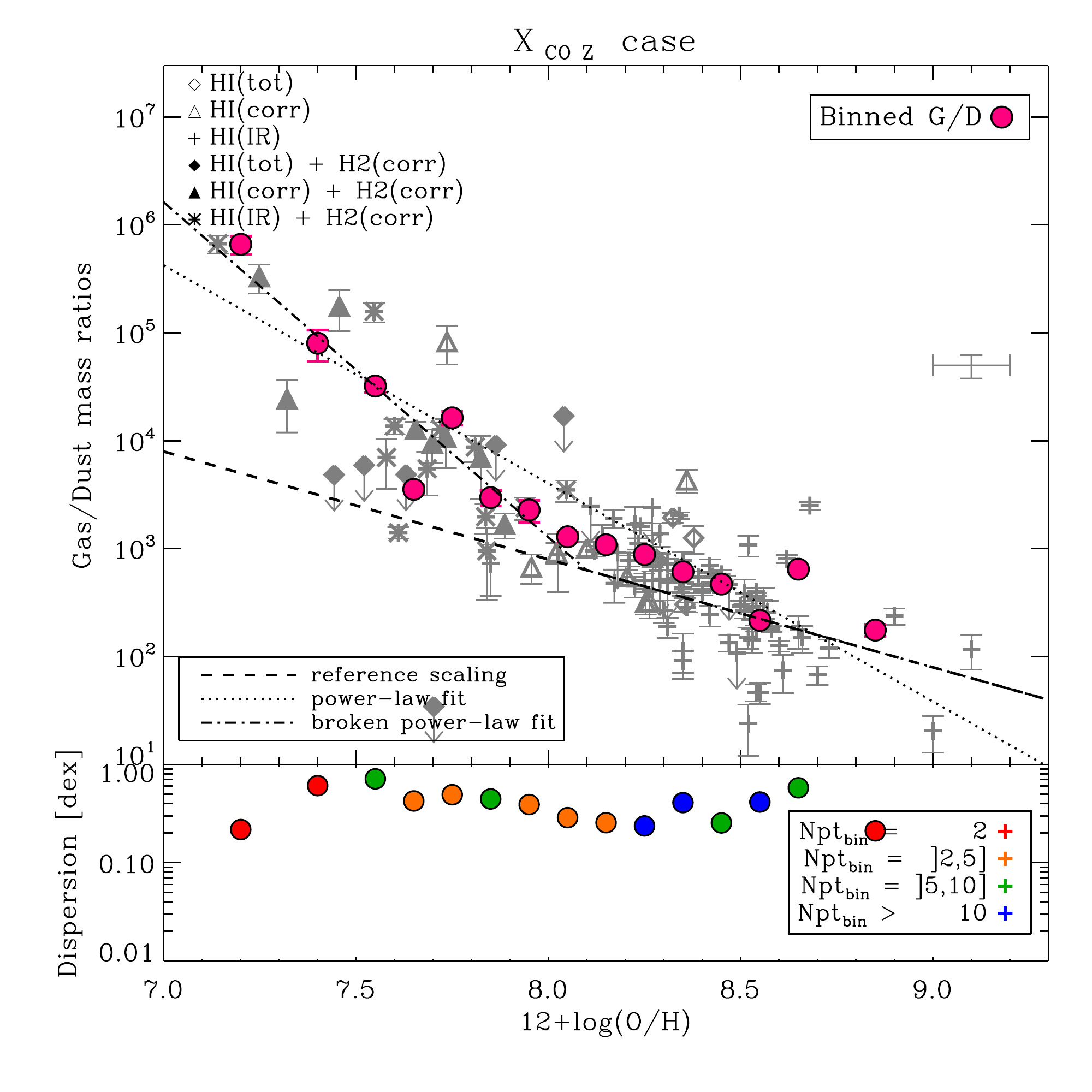} \\ 
 \end{tabular} 
\caption{{\it {\large (Top row):}} G/D as a function of metallicity for the 2 values of \xco: \xcogal\ ({\it a}) and \xcoz\ ({\it b}). The mean error for the data points is shown in grey on the right of the plots. The colours code the reliability of the point depending if the gas mass is uncertain (in blue), the dust mass is uncertain (in red) or if both are uncertain (in purple). The symbol traces the changes made in the \HI\ and H$_2$ masses (see text for details on the uncertainties and the changes on the gas masses). The dashed line represents the reference scaling of the G/D with metallicity (not fit to the data). The dotted and dash-dotted lines represent the best power-law and best broken power-law fits to the data.
{\it {\large (Bottom row):}} Same as top row for \xcogal\ ({\it c}) and \xcoz\ ({\it d}), where the binned G/D values (see text) have been added as pink filled circles. For clarity, the observed G/D values are now shown in grey. On the bottom panels the relative dispersion in each bins, in terms of standard deviation, is shown and the colours show the number of galaxies in each bin.
}
\label{G/Dline}
\end{center}
\end{figure*}

The G/D as a function of metallicity is presented in Fig.~\ref{G/Dline}~a\&b for the two cases: \xcogal\ or \xcoz. The average error on the observed G/D is $\sim$ 27\% in both \xco\ cases ($\sim$ 10\% for the total gas mass and $\sim$26\% for the dust mass). The dashed line indicates the reference scaling of the G/D with metallicity. The colours of the symbols indicate the reliability of the data points by tracing if the 
gas or dust masses determinations are uncertain. Blue symbols refer to \HI\ or H$_2$ non-detections or to the absence of H$_2$ observations for the galaxy. 
Red symbols indicate that the galaxy is not detected at wavelengths $\geq$ 160 \mic. The combination of both indications for the gas and dust masses is shown with the purple symbols. Black symbols indicate that both gas and dust masses have reliable measurements (67\% of the sample). 

The type of symbols indicates whether or not the \HI\ and/or H$_2$ masses have been corrected. For the \HI\ masses we distinguish three cases for the DGS galaxies: the \HI\ extent of the galaxy is unknown and we cannot correct the \HI\ mass (diamonds), the \HI\ extent is known and greater than the dust aperture and we correct the \HI\ mass (triangles) and the \HI\ extent is known and similar to the dust aperture, there is no need to correct the \HI\ mass (crosses). 
The galaxies with 12+log(O/H) $<$ 8.1 for which the H$_2$ masses have been corrected (either from upper limit or lack of measurements) are indicated as filled symbols (see paragraph above on H$_2$ masses).


\section{Analysis}\label{sec:analysis}

\subsection{Observed gas-to-dust mass ratio - metallicity relation and dispersion}\label{ssec:Obs}
To evaluate the general behaviour and scatter in the G/D values at different metallicities, we consider the error-weighted mean values of log(G/D) in metallicity bins (neglecting the upper/lower limits), 
with the bin sizes chosen to include at least two galaxies and to span at least 0.1 dex.
The result is overlaid as pink filled circles in Fig. \ref{G/Dline} c\&d. 
We also look at the dispersion of the G/D values in each metallicity bin (see bottom panels of Fig. \ref{G/Dline} c\&d), by computing the standard deviation of the log(G/D) values in each bin (also neglecting the upper/lower limits). 
The dispersion is $\sim$ 0.37 dex (i.e. a factor of 2.3) on average for all bins and for both \xco\ values. Additionally, in one bin the G/D vary on average by one order of magnitude. This confirms that the relation between G/D and metallicity is not trivial even at a given metallicity, and over the whole metallicity range. We also see that the dispersion in the observed G/D values does not depend on metallicity. 
This indicates that the scatter within each bin may be intrinsic and does not reflect systematic observational or correction errors. This also means that the metallicity is not the only driver for the observed G/D: other processes operating in galaxies can lead to large variation in the G/D values in a given metallicity range, throughout this range. However there might be a selection bias in our sample. Indeed our sample is mainly composed of star-forming gas-rich dwarf galaxies {\revised at low metallicities and spiral galaxies at high metallicity (see Fig. \ref{f:histosMorMsSFR}).} Thus we could wonder if gas-poor dwarf galaxies would show different, possibly lower, G/D than that observed in gas-rich dwarfs, thus possibly increasing the observed scatter at low metallicities. {\revised On the high-metallicity side, \cite{Smith2012a} showed that the 30 elliptical galaxies detected with \hersc\ in the \hersc\ Reference Survey \citep[HRS,][]{Boselli2010a} had a mean G/D of $\sim$ 120, which is slightly lower than what we find for our elliptical and spiral galaxies at moderate metallicities (see Fig. \ref{f:GDvsparam}, 150 - 300 on average for \xcogal\ and 270 - 500 on average for \xcoz). However the dust masses were estimated via a modified blackbody model, thus we will not go deeper into any further comparison. Nonetheless including more elliptical galaxies might also slightly increase the scatter at high metallicities.}

\subsection{Gas-to-dust mass ratio with other galactic parameters}
{\revised In this paragraph we want to see how the G/D depends on other galactic properties, namely the morphological type, the stellar mass and the star formation rate. The distribution for each of these parameters for the sample is presented in Appendix \ref{ap:galparam}. We perform the test by looking at the variation of the G/D as a function of these three parameters and the results are shown in Fig. \ref{f:GDvsparam}, where the galaxies are color coded by their metallicity. 

For the morphological types, the ``normal" type galaxies (i.e., elliptical and spirals) have lower G/D than irregular (dwarfs) galaxies or BCDs (Fig. \ref{f:GDvsparam}). As for the other two parameters, we found each time a correlation: galaxies with higher stellar masses or higher star formation rates have lower G/D than less massive or less active galaxies (Fig. \ref{f:GDvsparam}). However the correlation is weaker than with the metallicity: we have Spearman rank coefficients\footnote{The Spearman rank coefficient, $\rho$, indicates how well the relationship between X and Y can be described by a monotonic function: monotonically increasing: $\rho$ $>$ 0, or monotonically decreasing: $\rho$ $<$ 0.} $\rho \sim$-0.30 and -0.25 between G/D and the stellar masses and the star formation rates respectively, versus $\rho \sim$ -0.45 with the metallicity. 
On the absolute scale dwarf galaxies have lower stellar masses and lower absolute star formation rate than their metal-rich counterparts. The specific star formation rate (SSFR), defined as the SFR divided by the stellar mass, is more representative of the intrinsic star formation activity of the galaxy. When looking at the G/D as a function of SSFR, we find a even weaker correlation ($\rho \sim$ 0.16) between these two quantities (see Fig. \ref{f:GDvsSSFR}). 

As all of these parameters are themselves related to the metallicity of the galaxy, the observed weaker correlations are thus ``secondary'' correlations, resulting from the correlation between the metallicity and the other galactic parameters. This means that, as far as these five parameters are concerned (metallicity, stellar mass, SFR, SSFR and morphological type), metallicity is the fundamental parameter driving the observed G/D values. Thus in the following, we focus only on the relation between G/D and metallicity.
}

\subsection{Empirical relations and scatter}\label{ssec:Lines}
To investigate the variation of the G/D with metallicity, we first fit a power-law (dotted line in Fig. \ref{G/Dline}) through the observed G/D values (excluding the limits): G/D $\propto$ $(O/H)^\alpha$. 
The fit is performed with the IDL procedure {\sc mpfit}\footnote{http://www.physics.wisc.edu/~craigm/idl/idl.html} and is shown in Fig. \ref{G/Dline}. The fit is weighted by the individual errors bars of the G/D values and the number density of points to avoid being dominated by the more numerous high-metallicity galaxies. 
We get a slope for the power-law of $\alpha$ = -1.6 $\pm$ 0.3 for \xcogal\ and $\alpha$ = -2.0 $\pm$ 0.3 for \xcoz. In both cases, $\alpha$ is lower than -1, which corresponds to the slope of the reference relation. 

We also fit a broken power-law (dash-dotted line in Fig. \ref{G/Dline}), with two slopes $\alpha_L$ and $\alpha_H$ to describe the low- and high-metallicity slopes respectively, and with a transition metallicity, $x_t$, between the two regimes. 
Several studies \citep[e.g.][]{James2002, Draine2007, Galliano2008, Leroy2011} have shown that the G/D was well represented by a power-law with a slope of -1 at high-metallicities and down to 12+log(O/H) $\sim$ 8.0 - 8.2, and thus we fix $\alpha_H$ = -1. This gives us a low-metallicity slope, $\alpha_L$, of -3.1 $\pm$ 1.8 with a transition metallicity of 7.96 $\pm$ 0.47 for \xcogal\ and $\alpha_L$ = -3.1 $\pm$ 1.3 and a transition around a metallicity of 8.10 $\pm$ 0.43 for \xcoz. The low metallicity slopes, $\alpha_L$, are also for both cases lower than -1. 
The parameters for the different empirical relations are given in Table \ref{t:param}.


\begin{table}[h!tbp]
\begin{center}
\caption{Parameters for the three empirical relations between the G/D and metallicity: power-law (slope of -1 and free) and broken power-law for the two \xco\ values.}
\label{t:param}
 \begin{tabular}{lcc}
 \hline
 \hline
                   Parameters &         \xcogal\ case &            \xcoz\ case \\

\hline                                                                     
\multicolumn{3}{c}{{\bf Power law, slope fixed : y = a + (x$_\odot$ - x) (``reference" scaling)}} \\                                                  
\hline                                                                     
a$^{1,2}$                                &          2.21 &          2.21 \\
average logarithmic distance$^3$ [dex]   &          0.07 &          0.27 \\

\hline                                                                     
\multicolumn{3}{c}{{\bf Power-law, slope free : y = a + $\alpha$ (x$_\odot - x)$}} \\                                                                 
\hline                                                                     
a$^{1,2}$                                &          2.21 &          2.21 \\
$\alpha$                                 & 1.62$\pm$0.34 & 2.02$\pm$0.28 \\
average logarithmic distance$^3$ [dex]   &         -0.21 &         -0.19 \\

\hline                                                                     
\multicolumn{2}{c}{{\bf Broken power-law :}} & \\                                         
\multicolumn{3}{c}{{\bf y = a + $\alpha_H$ (x$_\odot - x)$ for x $>$ x$_t$}} \\           
\multicolumn{3}{c}{{\bf y = b + $\alpha_L$ (x$_\odot - x)$ for x $\leq$ x$_t$}} \\        
\hline                                                                     
a$^{1,2}$                                &          2.21 &          2.21 \\
$\alpha_H^1$                               & 1.00   & 1.00 \\
b                                        &          0.68 &          0.96 \\
$\alpha_L$                               & 3.08$\pm$1.76 & 3.10$\pm$1.33 \\
x$_t$                                    & 7.96$\pm$0.47 & 8.10$\pm$0.43 \\
average logarithmic distance$^3$ [dex]   &         -0.06 &          0.06 \\
\hline                                                                     
\end{tabular}
\end{center}
{\footnotesize
\noindent
{\bf Notes.} y = log(G/D), x = 12+ log(O/H) and x$_\odot$ = 8.69. 

\noindent
$^1$: Fixed parameters.

\noindent
$^{2}$: This correspond to the solar G/D: G/D$_\odot$ = 10$^a$ = 162 \citep{Zubko2004}.

\noindent
$^{3}$: Derived for all of the individual galaxies, neglecting the upper/lower limits on the G/D.
}

\end{table}

If we let $\alpha_H$ free in the broken power-law fit, we get similar results within errors for $\alpha_L$ and $x_t$ in both \xco\ cases. 
We get $\alpha_H$= -0.5 $\pm$ 0.9 and $\alpha_H$= -1.6 $\pm$ 0.6, for \xcogal\ and \xcoz\ respectively, which is coherent with a slope of -1 within errors. 
Note that we have imposed here that our fits go through the solar G/D determined by \cite{Zubko2004}. If we relax this condition (i.e. do not fix our ``a'' parameter in Table \ref{t:param}), we get values of the solar G/D ranging from (G/D)$_\odot$ = 90 to 240 within $\sim$ 60\% of the value from \cite{Zubko2004}.

\begin{figure*}[h!tbp]
\begin{center}
\includegraphics[width=8.8cm]{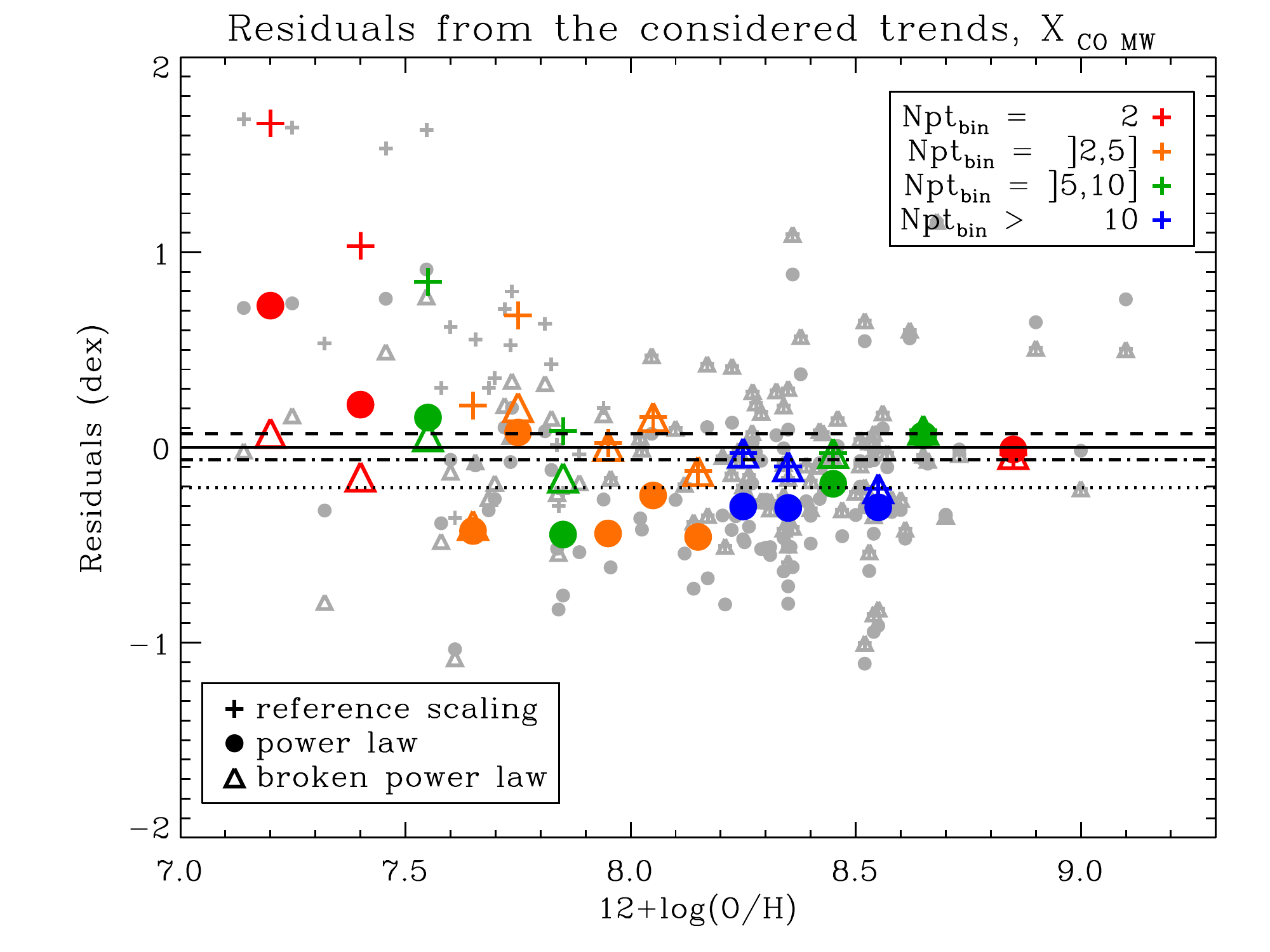}
\includegraphics[width=8.8cm]{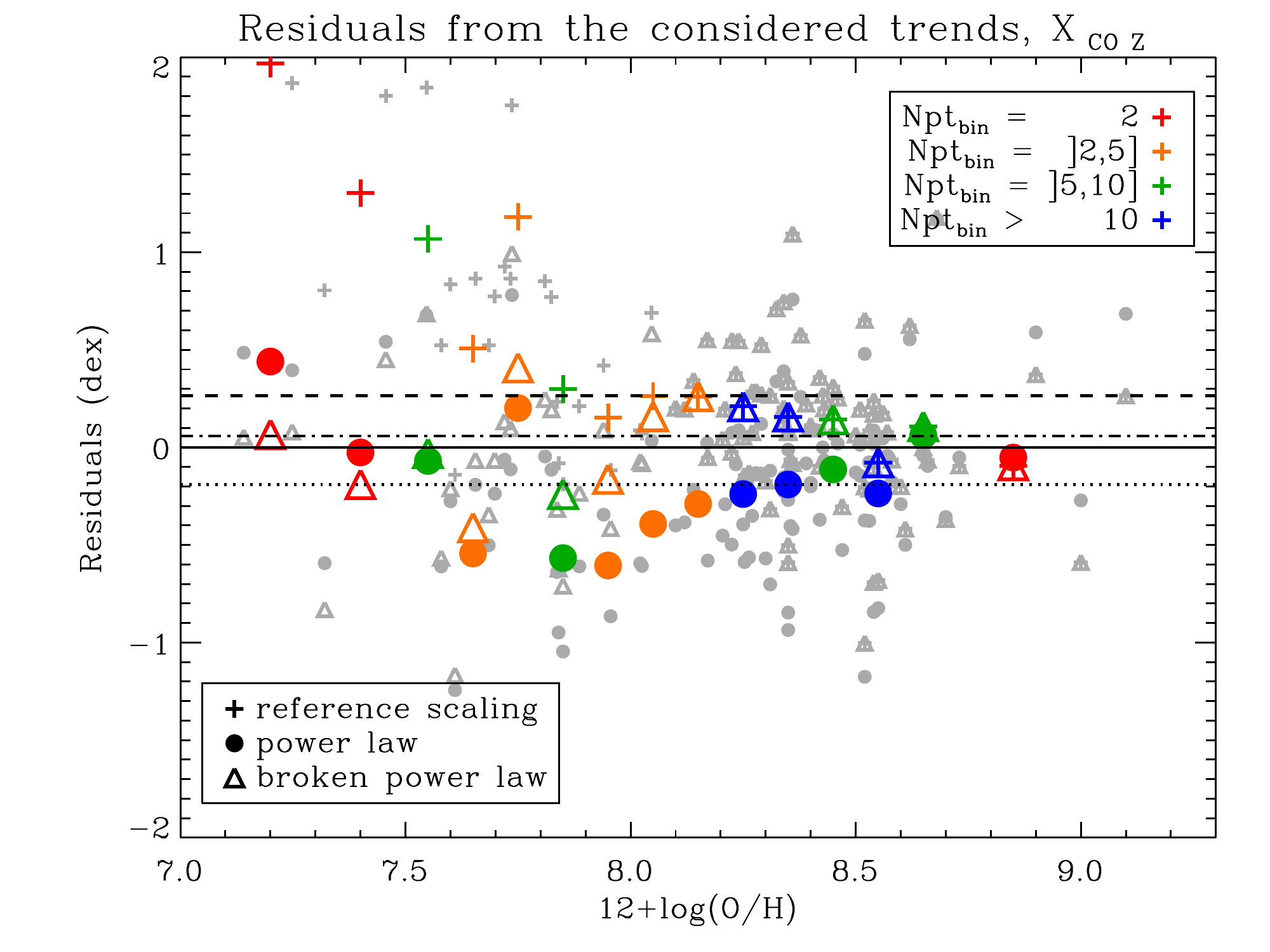}
\caption{Residuals (i.e. logarithmic distance) between the observed (and detected) G/D and predicted G/D for the three relations for \xcogal\ ({\it left}) and \xcoz\ ({\it right}): reference scaling of the G/D with metallicity (crosses), the best power-law fit (filled circles) and the best broken power-law fit (triangles). These residuals are shown in grey for the individual galaxies and in colour for the average residuals in each metallicity bin defined in Section \ref{ssec:Obs}. The colours show the number of galaxies in each bin.
The mean residual for all of the observed G/D values is shown by the dashed (reference scaling), dotted (power-law fit) and dash-dotted (broken power-law fit) lines for the three relations and are reported in Table \ref{t:param}.}
\label{G/Dresid}
\end{center}
\end{figure*}

Now we consider the deviation from each relation by looking at the logarithmic distance from the observed G/D values and the G/D values predicted by each of the three relations presented in Table \ref{t:param}. This is a way to look at the residuals from the two fits and the reference scaling, even though we did not actually fit the reference trend to the G/D values. These residuals are shown in Fig. \ref{G/Dresid}. Average residuals in each metallicity bin defined previously are also computed. For a given point, the best relation is the one giving the residual closest to zero.
From Fig.  \ref{G/Dresid} we have another confirmation that a reference scaling of the G/D with metallicity does not provide reliable estimates of the G/D at low metallicities 12+log(O/H) $\lesssim$ 8.4. 
We also note that, for the average residuals, the broken power-law gives the residuals that are the closest to zero for nearly all the metallicity bins in both \xcogal\ and \xcoz\ cases, and that this corresponds to a predicted G/D uncertain to a factor of 1.6.

Even though 30\% of our sample have metallicities below 1/5 \zsun, only seven galaxies have 12+log(O/H) $\leq$ 7.5 with two of them not detected in \HI\ (SBS1159+545 and Tol1214-277). The remaining five galaxies (IZw18, HS0822+3542, SBS0335-052, SBS1415+437 and UGC4483) are important constraints for the broken power-law fit. These five galaxies all present broad dust SEDs peaking at very short wavelengths ($\sim$ 40 \mic, and $\sim$ 70 \mic\ for HS0822+3542), indicating overall warmer dust with a wide range of dust grain temperatures, and subsequently very low dust masses; hence their high G/D. This peculiar SED shape had already been noted by \cite{RemyRuyer2013a}. 

Using \hersc\ data and a semi-empirical SED model, \cite{Sandstrom2012} looked at the G/D in a sub-sample of 26 KINGFISH galaxies, mostly spirals. They simultaneously derive \xco\ and G/D for their sample, taking advantage of the high spatial resolution of the KINGFISH gas and dust data. They found that the G/D for these galaxies follows the reference trend with the metallicity and shows small scatter. Their metallicity range is from 12+log(O/H) $\sim$ 8.1 to 8.8 and thus these results are in agreement with our findings. Moreover the small scatter (less than a factor of 2) observed by \cite{Sandstrom2012} can be due to the fact that they are probing very similar environments. In our case we have a wide variety of morphological types represented in our sample, that results in a larger scatter (a factor of $\sim$5 and 3 for \xcogal\ and \xcoz\ respectively for this metallicity range).

\subsection{Discussion}\label{ssec:discussion}

In the previous section, we have shown that the reference scaling relation between metallicity and G/D derived for metallicities above 12+log(O/H) $\sim$ 8.0 does not apply to objects with lower metallicity. We empirically derived a new scaling relation better described by a broken power-law with a transition metallicity around 12+log(O/H) $\sim$ 8.0, which confirms the importance of this value in low-metallicity dwarf galaxies.
As mentioned in the Introduction, this reference scaling relation arises from the hypothesis that the dust formation timescale and the dust destruction timescale behave similarly with time. Thus a possible interpretation of our results would be that the balance between formation and destruction of dust grains is altered at low metallicity, resulting in the observed steeper trend. 
Dwarf galaxies are subject to an overall harder ISRF than more metal-rich environments \citep{Madden2006}. The harder UV photons travel deeper into the ISM and photoprocess dust in much deeper regions in the clouds limiting the accretion and coagulation of the grains. The hard ISRF also 
affects the dust survivability in such extreme environments, especially carbonaceous dust: the dust destruction by hard UV photons is enhanced in low-metallicity galaxies for small carbon grains \citep[e.g.][]{Pilleri2012, Bocchio2012, Bocchio2013}. In dwarf galaxies, dust destruction by SN shocks is enhanced too compared to larger scale galaxies, as most of the ISM can be affected by the shock due to the small physical size of the dwarfs and the globally lower density of the ISM. 

In the following paragraphs we discuss the impact of several assumptions made to estimate the G/D on our results: the dust composition, the choice of the radiation field for the dust modelling, and the potential presence of a submm excess in some of our dwarf galaxies. 


\paragraph{Dust composition -} \cite{Galliano2011} demonstrated that a more emissive dust grain composition compared to that of the Galaxy, {\revised using amorphous carbon instead of graphite for the carbonaceous grains, }is more consistent for the low-metallicity Large Magellanic Cloud (LMC). This result has been confirmed by \cite{Galametz2013a} in a star-forming complex of the LMC with an updated version of the SPIRE calibration\footnote{This updated SPIRE calibration from September 2012 had the effect of decreasing the SPIRE flux densities by about 10\% compared to the \cite{Galliano2011} study.}. Changing accordingly the dust composition in our low-metallicity galaxies would give lower dust masses (by a factor of $\sim$ two to three in the case of the LMC) with more emissive dust grains and would also increase the G/D by the same factor, increasing the discrepancy at low metallicities between the observed G/D and the predicted G/D from the reference scaling relation. \\ 


\paragraph{Radiation field -} In the dust modelling, we use an ISRF with the spectral shape of the Galactic ISRF for all of our galaxies for consistency. However, the ISRF in low-metallicity dwarf galaxies is harder, so we could wonder if this spectral shape is appropriate for the modelling of dwarf galaxies. The shape of the radiation field determines the emission of out-of-equilibrium small grains. Increasing the hardness of the radiation field increases the maximum temperature the small grains can reach when they undergo stochastic heating. However, these very small grains only have a minor contribution to the total dust mass, and thus the assumed shape of the ISRF does not bias our estimation of the total dust mass for dwarf galaxies.

\paragraph{Submm excess -} A submm excess has been observed in the past in several low-metallicity galaxies that current dust SED models are unable to fully explain \citep{Galliano2003, Galliano2005, Dumke2004, Bendo2006, Zhu2009, Galametz2009, Bot2010, Grossi2010, Galametz2011}. Several hypothesis have been made to explain this excess among which the addition of a very cold dust (VCD) component in which most of the dust mass should reside. This VCD component would be in the form of very dense clumps in the ISM \citep{Galliano2003, Galliano2005}. Taking this additional VCD component into account in the DGS and KINGFISH galaxies presenting a submm excess can result in a drastic increase of the dust mass and thus in a lower G/D. However, \cite{Galliano2011} showed that for a strip of the LMC, the submm excess is more important in the diffuse regions, possibly in contradiction with the hypothesis of very cold dust in dense clumps. 
Other studies have suggested an enhanced fraction of very small grains with high emissivity \citep{Lisenfeld2002, Dumke2004, Bendo2006, Zhu2009}, ``spinning'' dust emission \citep{YsardVerstraete2010} or emission from magnetic nano-particles \citep{Draine2012} to explain the submm excess. \cite{Meny2007} proposed variations of the optical properties of the dust with the temperature which results in an enhanced emission of the dust at submm/mm wavelengths. 
{\revised \cite{Galametz2013a} demonstrated that an amorphous carbon dust composition did not lead to any submm excess in the LMC. This alternative dust composition is thus also a plausible explanation for the submm excess. As for the discussion on the dust composition, using dust masses estimated with amorphous carbon grains for galaxies presenting a submm excess would result in an increase of a factor of $\sim$ 2 for the G/D of these galaxies.}\\

\section{Chemical evolution models}\label{sec:Chemvol}
Chemical evolution models, under certain assumptions, can predict a possible evolution of the G/D as metallicity varies in a given galaxy. For example, in the disk of our Galaxy, chemical evolution models predict this ``reference'' scaling of the G/D with the metallicity \citep{Dwek1998}. 
We consider three different models here, from \cite{Galliano2008}, \cite{Asano2013} and Zhukovska et al. (in prep.) to interpret our data. However, we have to keep in mind during this comparison that, since we do not know the ages of these galaxies and that the same metallicity can be reached at very different time by different galaxies, our sample cannot be considered as the evolution (snapshots) of one single galaxy.

\subsection{A simple model to begin with}

\begin{figure*}[h!tbp]
\begin{center}
\includegraphics[width=8.8cm]{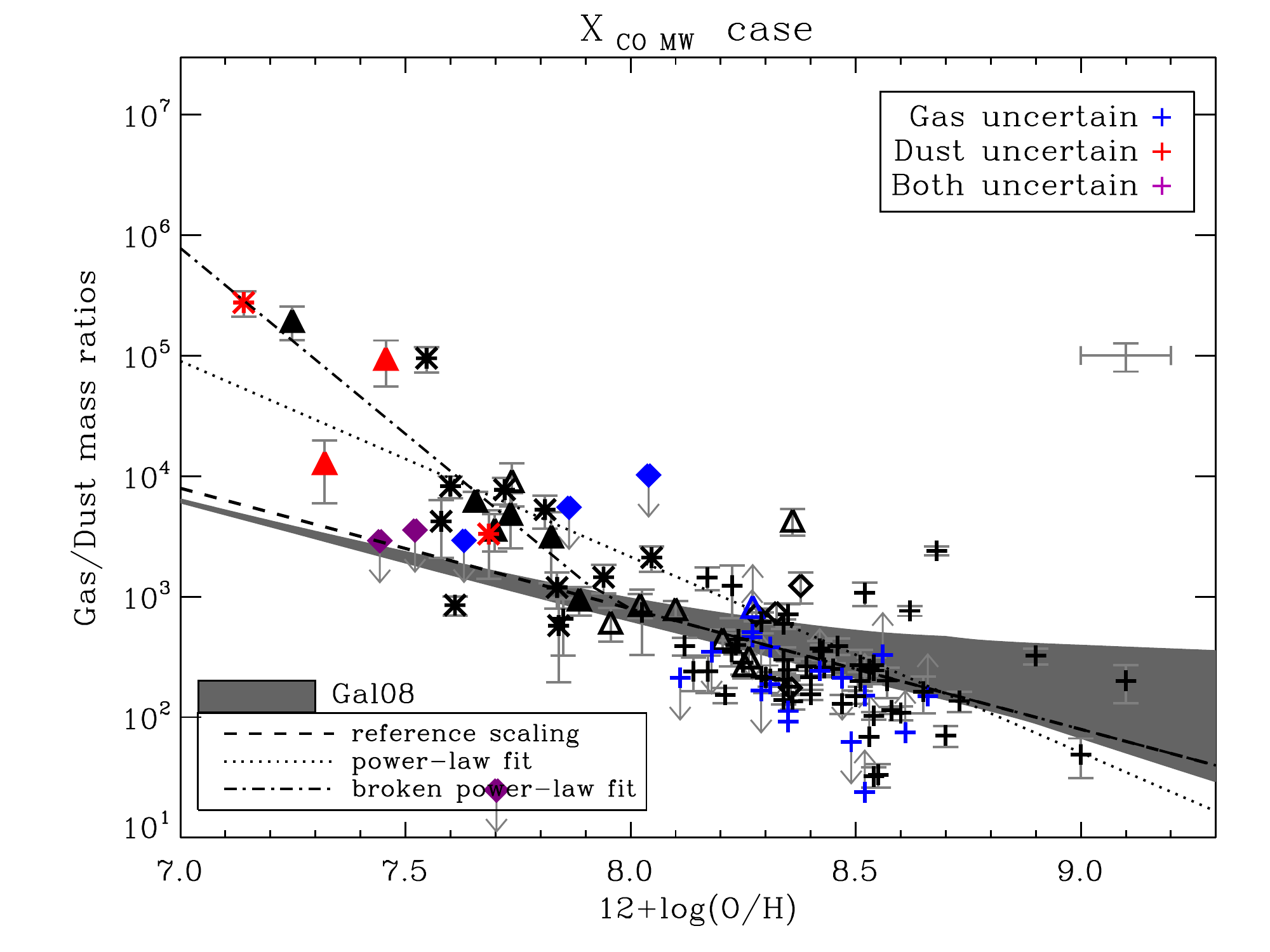}
\includegraphics[width=8.8cm]{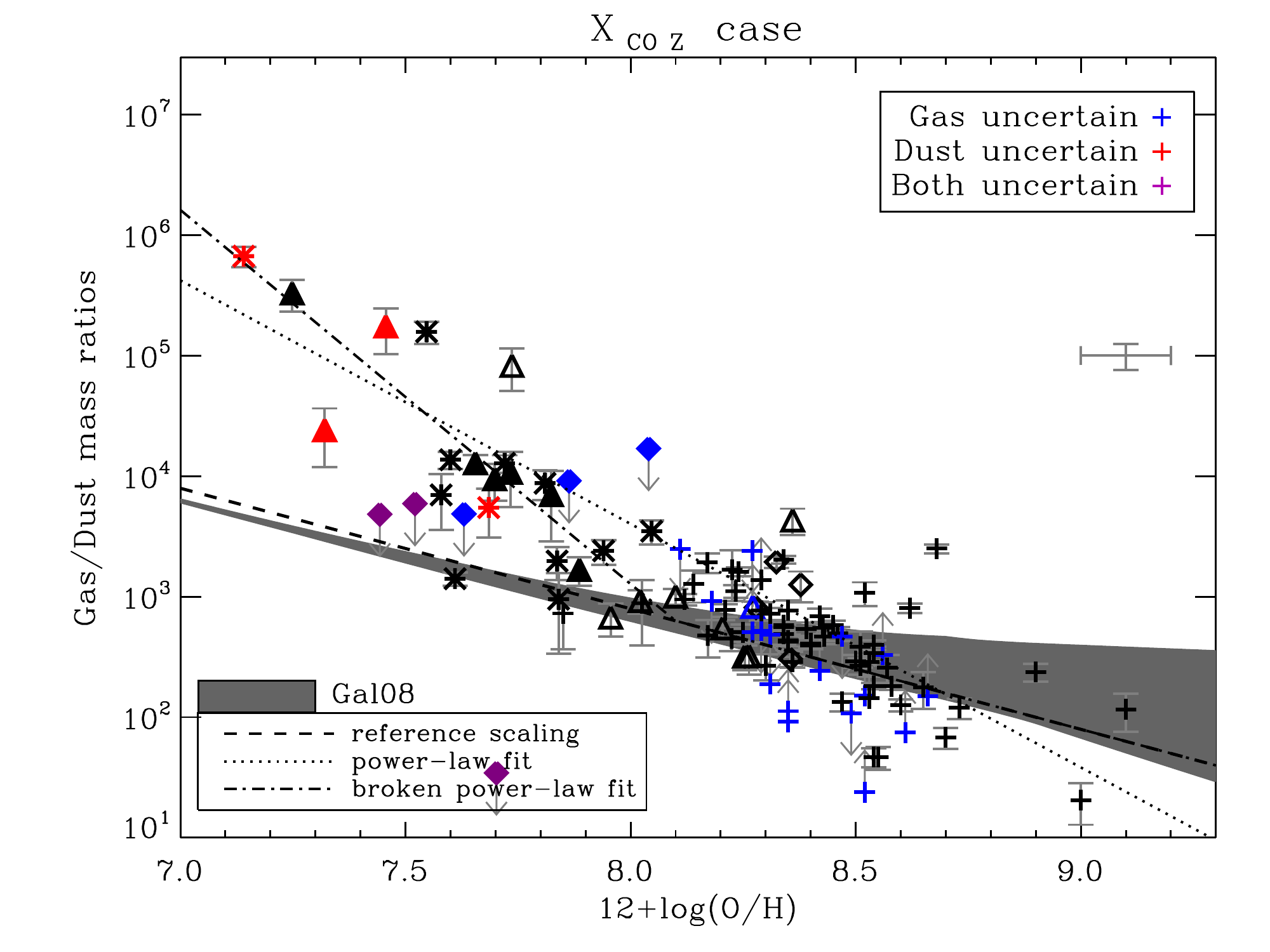}
\caption{G/D as a function of metallicity for the 2 values of \xco: \xcogal\ ({\it left}) and \xcoz\ ({\it right}) with the chemical evolution model of \cite{Galliano2008}. The colours and symbols are the same as for Fig.\ref{G/Dline}. 
The dark grey stripes show the range of values from the \cite{Galliano2008} chemical evolution model. The black dashed line represents the reference scaling of the G/D with metallicity (not fit to the data). The black dotted and dash-dotted lines represent the best power-law and best broken power-law fits to the data.}
\label{G/Dfred}
\end{center}
\end{figure*}

\cite{Galliano2008} developed a one-zone single-phase chemical evolution model, based on the model by \cite{Dwek1998}. They consider a closed-box model where the evolution of the dust content is regulated by balancing dust production by stars and dust destruction by star formation and SN blast waves. They assume the full condensation of the elements injected by Type II supernovae (SN{\sc II}) into dust and instantaneous mixing of the elements in the ISM. The model is shown on Fig. \ref{G/Dfred} for various SN destruction efficiencies as the dark grey zone. 

Two things can be noticed from Fig. \ref{G/Dfred}. First, the model is fairly consistent with the observed G/D at high metallicities within the scatter, and down to metallicities $\sim$ 0.5 \zsun. 
Second, the model does not work at low metallicities and systematically underestimates the G/D. This has already been noted by \cite{Galliano2008} for their test sample of galaxies and was attributed to the very crude assumptions made in the modelling, especially the instantaneous mixing of the SNII elements in the ISM. Another strong simplifying assumption made by \cite{Galliano2008} is that they did not take into account dust growth in the ISM as they assume full condensation of the grains. In the Galaxy, the typical timescale for dust formation by stars has been shown to be larger than the typical timescale for dust destruction \citep{JonesTielens1994, Jones1996}. Because we still observe dust in the ISM, we need to reach equilibrium between formation and destruction of the dust grains, either via high SN yields or dust growth processes in the ISM. In the following sections we thus look at models including dust growth in the ISM. 

\subsection{Including dust growth in the ISM}\label{ssec:As}
\label{ssec:asano}

\cite{Asano2013} propose a chemical evolution model, based on models from \cite{Hirashita1999} and \cite{Inoue2011}, taking into account the evolution of the metal content in the dust phase in addition to the evolution of the total amount of metals. The dust formation is regulated by asymptotic giant branch (AGB) stars, SN{\sc II} and dust growth, via accretion, in the ISM. The dust is destroyed by SN shocks. Inflows and outflows are not considered (closed-box model) and the total mass of the galaxy is constant and set to 10$^{10}$ \msun. Metallicity and age dependence of the various dust formation processes are taken into account. \cite{Asano2013} show that dust growth in the ISM becomes the main driver of the dust mass evolution, compared to the dust formation from metals produced and ejected into the ISM by stars, when the metallicity of the galaxy exceeds a certain ``critical'' metallicity. 
This critical metallicity increases with decreasing star formation timescale\footnote{The star formation timescale, $\tau_{{\rm SF}}$, is defined by the timescale during which star formation occurs: $\tau_{{\rm SF}}$=($M_{{\rm ISM}}$)/SFR, where SFR is the star formation rate \citep[see Eq. 5 of][]{Asano2013}. {\revised This is not to be mixed with the timescale, $\tau$, in exponentially decaying star formation histories going as $exp^{-(t/\tau)}$.}}. 
\cite{Asano2013} show that dust growth via accretion processes in the ISM is regulated by this critical metallicity over a large range of star formation timescales (for {\it $\tau_{{\rm SF}}$} = 0.5-5-50 Gyr). After reaching this critical metallicity the dust mass increases more rapidly, boosted by dust growth processes, before saturating when all of the metals available for dust formation are locked up in dust. The metallicity at which this saturation occurs thus also depends on the critical metallicity, which in turn depends on the star formation history of the galaxy.

\begin{figure*}[h!tbp]
\begin{center}
\includegraphics[width=8.8cm]{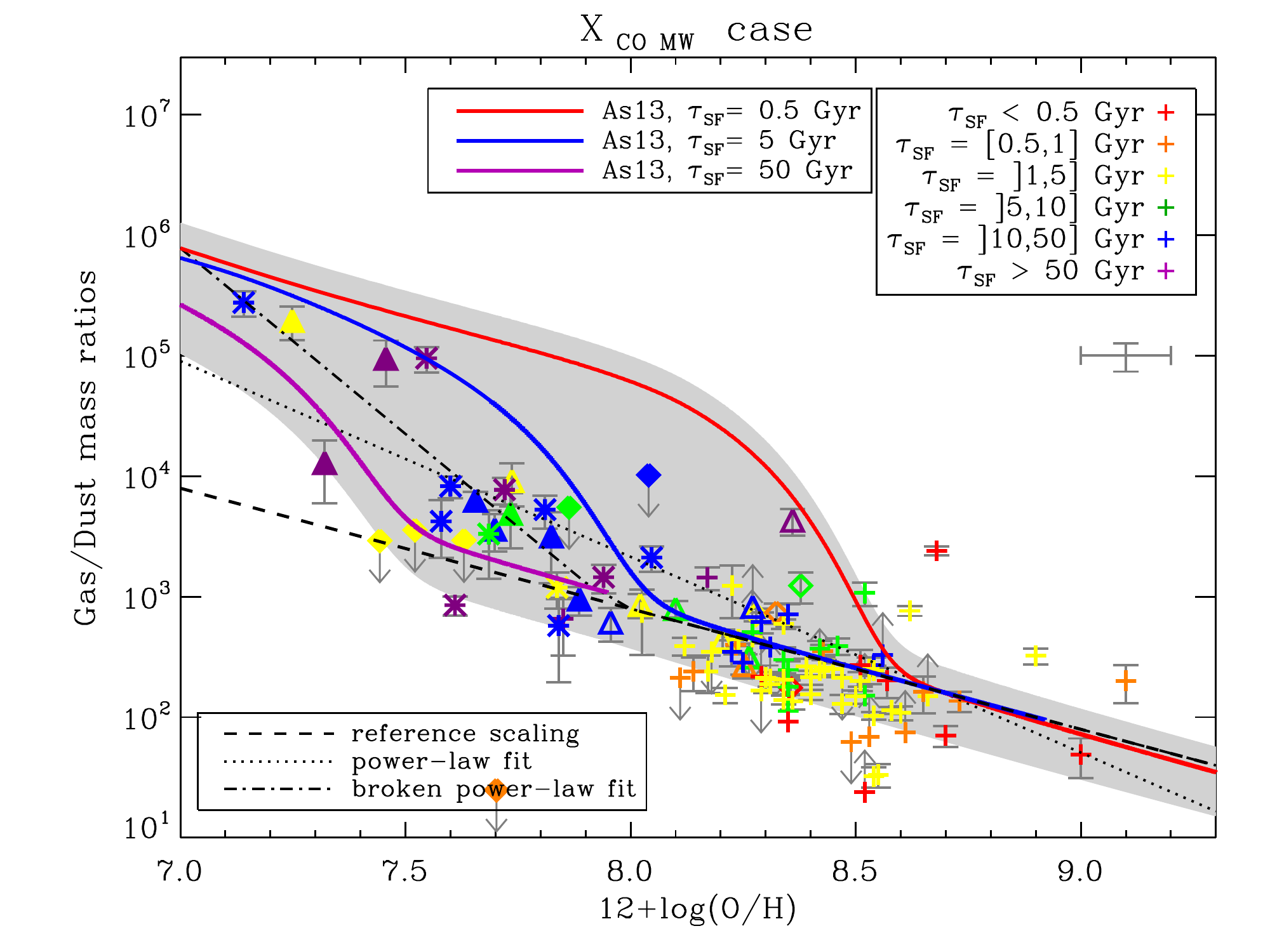}
\includegraphics[width=8.8cm]{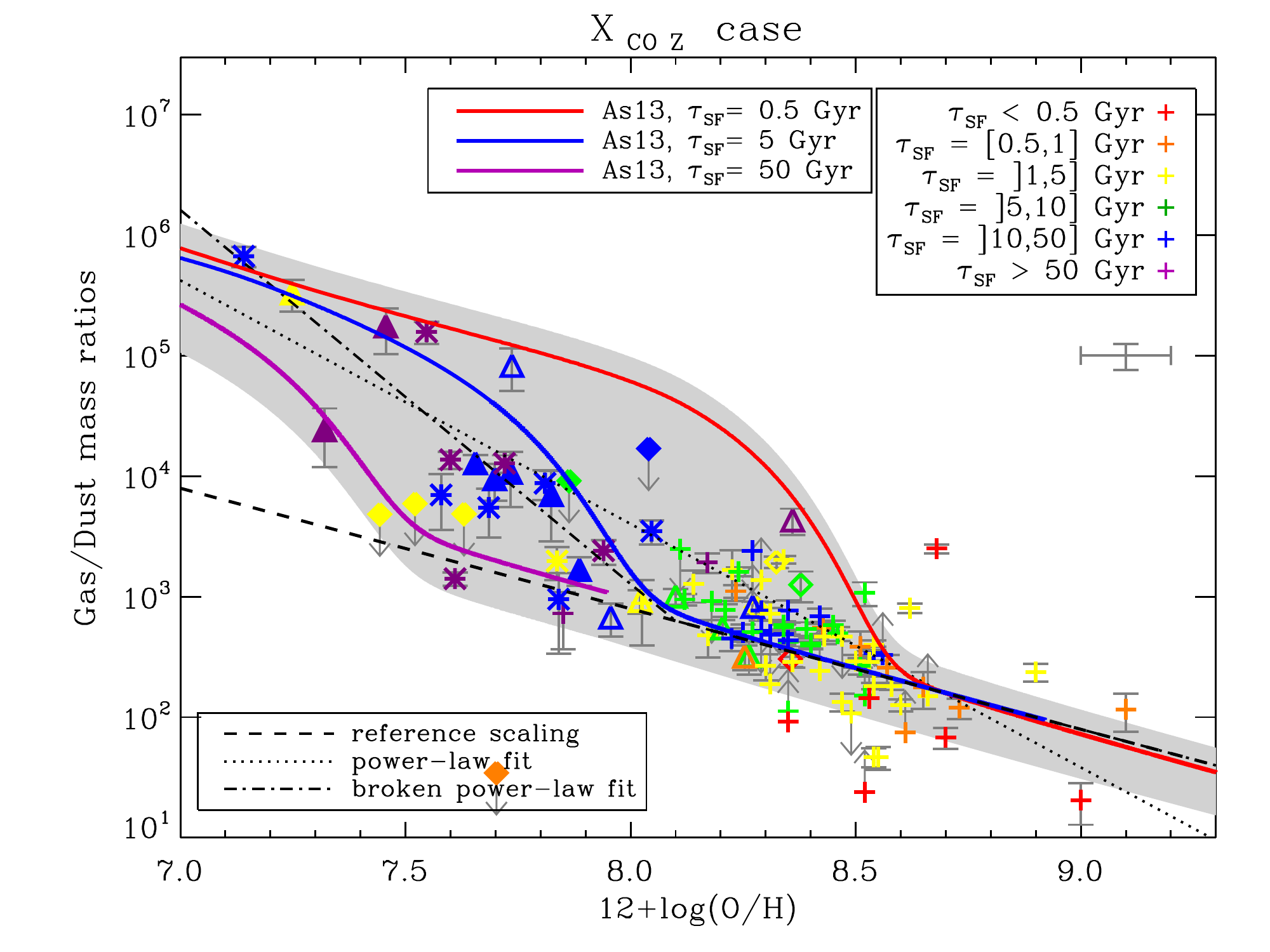}
\caption{G/D as a function of metallicity for the 2 values of \xco: \xcogal\ ({\it left}) and \xcoz\ ({\it right}) with the chemical evolution model of \cite{Asano2013}. The symbols are the same as for Fig. \ref{G/Dline}. The colours delineate ranges in star formation timescales {\it $\tau_{{\rm SF}}$}. The model from \cite{Asano2013} is overlaid on the points for various {\it $\tau_{{\rm SF}}$} = 0.5 (red), 5 (blue), 50 (purple) Gyr. The black dashed line represents the reference scaling of the G/D with metallicity (not fit to the data). The black dotted and dash-dotted lines represent the best power-law and best broken power-law fits to the data.}
\label{G/Dasano}
\end{center}
\end{figure*}

Figure \ref{G/Dasano} shows the models of \cite{Asano2013} (for {\it $\tau_{{\rm SF}}$} = 0.5-5-50 Gyr) overlaid on the observed G/D values. The models were originally on an arbitrary scale and they are thus normalised at the (G/D)$_\odot$ value. We assume an error on this value of $\sim$ 60\%, from the range of values determined from the fits in Section \ref{ssec:Lines}, to have a tolerance range around the model (shown by the shaded grey area on Fig. \ref{G/Dasano}).
The three models show similar evolution with metallicity and indeed are homologous to each other when normalised by their respective critical metallicities \citep[see Fig. 3 of][]{Asano2013}. We clearly see the influence of the critical metallicity on the dust mass evolution: at low metallicities the range of possible G/D values (illustrated by the grey area on Fig. \ref{G/Dasano}) becomes wider around 12+log(O/H) $\sim$ 7.2 -7.3 before narrowing down around 12+log(O/H) $\sim$ 8.6. This broadening is due to the fact that in this range of metallicities, galaxies with high star formation timescales have already reached their critical metallicity and have a rapidly increasing dust mass (and thus a low G/D at a given metallicity), compared to galaxies with lower star formation timescales which have not yet reached this critical metallicity and with a dust mass still regulated by stars (thus with a higher G/D at the same metallicity). Galaxies with high star formation timescales then reach saturation at moderate metallicities as they started their ``active dust growth'' phase at a lower critical metallicity (i.e. earlier in their evolution), while, at the same metallicity, galaxies with low star formation timescales are still in the ``active dust growth'' phase. 
Then when these galaxies also reach saturation, because the dust growth in the ISM becomes ineffective, the range of possible G/D values narrows down.


From Fig. \ref{G/Dasano} we see that the models from \cite{Asano2013} are consistent with the G/D from {\it both} \xco\ values.
We note that below 12+log (O/H) $\sim$ 7.5, even the {\it $\tau_{{\rm SF}}$} = 50 Gyr model does not agree anymore with the reference scaling of the G/D with metallicity (and below 12+log(O/H) $\sim$ 8.0 for {\it $\tau_{{\rm SF}}$} = 5 Gyr). 
The other two empirical relations (our best power-law and broken power-law fits) are consistent with the models of \cite{Asano2013} {\it within the considered metallicity range}: from 12+log(O/H) $\sim$ 7.0 to 9.1. Comparing with the shape of the \cite{Asano2013} models, our best broken power-law fit may overestimate the G/D for 12+log(O/H) $\leq$ 7.0. However, \cite{Izotov2012} recently suggested that there seems to be a metallicity floor around 12+log(O/H)~$\sim$ 6.9, below which no galaxies are found in the local Universe, as already proposed by \cite{Kunth1986}. Thus our metallicity range is close to being the largest achievable in the local Universe as far as low metallicities are concerned.

The galaxies from the DGS, KINGFISH and G11 samples are colour coded in Fig. \ref{G/Dasano} by an approximation of their star formation timescale {\it $\tau_{{\rm SF}}$}, estimated from {\it $\tau_{{\rm SF}}$} = (\mgas\ + \mdust)/SFR, where the star formation rates (SFR) have been estimated from L$_{TIR}$ (obtained by integrating over the modelled SEDs between 1 and 1000 \mic, R\'emy-Ruyer et al., (in prep.)). The {\it $\tau_{{\rm SF}}$} values are roughly consistent with the models from \cite{Asano2013}. The median value of {\it $\tau_{{\rm SF}}$} is $\sim$ 3.0 (\xcogal) and 5.5 (\xcoz) Gyr, but with a large dispersion of $\sim$ 20 Gyr around this value. 
As the models from \cite{Asano2013} encompass most of the observed G/D values, the dispersion seen in the G/D values can be due to the wide range of star formation timescales in the considered galaxies. This is consistent with the large dispersion in the approximated star-formation timescales in our sample. 

In \cite{Asano2013}, the star formation is assumed to be continuous over the star formation timescale. However, star formation histories of many dwarf galaxies derived from colour-magnitude diagrams reconstruction show distinct episodes of star formation separated by more quiescent phases \citep[e.g.][and references therein]{Tolstoy2009}. For example, \cite{Legrand2000} suggested for IZw18 a star formation history made of bursts of star formation in between more quiescent phases, following the suggestion of \cite{SearleSargent1972}. Episodic star formation histories have also been suggested in Nbody/Smoothed Particle Hydrodynamics simulations of dwarf galaxy evolution \citep[e.g.][]{Valcke2008, RevazJablonka2012}. As we saw that the scatter in Fig. \ref{G/Dasano} seems to be due to the range of star formation timescales probed by our sample, and assuming a continuous star formation, we thus need to consider the influence of the continuous vs. episodic star-formation modes. 

\subsection{Episodic versus continuous star formation}\label{ssec:Zhuk}

\begin{figure*}[h!tbp]
\begin{center}
\includegraphics[width=8.8cm]{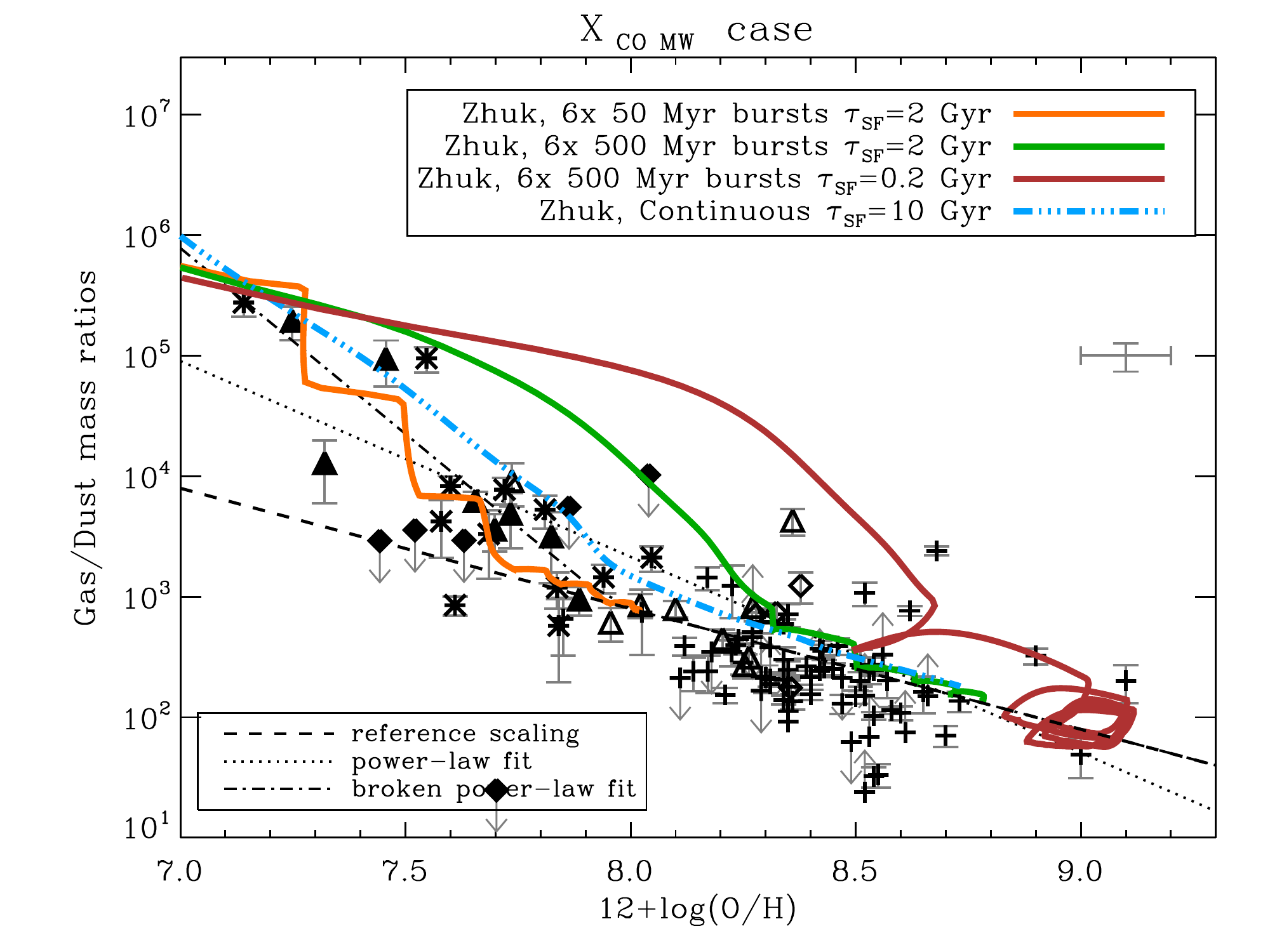}
\includegraphics[width=8.8cm]{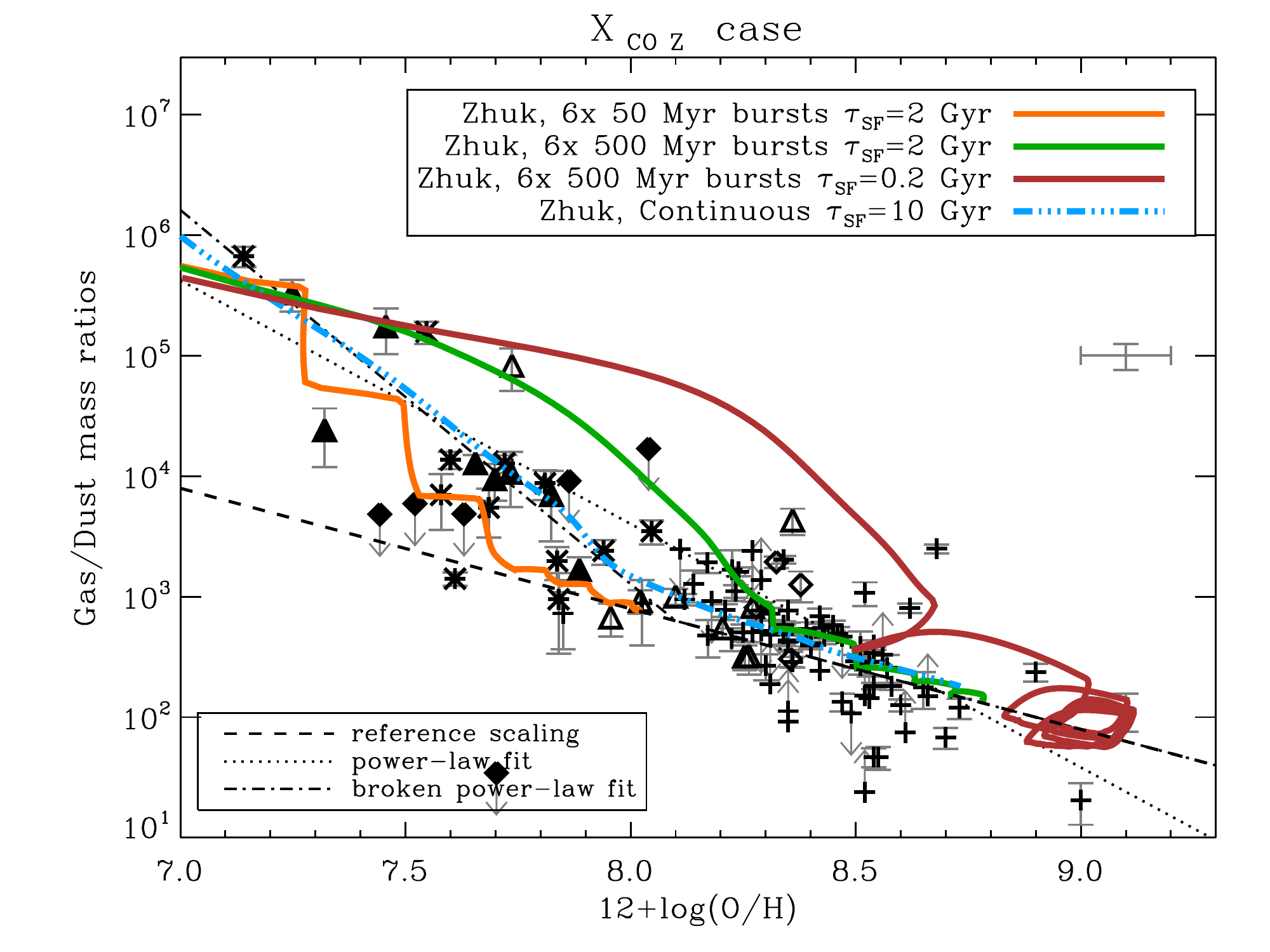}
\caption{G/D as a function of metallicity for the 2 values of \xco: \xcogal\ ({\it left}) and \xcoz\ ({\it right}) with the chemical evolution model of Zhukovska et al. (in prep.). The symbols are the same as for Fig. \ref{G/Dline}. The model from Zhukovska et al. (in prep.) is shown for various star formation histories: episodic with 6 bursts of 50 Myr and star formation timescale of {\it $\tau_{{\rm SF}}$} = 2 Gyr (orange), episodic with 6 bursts of 500 Myr and {\it $\tau_{{\rm SF}}$} = 2 Gyr (green), episodic with 6 bursts of 500 Myr and {\it $\tau_{{\rm SF}}$} = 0.2 Gyr (more intense star formation, brown) and continuous with a star formation timescale {\it $\tau_{{\rm SF}}$} = 10 Gyr (cyan dash-3 dots line). The black dashed line represents the reference scaling of the G/D with metallicity (not fit to the data). The black dotted and dash-dotted lines represent the best power-law and best broken power-law fits to the data.}
\label{f:G/Dzhuk}
\end{center}
\end{figure*}

In the following we compare the observationally derived G/D with results of dust evolution models in dwarf galaxies with episodic star formation history from Zhukovska et al. (in prep.). These models were originally introduced to study the lifecycle of dust species from different origins in the Solar neighbourhood \citep{Zhukovska2008}. The model of Zhukovska et al. (in prep.) is based on the \cite{Zhukovska2008} model that has been adapted to treat dwarf galaxies, specifically by considering episodic star-formation. In Zhukovska et al. (in prep.), the equations describing the evolution of the galaxy are now normalised to the total galactic masses $M_{tot}$ (instead of surface densities) because dwarf galaxies are smaller in size and thus assumed to have a well mixed ISM. As in \cite{Zhukovska2008}, the modelled dwarf galaxy is formed by gas infall starting from $M_{tot}$=0 and reach its total mass $M_{tot}$ on the infall timescale. The assumed value of the infall timescale in Zhukovska et al. (in prep.) is set to a much shorter value for dwarf galaxies than for the Solar neighbourhood. Since the G/D is the ratio of the gas and dust masses, it does not depend on the normalisation by the total mass $M_{tot}$, and is determined by the star formation history and infall timescale. We refer the reader to Zhukovska et al. (in prep.), for more details on the modelling.

Similarly to the models from \cite{Asano2013}, Zhukovska et al. (in prep.) include dust formation in AGB stars, SN II and dust growth by mantle accretion in the ISM. The main difference between these models is in the treatment of dust growth. Zhukovska et al. (in prep.) assume a two-phase ISM consisting of clouds and an intercloud medium, where clouds are characterised by temperature, density, mass fraction, and lifetime. Dust growth by accretion in their model takes place only in the dense gas and also critically depends on the metallicity \citep[see][]{Zhukovska2008PhD}.  

In this paper we consider three models from Zhukovska et al. (in prep.), which differ only in duration and intensity of the star formation bursts. All models consider six bursts of star formation starting at instants $t =$ 0.5, 1, 2, 5, 7, and 11~Gyr. In the first and the second model, the burst duration is 50~Myr and 500~Myr, respectively, and the {\it $\tau_{{\rm SF}}$} during bursts is 2~Gyr. The first model is typical of a low-metallicity dwarf galaxy. In the third model, the burst duration is 500~Myr but the value of {\it $\tau_{{\rm SF}}$} is much shorter, 0.2~Gyr. During the quiescence phases {\it $\tau_{{\rm SF}}$}  is set to be 200~Gyr. We also consider a model with continuous star formation on a 10 Gyr timescale, for comparison. 
In all models, the infall timescale is 0.3~Gyr and there are no galactic outflows. The initial metallicity $Z$ of the infalling gas is set to be $10^{-4}$ with SNII like enhanced $\rm [\alpha/Fe]$ ratio. 

The models from Zhukovska et al. (in prep.) are presented in Fig. \ref{f:G/Dzhuk} and reproduce the broadening of the observed G/D values at low metallicities (12+log(O/H) $\lesssim$ 8.3), and also converge around 12+log(O/H) $\sim$ 7.2, similar to the models of continuous star formation of \cite{Asano2013}. Note how the star formation history impacts the shape of the modelled G/D: the most extreme G/D values are obtained by the three models with episodic bursts of star formation. For the model with more intense star formation (brown curve on Fig. \ref{f:G/Dzhuk}), 12+log(O/H) = 8.6 is reached during the first burst, and very high values of the G/D are quickly reached, up to two orders of magnitude above the reference scaling relation at moderate metallicities (12+log(O/H) $\sim$ 8.2 - 8.3). It also presents an interesting scatter of G/D values near 12+log(O/H) = 9.0 that is due to dust destruction during the SF bursts, and consistent with the scatter predicted by the \cite{Galliano2008} model (see Fig. \ref{f:G/Dall}).
The fact that the low-metallicity slope of the broken power-law is consistent with the continuous star formation model at low metallicities for the \xcoz\ case 
indicates that this broken power-law can provide a fairly good empirical way of estimating the G/D for a given metallicity.

\subsection{Explaining the observed scatter in G/D values}

\begin{figure*}
\begin{center}
\includegraphics[width=8.8cm]{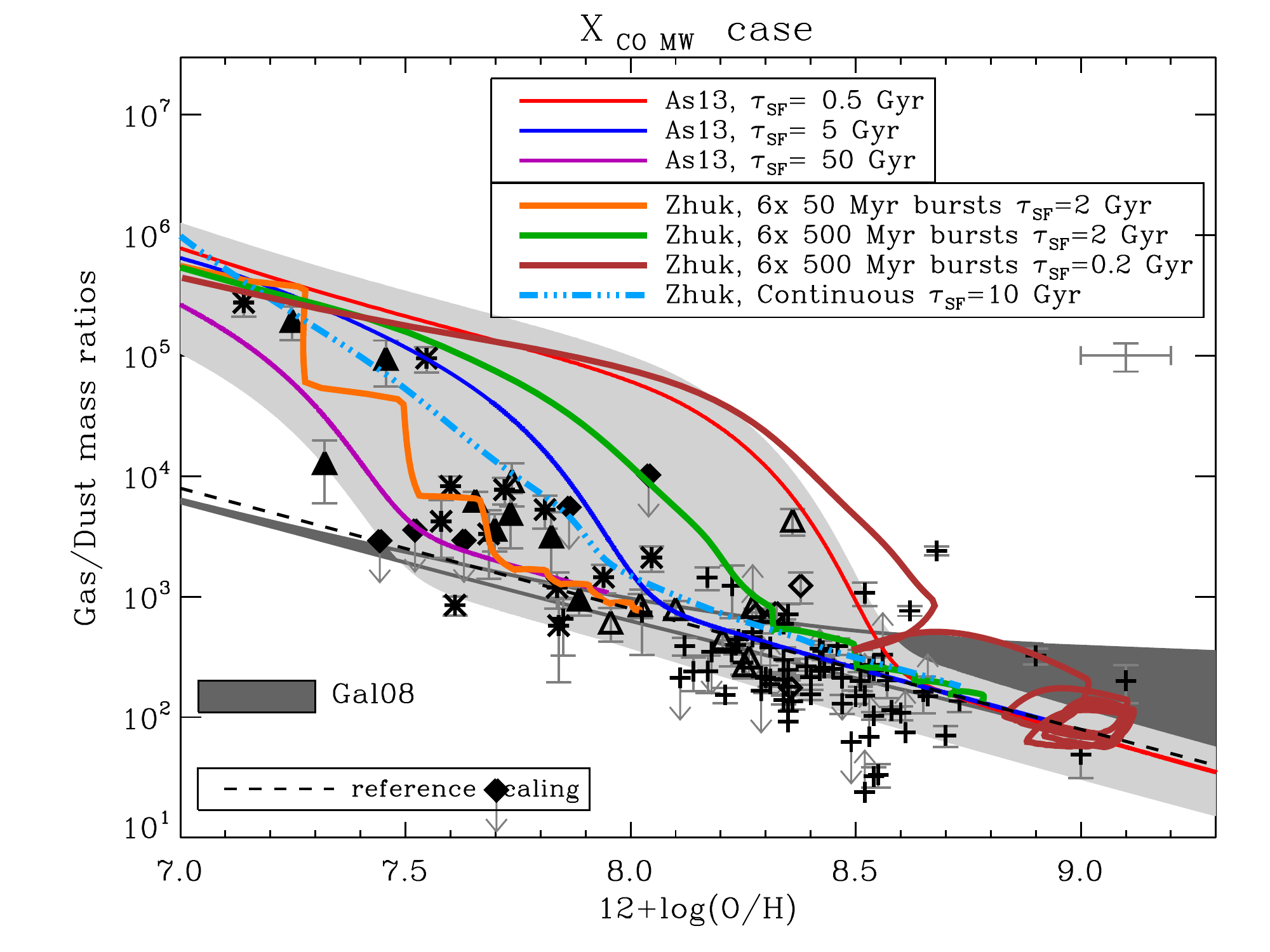}
\includegraphics[width=8.8cm]{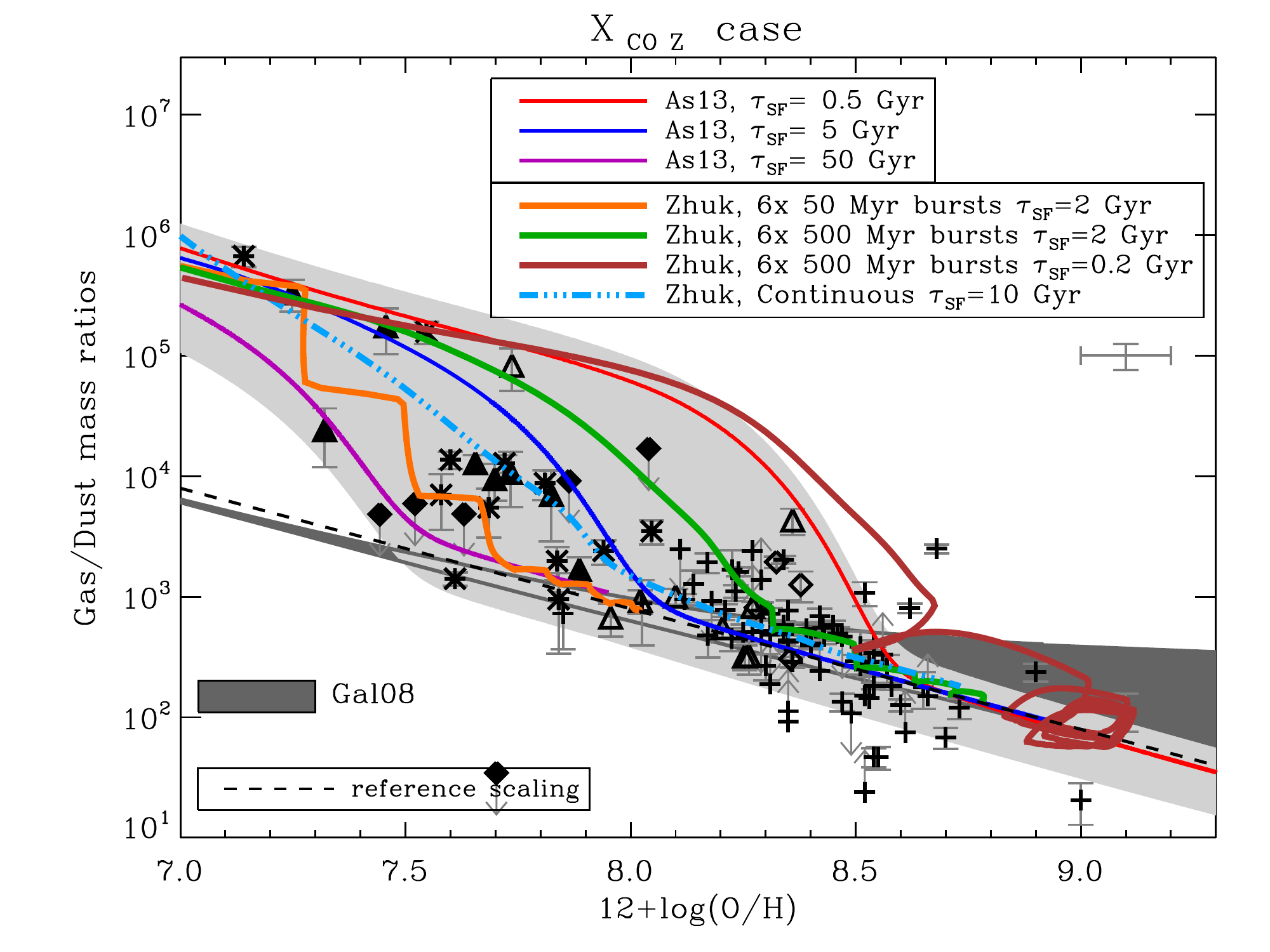}
\caption{G/D as a function of metallicity for the 2 values of \xco: \xcogal\ ({\it left}) and \xcoz\ ({\it right}) with the three chemical evolution model considered in Section \ref{sec:Chemvol}. The symbols are the same as for Fig. \ref{G/Dline}. The model ranges from \cite{Galliano2008} are delineated by the dark grey stripe. The models from \cite{Asano2013} are shown with the red, blue and purple lines. The models from Zhukovska et al. (in prep.) are shown with the orange, green brown solid lines and cyan dash-3 dots line. The black dashed line represents the reference scaling of the G/D with metallicity (not fit to the data).}
\label{f:G/Dall}
\end{center}
\end{figure*}
 
Figure \ref{f:G/Dall} shows the three models overlaid on the observed G/D values. The models from \cite{Asano2013} and Zhukovska et al. (in prep.) provide trends that are consistent with each other and with the data and its scatter. More dust observations of extremely low-metallicity galaxies with 12+log(O/H) $<$ 7.5 are nonetheless needed to confirm this agreement between the models and the data at very low metallicities. The model of \cite{Galliano2008} fails to reproduce the observed G/D at low metallicities, but provides a good complement to explain the scatter seen at high metallicities, consistent with the predictions of the third bursty model by Zhukovska et al. (in prep.). We thus conclude that the observed scatter at low-metallicities in the G/D values is due to the wide variety of environments we are probing, and especially to the different star formation histories. The observed scatter at higher metallicity seems to be due to different timescales for dust destruction by SN blast waves in the different environments and to the efficiency of dust shattering in the ISM. 

We investigated here two different parameters to explain the scatter in the G/D values: star formation histories and efficiency of dust destruction, but other processes could also give rise to the observed scatter. 
In our dust modelling we allow the mass fraction of small grains compared to big grains to vary from galaxy to galaxy (controlled by the $f_{vsg}$ parameter), to account for potential variations in the grain size distribution. This had already been done in one low-metallicity galaxy by \cite{Lisenfeld2002}. On the theoretical side, \cite{HirashitaKuo2011} showed that the dust grain size distribution can have an important impact on the dust growth process in the ISM by regulating the grain growth rate. They also showed that the critical metallicity mentioned in Section \ref{ssec:asano}, for which grain growth becomes dominant, also depends on the grain size distribution. Additionally the grain size distribution varies as the galaxy evolves and this evolution is controlled by different dust formation processes at different ages \citep{Asano2013}. Thus the observed scatter can also be due to variations of the grain size distribution between the galaxies, the effect of which can be related to the star-formation history.

Second, as discussed in Section \ref{ssec:discussion} we use a dust model with the dust composition and optical properties representative of dust in the Milky Way for all of the galaxies. Another explanation for the scatter seen at all metallicities could be that the dust composition in fact varies between the galaxies, leading to important variations in the emissivity of the dust grains \citep{Jones2012}. This would then imply dust masses relatively similar at a given metallicity but large variations in the emissivity properties of dust from one galaxy to another. With our fixed emissivity (due to our fixed dust composition in our dust model) this effect would be seen through the variations in the derived dust masses (and then in the G/D values) seen at a given metallicity, thus giving rise to the observed scatter.

Another aspect not taken into account in the chemical models we use here is that the mass of the galaxy plays an important role in its chemical evolution. 
According to the chemical downsizing scenario, galaxies with different masses have different star-formation efficiencies \citep[e.g.][]{Brooks2007}. The massive galaxies form stars before low-mass galaxies in the history of the Universe \citep{Cowie1996}. The exact reasons are not well known, some models would argue this is a feedback effect, with metals being lost in outflows in less massive galaxies \citep{Frye2002}, and others would argue that the local density and the ISM pressure is responsible \citep[e.g.][]{CenOstriker1999}: if the ISM density is systematically lower in a dwarf galaxy, for instance, then the star formation activity will be low on average, impacting the G/D value. A massive galaxy will not show the same behaviour, and this difference will also introduce some scatter in the observed G/D at a given metallicity.

Additionally external processes such as outflows or interactions and mergers, not considered here either in our models, can also be responsible for the scatter in the G/D values.



\subsection{Implications for the observed G/D in galaxies}

We saw in Sections \ref{ssec:As} and \ref{ssec:Zhuk} that the broken power-law relation in the \xcoz\ case is consistent with the predictions from both chemical evolution models. Even though there are only five galaxies with 12+log(O/H) $\sim$ 7.5 to constrain the observed G/D at extremely low metallicities, this broken power-law relation is the best empirical estimate of the observed G/D for local galaxies we have at our disposal so far. Thus, we advise to use this empirical prescription to estimate the G/D based on a metallicity value for local galaxies, keeping in mind the large scatter in the observed G/D values and the uncertainties on the broken power-law parameters: the estimated G/D would be accurate to a factor of $\sim$ 1.6. Note also that this empirical relation has been derived for metallicities estimated with a strong emission line method and the calibration from \cite{PilyuginThuan2005}. Thus the estimation of the G/D with this empirical relation should be done from a metallicity derived {\it with the same method}.
Additionally, more observations of the dust in extremely low-metallicity galaxies are needed in order to place more constraints on this empirical relation at extremely low metallicities. 



\section{Conclusions}
We present here the evolution of gas-to-dust mass ratios for a sample of 126 galaxies over a 2 dex metallicity range: from 12+log(O/H) = 7.14 to 9.1. We use two different values for the \xco\ factor: the Galactic value, \xcogal, and a metallicity dependent value, \xcoz ($\propto Z^{-2}$), to determine the G/D relation with metallicity. We consider several empirical trends to describe the G/D as a function of metallicity: power-law with a slope fixed to -1 (``reference'' relation) and a slope left free, and a broken power-law. We also compare the observed G/D evolution with metallicity with predictions from chemical evolution models from \cite{Galliano2008, Asano2013} and Zhukovska et al. (in prep.). 
We show that:

\begin{itemize}
\item {\revised The G/D correlates with metallicity, stellar mass, SFR and SSFR. However, the correlation is weaker with the last three parameters, which themselves depend on metallicity. Thus metallicity is the main physical property of the galaxy driving the observed G/D values.}

\item The observed G/D vs metallicity relation cannot be represented by a power-law with a slope of -1 at all metallicities. The observed trend is steeper at low-metallicities and could be due to the harder ISRF in low-metallicity dwarf galaxies that affects the balance between dust formation and destruction by limiting the accretion or enhancing the destruction of the dust grains.

\item There is a large scatter in the G/D values for a given metallicity throughout the metallicity range: in metallicity bins of $\sim$ 0.1 dex, the dispersion is $\sim$ 0.37 dex on average for all bins and for both \xco values. This scatter does not depend on the metallicity indicating that metallicity is not the only driver of the observed G/D. 

\item A power-law fit to the data gives a slope $\alpha$ of -1.6 $\pm$ 0.3 (\xcogal) or -2.0 $\pm$ 0.3 (\xcoz). A broken power-law fit, with a fixed high-metallicity slope $\alpha_H$=-1, gives a low-metallicity slope, $\alpha_L$, of -3.1 $\pm$ 1.8 (\xcogal) or -3.1 $\pm$ 1.3 (\xcoz), 
with a transition metallicity of 7.96 $\pm$ 0.47 (\xcogal) or 8.10 $\pm$ 0.43 (\xcoz). If left free, the high-metallicity slope is consistent within errors with the slope of -1 observed by other studies. On average, the broken power-law reproduces best the observed G/D compared to the two power laws ($\alpha$ = -1 or free) and provides estimates of the G/D that are accurate to a factor of 1.6. 

\item We recommend using the best broken power-law fit from the \xcoz\ case to empirically estimate a G/D from a metallicity, though keeping in mind the large uncertainty on the estimated G/D value, and the differences existing between the different methods to estimate metallicities.

\item Chemical evolution models from \cite{Asano2013} (including dust growth in the ISM as a source of dust production and continuous star formation) and Zhukovska et al. (in prep.) (testing episodic star formation histories and effects of dust growth), are consistent with the observed G/D trend with metallicity for both \xco values, implying that dust growth in the ISM and the global star formation history are both important in the evolution of the G/D. The model from \cite{Galliano2008}, while it well reproduces the observations at higher metallicity, does not reproduce the observed trend at low-metallicities because of too-simple assumptions in their model. The three models confirm that the reference scaling of the G/D with metallicity is not plausible at low metallicities below 12+ log(O/H) $\sim$ 7.7. Our best power-law fit with a free slope, and broken power-law fit are consistent with the \cite{Asano2013} and Zhukovska et al. (in prep.) models.

\item The scatter present at all metallicities is due to the variety of the considered environments, and is consistent with predictions of models considered for this study. Variation in the star formation histories of the galaxies and in the dust destruction efficiency are explored with the different models, and can explain the observed scatter in the G/D values at all metallicities. Variations in the dust properties: grain size distribution and chemical composition can also be invoked, as well as interactions of the galaxies with their surroundings. 
However, all these effects are highly affected by numerous degeneracies, which cannot be disentangled at this stage.
\end{itemize}


\begin{acknowledgements}

{\revised The authors would like to thank the anonymous referee for his/her helpful comments.} The authors would like to thank H. Hirashita for interesting discussions on the chemical evolution models. ARR is supported by a CFR grant from the AIM laboratory (Saclay, France). This research was, in part, made possible through the financial support of the Agence Nationale de la Recherche (ANR) through the programme SYMPATICO (Program Blanc Projet ANR-11-BS56-0023). SZ acknowledges support by the \textit{Deutsche Forschungsgemeinschaft} through SPP 1573: ``Physics of the Interstellar MediumÕÕ. IDL is a postdoctoral researcher of the FWO-Vlaanderen (Belgium). \\

PACS has been developed by MPE (Germany); UVIE (Austria); KU Leuven, CSL, IMEC (Belgium); CEA, LAM (France); MPIA (Germany); INAF-IFSI/OAA/OAP/OAT, LENS, SISSA (Italy); IAC (Spain). This development has been supported by BMVIT (Austria), ESA-PRODEX (Belgium), CEA/CNES (France), DLR (Germany), ASI/INAF (Italy), and CICYT/MCYT (Spain). SPIRE has been developed by Cardiff University (UK); Univ. Lethbridge (Canada); NAOC (China); CEA, LAM (France); IFSI, Univ. Padua (Italy); IAC (Spain); SNSB (Sweden); Imperial College London, RAL, UCL-MSSL, UKATC, Univ. Sussex (UK) and Caltech, JPL, NHSC, Univ. Colorado (USA). This development has been supported by CSA (Canada); NAOC (China); CEA, CNES, CNRS (France); ASI (Italy); MCINN (Spain); Stockholm Observatory (Sweden); STFC (UK); and NASA (USA).

SPIRE has been developed by a consortium of institutes led by Cardiff Univ. (UK) and including: Univ. Lethbridge (Canada); NAOC (China); CEA, LAM (France); IFSI, Univ. Padua (Italy); IAC (Spain); Stockholm Observatory (Sweden); Imperial College London, RAL, UCL-MSSL, UKATC, Univ. Sussex (UK); and Caltech, JPL, NHSC, Univ. Colorado (USA). This development has been supported by national funding agencies: CSA (Canada); NAOC (China); CEA, CNES, CNRS (France); ASI (Italy); MCINN (Spain); SNSB (Sweden); STFC, UKSA (UK); and NASA (USA).

\end{acknowledgements}


\bibliographystyle{aa}
\bibliography{GDletter_biblio}


\appendix
\section{Data Tables}
\begin{landscape}                                                                                   
\begin{table}[h!tbp]                                                                                
\begin{center}                                                                                      
\caption{Gas masses for the DGS, KINGFISH and G11 samples.}                                         
\label{t:Gas}                                                                                       
{\footnotesize                                                                                      
 \begin{tabular}{lccc | ccc | c}                                                                    
\hline                                                                                              
\hline                                                                                              
  & \multicolumn{3}{ c |}{{\bf \matom} [\msun]} & \multicolumn{3}{ c |}{{\bf \mmol} [\msun]} &  \\                                                                                                      
\hline                                                                                              
Name          &                \matom\ (Ref) &       Uncertainty (\%) &    HIarea (arcmin$^2$) &           12+log(O/H) (Ref) &              \mmolgal\ (Ref) &                      \mmolz\ & $\mu_{gal}$ \\
(1)                &			(2)		&		(3)		  &		(4)			&		(5)			 &    			(6)		& 		(7)		  &	(8) \\
\hline                                                                                              
{\bf {\small DGS}} & & & & & & & \\                                                                 
Haro11          &       5.01$\times 10^{  8 a,e}$ (1)  &                  50.00 &                    - &    8.36      (1) &       2.51$\times 10^{  8   }$ (1) &       1.17$\times 10^{  9       }$ & 1.38 \\
Haro2           &       3.77$\times 10^{  8   c}$ (2)  &                  16.08 &                    1.6 &    8.23      (1) &       1.26$\times 10^{  8   }$ (2) &       1.03$\times 10^{  9       }$ & 1.38 \\
Haro3           &       1.13$\times 10^{  9   a}$ (3)  &                   2.13 &                    - &    8.28      (1) &       3.02$\times 10^{  7   }$ (3) &       2.01$\times 10^{  8       }$ & 1.38 \\
He2-10          &       3.10$\times 10^{  8   c}$ (4)  &                   6.45 &                    3.2 &    8.43      (1) &       9.55$\times 10^{  7   }$ (4) &       3.21$\times 10^{  8       }$ & 1.39 \\
HS0017+1055     & $\leq$1.74$\times 10^{  8   a}$ (5)  &                    - &                    - &    7.63      (1) & $\leq$2.15$\times 10^{  6  d}$ (-) & $\leq$1.18$\times 10^{  8       }$ & 1.37 \\
HS0052+2536     & $\leq$4.82$\times 10^{ 10   a}$ (6)  &                    - &                    - &    8.04      (1) & $\leq$5.94$\times 10^{  8  d}$ (-) & $\leq$3.26$\times 10^{ 10       }$ & 1.38 \\
HS0822+3542     &       5.75$\times 10^{  7   b}$ (7)  &                  22.22 &                    0.6 &    7.32      (1) &       9.48$\times 10^{  5  d}$ (-) &       5.21$\times 10^{  7       }$ & 1.37 \\
HS1222+3741     &       - & - & - &    7.79      (1) &      - & - & 1.37 \\
HS1236+3937     &       - & - & - &    7.72      (1) &      - & - & 1.37 \\
HS1304+3529     &       - & - & - &    7.93      (1) &      - & - & 1.38 \\
HS1319+3224     &       - & - & - &    7.81      (1) &      - & - & 1.37 \\
HS1330+3651     &       - & - & - &    7.98      (1) &      - & - & 1.38 \\
HS1442+4250     &       3.10$\times 10^{  8   a}$ (5)  &                   2.27 &                    - &    7.60      (1) &       3.82$\times 10^{  6  d}$ (-) &       2.10$\times 10^{  8       }$ & 1.37 \\
HS2352+2733     & $\leq$2.30$\times 10^{ 10   a}$ (6)  &                    - &                    - &    8.40      (1) & - & - & 1.38 \\
IZw18           &       1.00$\times 10^{  8  c2}$ (8)  &                  16.08 &                    - &    7.14      (1) &       2.71$\times 10^{  6  d}$ (5) &       1.49$\times 10^{  8       }$ & 1.37 \\
IC10            &       4.41$\times 10^{  7  c2}$ (9)  &                   4.52 &                    - &    8.17      (1) &       4.84$\times 10^{  6   }$ (6) &       5.27$\times 10^{  7       }$ & 1.38 \\
IIZw40          &       5.66$\times 10^{  8  c}$ (10)  &                  16.08 &                    0.3 &    8.23      (1) &       2.83$\times 10^{  7   }$ (7) &       2.41$\times 10^{  8       }$ & 1.38 \\
Mrk1089         &       1.47$\times 10^{ 10  b}$ (11)  &                   7.91 &                   26.4 &    8.10      (1) &       2.84$\times 10^{  8   }$ (1) &       4.31$\times 10^{  9       }$ & 1.38 \\
Mrk1450         &       4.30$\times 10^{  7  c}$ (12)  &                  16.08 &                    0.1 &    7.84      (1) &       5.30$\times 10^{  5  d}$ (-) &       2.91$\times 10^{  7       }$ & 1.37 \\
Mrk153          & $\leq$6.81$\times 10^{  8  a}$ (13)  &                    - &                    - &    7.86      (1) & $\leq$8.39$\times 10^{  6  d}$ (-) & $\leq$4.61$\times 10^{  8       }$ & 1.37 \\
Mrk209          &       2.76$\times 10^{  7  b}$ (13)  &                   6.58 &                   12.2 &    7.74      (1) &       3.09$\times 10^{  6   }$ (8) &       2.50$\times 10^{  8       }$ & 1.37 \\
Mrk930          &       3.19$\times 10^{  9  c}$ (14)  &                  11.06 &                    0.1 &    8.03      (1) &       3.07$\times 10^{  7   }$ (1) &       6.57$\times 10^{  8       }$ & 1.38 \\
NGC1140         &       3.50$\times 10^{  9  a}$ (15)  &                  26.71 &                    - &    8.38      (1) &       1.91$\times 10^{  7   }$ (9) &       8.02$\times 10^{  7       }$ & 1.38 \\
NGC1569         &       1.77$\times 10^{  8  b}$ (16)  &                  16.08 &                   24.0 &    8.02      (1) &       7.08$\times 10^{  5  }$ (10) &       1.54$\times 10^{  7       }$ & 1.38 \\
NGC1705         &       7.67$\times 10^{  7  b}$ (17)  &                  12.50 &                   13.0 &    8.27      (1) & - & - & 1.38 \\
NGC2366         &       2.93$\times 10^{  8  b}$ (18)  &                   6.76 &                  420.0 &    7.70      (1) &       9.12$\times 10^{  6 d}$ (11) &       5.01$\times 10^{  8       }$ & 1.37 \\
NGC4214         &       3.76$\times 10^{  8  b}$ (16)  &                  16.08 &                  156.0 &    8.26      (1) &       1.62$\times 10^{  6  }$ (12) &       1.16$\times 10^{  7       }$ & 1.38 \\
NGC4449         &       9.55$\times 10^{  8  b}$ (16)  &                  16.08 &                  144.0 &    8.20      (1) &       2.67$\times 10^{  7  }$ (13) &       2.50$\times 10^{  8       }$ & 1.38 \\
NGC4861         &       4.12$\times 10^{  8  b}$ (19)  &                   6.27 &                   34.3 &    7.89      (1) &       5.90$\times 10^{  6  d}$ (1) &       3.24$\times 10^{  8       }$ & 1.37 \\
NGC5253         &       1.06$\times 10^{  8  b}$ (20)  &                   5.66 &                   77.0 &    8.25      (1) &       3.02$\times 10^{  6  }$ (14) &       2.26$\times 10^{  7       }$ & 1.38 \\
NGC625          &       1.10$\times 10^{  8  c}$ (21)  &                  18.18 &                   20.0 &    8.22      (1) &       4.36$\times 10^{  6   }$ (1) &       3.72$\times 10^{  7       }$ & 1.38 \\
NGC6822         &       1.04$\times 10^{  8  b}$ (22)  &                  10.45 &                  676.0 &    7.96      (1) &       3.58$\times 10^{  5  }$ (15) &       1.05$\times 10^{  7       }$ & 1.38 \\
Pox186          & $\leq$2.00$\times 10^{  6a,e}$ (23)  &                    - &                    - &    7.70      (1) & $\leq$2.47$\times 10^{  4  d}$ (-) & $\leq$1.35$\times 10^{  6       }$ & 1.37 \\
SBS0335-052     &       4.37$\times 10^{  8  b}$ (24)  &                  10.27 &                    0.2 &    7.25      (1) &       5.67$\times 10^{  6 d}$ (16) &       3.12$\times 10^{  8       }$ & 1.37 \\
SBS1159+545     & $\leq$6.30$\times 10^{  7   a}$ (5)  &                    - &                    - &    7.44      (1) & $\leq$7.76$\times 10^{  5  d}$ (-) & $\leq$4.26$\times 10^{  7       }$ & 1.37 \\
SBS1211+540     &       5.60$\times 10^{  7  c}$ (12)  &                  16.08 &                    0.1 &    7.58      (1) &       6.90$\times 10^{  5  d}$ (-) &       3.79$\times 10^{  7       }$ & 1.37 \\
SBS1249+493     &       1.00$\times 10^{  9  c}$ (25)  &                  48.15 &                    0.0 &    7.68      (1) &       1.23$\times 10^{  7  d}$ (-) &       6.77$\times 10^{  8       }$ & 1.37 \\
SBS1415+437     &       4.37$\times 10^{  8  c}$ (12)  &                  16.08 &                    0.4 &    7.55      (1) &       5.38$\times 10^{  6  d}$ (-) &       2.96$\times 10^{  8       }$ & 1.37 \\
SBS1533+574     &       3.00$\times 10^{  9  c}$ (14)  &                  11.29 &                    0.2 &    8.05      (1) &       3.70$\times 10^{  7  d}$ (-) &       2.03$\times 10^{  9       }$ & 1.38 \\
Tol0618-402     &       - & - & - &    8.09      (1) &      - & - & 1.38 \\
Tol1214-277     & $\leq$3.22$\times 10^{  8   a}$ (5)  &                    - &                    - &    7.52      (1) & $\leq$3.97$\times 10^{  6  d}$ (-) & $\leq$2.18$\times 10^{  8       }$ & 1.37 \\
UGC4483         &       2.52$\times 10^{  7  b}$ (26)  &                   5.00 &                   14.4 &    7.46      (1) &       4.06$\times 10^{  5 d}$ (14) &       2.23$\times 10^{  7       }$ & 1.37 \\
UGCA20          &       6.92$\times 10^{  6  b}$ (27)  &                  13.82 &                    9.8 &    7.50      (1) &       2.46$\times 10^{  5  d}$ (-) &       1.35$\times 10^{  7       }$ & 1.37 \\
\hline                                                                                              
\end{tabular}                                                                                       
}                                                                                                   
\end{center}                                                                                        
\end{table}                                                                                         
\end{landscape}                                                                                     
\begin{landscape}                                                                                   
\addtocounter{table}{-1}                                                                            
\begin{table}[h!tbp]                                                                                
\begin{center}                                                                                      
\caption{(continued) Gas masses for the DGS, KINGFISH and G11 samples.}                             
{\footnotesize                                                                                      
 \begin{tabular}{lccc | ccc | c}                                                                    
\hline                                                                                              
\hline                                                                                              
  & \multicolumn{3}{ c |}{{\bf \matom} [\msun]} & \multicolumn{3}{ c |}{{\bf \mmol} [\msun]} &  \\                                                                                                      
\hline                                                                                              
Name          &                \matom\ (Ref) &       Uncertainty (\%) &    HIarea (arcmin$^2$) &           12+log(O/H) (Ref) &              \mmolgal\ (Ref) &                      \mmolz\ & $\mu_{gal}$ \\
(1)                &			(2)		&		(3)		  &		(4)			&		(5)			 &    			(6)		& 		(7)		  &	(8) \\
\hline                                                                                              
UM133           &       2.15$\times 10^{  8  b}$ (28)  &                   4.38 &                    6.2 &    7.82      (1) &       4.93$\times 10^{  6  d}$ (-) &       2.71$\times 10^{  8       }$ & 1.37 \\
UM311           &       6.09$\times 10^{ 10  b}$ (29)  &                  16.08 &                   21.2 &    8.36      (1) &       9.37$\times 10^{  7   }$ (1) &       4.28$\times 10^{  8       }$ & 1.38 \\
UM448           &       6.00$\times 10^{  9  a}$ (15)  &                  27.63 &                    - &    8.32      (1) &       3.55$\times 10^{  9   }$ (3) &       1.92$\times 10^{ 10       }$ & 1.38 \\
UM461           &       7.31$\times 10^{  7  b}$ (26)  &                   5.00 &                    2.6 &    7.73      (1) &       1.66$\times 10^{  6  d}$ (3) &       9.09$\times 10^{  7       }$ & 1.37 \\
VIIZw403        &       3.25$\times 10^{  7   b}$ (2)  &                   8.28 &                   10.4 &    7.66      (1) &       6.41$\times 10^{  5 d}$ (17) &       3.52$\times 10^{  7       }$ & 1.37 \\
& & & & & & &  \\                                                                                   
\hline                                                                                              
{\bf {\small KINGFISH}} & & & & & & &  \\                                                           
NGC0337         &       3.34$\times 10^{  9   }$ (27)  &                  11.97 &                  - &    8.18      (2) & $\leq$6.98$\times 10^{  8  }$ (18) & $\leq$7.31$\times 10^{  9       }$ & 1.38 \\
NGC0584         &       1.58$\times 10^{  8   }$ (27)  &                  35.92 &                  - &    8.43      (2) & $\leq$4.78$\times 10^{  8  }$ (18) & $\leq$1.58$\times 10^{  9       }$ & 1.39 \\
NGC0628         &       3.72$\times 10^{  9   }$ (30)  &                  16.08 &                  - &    8.35      (2) &       8.71$\times 10^{  8  }$ (19) &       4.17$\times 10^{  9       }$ & 1.38 \\
NGC0855         &       1.32$\times 10^{  8   }$ (27)  &                  29.47 &                  - &    8.29      (2) & $\leq$8.91$\times 10^{  7  }$ (18) & $\leq$5.62$\times 10^{  8       }$ & 1.38 \\
NGC0925         &       4.56$\times 10^{  9   }$ (30)  &                  16.08 &                  - &    8.25      (2) &       6.14$\times 10^{  8  }$ (19) &       4.66$\times 10^{  9       }$ & 1.38 \\
NGC1097         &       7.56$\times 10^{  9   }$ (30)  &                  16.08 &                  - &    8.47      (2) &       1.67$\times 10^{  8  }$ (20) &       4.59$\times 10^{  8       }$ & 1.39 \\
NGC1266         &       -  &                    - &                  - &    8.29      (2) &       1.33$\times 10^{  9  }$ (21) &       8.39$\times 10^{  9       }$ & 1.38 \\
NGC1291         &       1.78$\times 10^{  9   }$ (30)  &                  16.08 &                  - &    8.52      (2) & - & - & 1.39 \\
NGC1316         & $\leq$4.73$\times 10^{  8   }$ (30)  &                    - &                  - &    8.77      (2) & - & - & 1.40 \\
NGC1377         &       -  &                    - &                  - &    8.29      (2) & - & - & 1.38 \\
NGC1404         &       -  &                    - &                  - &    8.54      (2) & - & - & 1.39 \\
IC0342          &       9.49$\times 10^{  9   }$ (31)  &                  16.08 &                  - &    8.46      (3) &       1.60$\times 10^{  9  }$ (20) &       4.60$\times 10^{  9       }$ & 1.39 \\
NGC1482         & $\leq$8.01$\times 10^{  8   }$ (30)  &                    - &                  - &    8.11      (2) &       3.11$\times 10^{  9  }$ (19) &       4.49$\times 10^{ 10       }$ & 1.38 \\
NGC1512         &       7.33$\times 10^{  9   }$ (30)  &                  16.08 &                  - &    8.56      (2) & - & - & 1.39 \\
NGC2146         &       3.89$\times 10^{  9   }$ (27)  &                  22.10 &                  - &    8.68      (2) &       7.77$\times 10^{ 10  }$ (22) &       8.13$\times 10^{ 10       }$ & 1.40 \\
HoII            &       4.21$\times 10^{  8   }$ (27)  &                  23.95 &                  - &    7.72      (2) &       5.18$\times 10^{  6 d}$ (18) &       2.85$\times 10^{  8       }$ & 1.37 \\
DDO053          &       6.16$\times 10^{  7   }$ (30)  &                  16.08 &                  - &    7.60      (2) &       7.60$\times 10^{  5 d}$ (11) &       4.17$\times 10^{  7       }$ & 1.37 \\
NGC2798         &       2.13$\times 10^{  9   }$ (30)  &                  16.08 &                  - &    8.34      (2) &       3.21$\times 10^{  9  }$ (19) &       1.61$\times 10^{ 10       }$ & 1.38 \\
NGC2841         &       8.63$\times 10^{  9   }$ (30)  &                  16.08 &                  - &    8.54      (2) &       2.92$\times 10^{  9  }$ (19) &       5.82$\times 10^{  9       }$ & 1.39 \\
NGC2915         &       3.52$\times 10^{  8   }$ (30)  &                  16.08 &                  - &    7.94      (2) &       4.34$\times 10^{  6  d}$ (-) &       2.39$\times 10^{  8       }$ & 1.38 \\
HoI             &       1.45$\times 10^{  8   }$ (16)  &                  16.08 &                  - &    7.61      (2) &       1.79$\times 10^{  6 d}$ (18) &       9.85$\times 10^{  7       }$ & 1.37 \\
NGC2976         &       1.27$\times 10^{  8   }$ (30)  &                  16.08 &                  - &    8.36      (2) &       5.81$\times 10^{  7  }$ (19) &       2.66$\times 10^{  8       }$ & 1.38 \\
NGC3049         &       1.21$\times 10^{  9   }$ (30)  &                  16.08 &                  - &    8.53      (2) &       1.80$\times 10^{  8  }$ (18) &       3.75$\times 10^{  8       }$ & 1.39 \\
NGC3077         &       8.71$\times 10^{  8   }$ (16)  &                  16.08 &                  - &    8.52      (4) &       3.80$\times 10^{  6   }$ (9) &       8.32$\times 10^{  6       }$ & 1.39 \\
M81dwB          &       1.16$\times 10^{  7   }$ (16)  &                  16.08 &                  - &    7.84      (2) &       1.43$\times 10^{  5 d}$ (18) &       7.85$\times 10^{  6       }$ & 1.37 \\
NGC3190         &       4.27$\times 10^{  8   }$ (27)  &                  36.84 &                  - &    8.49      (2) & $\leq$3.93$\times 10^{  8  }$ (18) & $\leq$9.87$\times 10^{  8       }$ & 1.39 \\
NGC3184         &       3.37$\times 10^{  9   }$ (30)  &                  16.08 &                  - &    8.51      (2) &       1.19$\times 10^{  9  }$ (19) &       2.73$\times 10^{  9       }$ & 1.39 \\
NGC3198         &       6.85$\times 10^{  9   }$ (27)  &                  26.71 &                  - &    8.34      (2) &       1.30$\times 10^{  9  }$ (18) &       6.53$\times 10^{  9       }$ & 1.38 \\
IC2574          &       1.32$\times 10^{  9   }$ (30)  &                  16.08 &                  - &    7.85      (2) &       3.16$\times 10^{  6  }$ (19) &       1.51$\times 10^{  8       }$ & 1.37 \\
NGC3265         &       1.75$\times 10^{  8   }$ (27)  &                  38.68 &                  - &    8.27      (2) & $\leq$4.39$\times 10^{  8  }$ (18) & $\leq$3.04$\times 10^{  9       }$ & 1.38 \\
NGC3351         &       1.02$\times 10^{  9   }$ (30)  &                  16.08 &                  - &    8.60      (2) &       4.77$\times 10^{  8  }$ (19) &       7.23$\times 10^{  8       }$ & 1.39 \\
NGC3521         &       8.71$\times 10^{  9   }$ (30)  &                  16.08 &                  - &    8.39      (2) &       4.77$\times 10^{  9  }$ (19) &       1.90$\times 10^{ 10       }$ & 1.38 \\
NGC3621         &       6.99$\times 10^{  9   }$ (30)  &                  16.08 &                  - &    8.27      (2) & - & - & 1.38 \\
NGC3627         &       8.46$\times 10^{  8   }$ (30)  &                  16.08 &                  - &    8.34      (2) &       3.21$\times 10^{  9  }$ (19) &       1.61$\times 10^{ 10       }$ & 1.38 \\
NGC3773         &       9.03$\times 10^{  7   }$ (30)  &                  16.08 &                  - &    8.43      (2) &       4.95$\times 10^{  7  }$ (18) &       1.64$\times 10^{  8       }$ & 1.39 \\
NGC3938         &       8.00$\times 10^{  9   }$ (30)  &                  16.08 &                  - &    8.42      (2) &       4.38$\times 10^{  9  }$ (19) &       1.52$\times 10^{ 10       }$ & 1.38 \\
NGC4236         &       2.69$\times 10^{  9   }$ (30)  &                  16.08 &                  - &    8.17      (2) &       9.33$\times 10^{  7  }$ (18) &       1.02$\times 10^{  9       }$ & 1.38 \\
\hline                                                                                              
\end{tabular}                                                                                       
}                                                                                                   
\end{center}                                                                                        
\end{table}                                                                                         
\end{landscape}                                                                                     
\begin{landscape}                                                                                   
\addtocounter{table}{-1}                                                                            
\begin{table}[h!tbp]                                                                                
\begin{center}                                                                                      
\caption{(continued) Gas masses for the DGS, KINGFISH and G11 samples.}                             
{\footnotesize                                                                                      
 \begin{tabular}{lccc | ccc | c}                                                                    
\hline                                                                                              
\hline                                                                                              
  & \multicolumn{3}{ c |}{{\bf \matom} [\msun]} & \multicolumn{3}{ c |}{{\bf \mmol} [\msun]} &  \\                                                                                                      
\hline                                                                                              
Name          &                \matom\ (Ref) &       Uncertainty (\%) &    HIarea (arcmin$^2$) &           12+log(O/H) (Ref) &              \mmolgal\ (Ref) &                      \mmolz\ & $\mu_{gal}$ \\
(1)                &			(2)		&		(3)		  &		(4)			&		(5)			 &    			(6)		& 		(7)		  &	(8) \\
\hline                                                                                              
NGC4254         &       3.76$\times 10^{  9   }$ (30)  &                  16.08 &                  - &    8.45      (2) &       6.82$\times 10^{  9  }$ (19) &       2.06$\times 10^{ 10       }$ & 1.39 \\
NGC4321         &       2.39$\times 10^{  9   }$ (30)  &                  16.08 &                  - &    8.50      (2) &       5.34$\times 10^{  9  }$ (19) &       1.28$\times 10^{ 10       }$ & 1.39 \\
NGC4536         &       1.73$\times 10^{  9   }$ (30)  &                  16.08 &                  - &    8.21      (2) &       1.72$\times 10^{  9  }$ (19) &       1.57$\times 10^{ 10       }$ & 1.38 \\
NGC4559         &       4.09$\times 10^{  9   }$ (30)  &                  16.08 &                  - &    8.29      (2) &       1.92$\times 10^{  8  }$ (18) &       1.21$\times 10^{  9       }$ & 1.38 \\
NGC4569         &       1.55$\times 10^{  8   }$ (30)  &                  16.08 &                  - &    8.58      (2) &       1.14$\times 10^{  9  }$ (19) &       1.90$\times 10^{  9       }$ & 1.39 \\
NGC4579         &       5.47$\times 10^{  8   }$ (30)  &                  16.08 &                  - &    8.54      (2) &       1.89$\times 10^{  9  }$ (19) &       3.77$\times 10^{  9       }$ & 1.39 \\
NGC4594         &       2.60$\times 10^{  8   }$ (30)  &                  16.08 &                  - &    8.54      (2) &       2.16$\times 10^{  8  }$ (19) &       4.31$\times 10^{  8       }$ & 1.39 \\
NGC4625         &       1.00$\times 10^{  9   }$ (30)  &                  16.08 &                  - &    8.35      (2) &       2.09$\times 10^{  7  }$ (18) &       1.00$\times 10^{  8       }$ & 1.38 \\
NGC4631         &       8.77$\times 10^{  9   }$ (30)  &                  16.08 &                  - &    8.12      (2) &       1.10$\times 10^{  9  }$ (19) &       1.52$\times 10^{ 10       }$ & 1.38 \\
NGC4725         &       3.59$\times 10^{  9   }$ (30)  &                  16.08 &                  - &    8.35      (2) &       2.16$\times 10^{  9  }$ (19) &       1.03$\times 10^{ 10       }$ & 1.38 \\
NGC4736         &       4.03$\times 10^{  8   }$ (30)  &                  16.08 &                  - &    8.31      (2) &       4.41$\times 10^{  8  }$ (19) &       2.54$\times 10^{  9       }$ & 1.38 \\
DDO154          &       3.55$\times 10^{  8   }$ (16)  &                  16.08 &                  - &    7.54      (2) &       4.37$\times 10^{  6 d}$ (23) &       2.40$\times 10^{  8       }$ & 1.37 \\
NGC4826         &       2.77$\times 10^{  8   }$ (30)  &                  16.08 &                  - &    8.54      (2) &       4.80$\times 10^{  8  }$ (19) &       9.58$\times 10^{  8       }$ & 1.39 \\
DDO165          &       1.12$\times 10^{  8   }$ (27)  &                  14.74 &                  - &    7.63      (2) &       1.38$\times 10^{  6 d}$ (18) &       7.60$\times 10^{  7       }$ & 1.37 \\
NGC5055         &       5.59$\times 10^{  9   }$ (30)  &                  16.08 &                  - &    8.40      (2) &       2.73$\times 10^{  9  }$ (19) &       1.04$\times 10^{ 10       }$ & 1.38 \\
NGC5398         &       2.48$\times 10^{  8   }$ (27)  &                  11.97 &                  - &    8.35      (2) & - & - & 1.38 \\
NGC5408         &       3.22$\times 10^{  8   }$ (30)  &                  16.08 &                  - &    7.81      (2) &       3.97$\times 10^{  6  d}$ (-) &       2.18$\times 10^{  8       }$ & 1.37 \\
NGC5457         &       1.16$\times 10^{ 10   }$ (16)  &                  16.08 &                  - &    8.42      (5) & - & - & 1.38 \\
NGC5474         &       9.71$\times 10^{  8   }$ (27)  &                  24.87 &                  - &    8.31      (2) & $\leq$5.88$\times 10^{  7  }$ (18) & $\leq$3.38$\times 10^{  8       }$ & 1.38 \\
NGC5713         &       5.51$\times 10^{  9   }$ (30)  &                  16.08 &                  - &    8.24      (2) &       3.39$\times 10^{  9  }$ (19) &       2.69$\times 10^{ 10       }$ & 1.38 \\
NGC5866         & $\leq$2.85$\times 10^{  8   }$ (30)  &                    - &                  - &    8.47      (2) &       6.38$\times 10^{  8  }$ (19) &       1.76$\times 10^{  9       }$ & 1.39 \\
NGC6946         &       3.58$\times 10^{  9   }$ (30)  &                  16.08 &                  - &    8.40      (2) &       4.50$\times 10^{  9  }$ (19) &       1.71$\times 10^{ 10       }$ & 1.38 \\
NGC7331         &       8.85$\times 10^{  9   }$ (30)  &                  16.08 &                  - &    8.34      (2) &       6.70$\times 10^{  9  }$ (19) &       3.36$\times 10^{ 10       }$ & 1.38 \\
NGC7793         &       8.75$\times 10^{  8   }$ (30)  &                  16.08 &                  - &    8.31      (2) & - & - & 1.38 \\
& & & & & & &  \\                                                                                   
\hline                                                                                              
{\bf {\small G11}} & & & & & & & \\                                                                 
M83             &       5.13$\times 10^{  9   }$ (31)  &                   5.00 &                  - &    8.62      (6) &       8.65$\times 10^{  8  }$ (20) &       1.19$\times 10^{  9       }$ & 1.39 \\
NGC1808         &       1.78$\times 10^{  9   }$ (31)  &                   9.46 &                  - &    9.10      (7) &       1.75$\times 10^{  9  }$ (20) &       2.64$\times 10^{  8       }$ & 1.44 \\
NGC7552         &       4.79$\times 10^{  9   }$ (31)  &                  16.08 &                  - &    8.35      (8) & - & - & 1.38 \\
M82             &       8.91$\times 10^{  8   }$ (31)  &                  10.29 &                  - &    8.51      (8) &       4.43$\times 10^{  8  }$ (20) &       1.01$\times 10^{  9       }$ & 1.39 \\
NGC1068         &       2.24$\times 10^{  9   }$ (31)  &                  18.35 &                  - &    9.00      (7) &       7.26$\times 10^{  9  }$ (20) &       1.74$\times 10^{  9       }$ & 1.43 \\
NGC0891         &       7.59$\times 10^{  9   }$ (31)  &                  16.08 &                  - &    8.90      (7) &       5.75$\times 10^{  9  }$ (20) &       2.19$\times 10^{  9       }$ & 1.42 \\
MGC+02-04-025   &       1.74$\times 10^{  9   }$ (31)  &                  16.08 &                  - &    8.52      (9) & - & - & 1.39 \\
NGC7469         &       1.51$\times 10^{  9   }$ (31)  &                  16.08 &                  - &    8.70     (10) &       6.28$\times 10^{  9  }$ (20) &       6.00$\times 10^{  9       }$ & 1.40 \\
NGC5256         &       -  &                    - &                  - &    8.53      (9) &       9.92$\times 10^{  9  }$ (20) &       2.07$\times 10^{ 10       }$ & 1.39 \\
NGC5953         &       5.75$\times 10^{  8   }$ (31)  &                  16.08 &                  - &    8.73      (7) &       1.62$\times 10^{  9  }$ (20) &       1.34$\times 10^{  9       }$ & 1.40 \\
M51             &       5.01$\times 10^{  9   }$ (31)  &                   5.63 &                  - &    8.55      (8) &       3.95$\times 10^{  9  }$ (20) &       7.53$\times 10^{  9       }$ & 1.39 \\
NGC3995         &       6.17$\times 10^{  9   }$ (31)  &                  16.08 &                  - &    8.66      (7) & - & - & 1.40 \\
NGC3994         &       2.82$\times 10^{  9   }$ (31)  &                  16.08 &                  - &    8.61      (7) & - & - & 1.39 \\
NGC6052         &       3.80$\times 10^{  9   }$ (31)  &                  16.08 &                  - &    8.65      (7) &       3.00$\times 10^{  9  }$ (20) &       3.61$\times 10^{  9       }$ & 1.40 \\
NGC1222         &       1.20$\times 10^{  9   }$ (31)  &                  16.08 &                  - &    8.57      (7) &       7.23$\times 10^{  8  }$ (21) &       1.26$\times 10^{  9       }$ & 1.39 \\
NGC7674         &       1.07$\times 10^{ 10   }$ (31)  &                  16.08 &                  - &    8.14      (9) &       6.34$\times 10^{  9  }$ (20) &       7.99$\times 10^{ 10       }$ & 1.38 \\
NGC4670         &       1.66$\times 10^{  8   }$ (31)  &                  16.08 &                  - &    8.30      (7) &       9.00$\times 10^{  6  }$ (20) &       5.42$\times 10^{  7       }$ & 1.38 \\
\hline                                                                                              
\end{tabular}                                                                                       
}                                                                                                   
\end{center}                                                                                        
\end{table}                                                                                         
\end{landscape}                                                                                     
\begin{landscape}                                                                                   
\addtocounter{table}{-1}                                                                            
\begin{table}[h!tbp]                                                                                
\begin{center}                                                                                      
\begin{tabular}{lccc | ccc | c}                                                                     
\end{tabular}                                                                                       
\end{center}                                                                                        
{\footnotesize                                                                                      
\noindent                                                                                           
(1) : Galaxy name \\
(2) : Final HI mass in solar masses. (The HI masses for the DGS have been corrected to match the extent of the dust aperture.) \\
\noindent                                                                                           
$^a$ For these galaxies, no information was available on the extent of the \HI\ observations. The quoted \HI\ mass is thus the total \HI\ mass (diamonds in Fig. \ref{G/Dline}). \\                                                                                                                         
\noindent                                                                                           
$^b$ For these galaxies, we were able to correct the \HI\ mass and to rescale it to the \HI\ mass contained in the IR aperture (triangles in Fig. \ref{G/Dline}).\\                                                                                                                                         
\noindent                                                                                           
 $^c$ For these galaxies, the \HI\ extent is similar to the IR aperture size and we consider that the total \HI\ mass corresponds to the \HI\ mass within the IR aperture (crosses in Fig. \ref{G/Dline}).\\                                                                                                
\noindent                                                                                           
 $^{c2}$ For these galaxies, the \HI\ mass corresponding to the dust emitting region is given in the reference, and is the mass quoted here.\\            
 (3) : Uncertainty on HI mass in \%. \\
 (4) : HI area used to correct the DGS HI masses, in arcmin$^2$, assuming an exponential mass distribution profile. \\
 (5) : Metallicity in term of 12+log(O/H). \\
 (6) : Final H$_2$ mass derived with \xcogal\ in solar masses. \\          
 (7) : Final H$_2$ mass derived with \xcoz, going as: \xcoz\ $\propto$ Z$^{-2}$, in solar masses. \\
 \noindent                                                                                           
$^d$ For these galaxies, the H$_2$ mass correction was applied : i.e. non-detections or absence of data were replaced by 0.012$\times$\matom\ for \mmolgal\ and 0.68$\times$\matom\ for \mmolz\ for galaxies with 12+log(O/H) $\leq$ 8.1.\\  
 (8) : Mean atomic weight  of the galaxy, computed as $\mu_{gal}$ = 1/(1-\ysun - Z$_{{\rm gal}}$). The total gas mass can be obtained using Eq. \ref{eq:totgas} (i.e., column 8 $\times$ (column 2 + column 6 or 7)). \\                                                                                                                                                                                   
\noindent                                                                                           
$^e$ For these 2 galaxies, the H{\sc ii} mass has a significant contribution to the total gas mass : \mion~$\sim$~1.2~$\times$~\matom\ for Haro 11 from \cite{Cormier2012} and \mion~$\sim$~\matom\ for Pox186 from \cite{dePaz2003} in the \xcogal\ case, and have to be taken into account to estimate the total gas mass. \\

{\bf References for metallicities}                                                                  
(1) \cite{Madden2013} ; 
(2) \cite{Kennicutt2011} ; 
(3) \cite{McCall1985} ; 
(4) \cite{McQuade1995} ; 
(5) \cite{Kennicutt2003b} ; 
(6) \cite{Pilyugin2006} ; 
(7) \cite{Galametz2011} ; 
(8) \cite{Moustakas2010} ; 
(9) \cite{Veilleux1995} ; 
(10) \cite{BonattoPastoriza1990} \\
\noindent                                                                                           
{\bf References for HI masses} :                                                                    
(1) Cormier et al., A\&A, in press ; 
(2) \cite{Thuan2004} ; 
(3) \cite{Gordon1981} ; 
(4) \cite{Sauvage1997} ; 
(5) \cite{Pustilnik2007} ; 
(6) MacHattie et al., in prep. ; 
(7) \cite{Chengalur2006} ; 
(8) \cite{Lelli2012} ; 
(9) \cite{Huchtmeier1988} ; 
(10) \cite{Bettoni2003} ; 
(11) \cite{Williams1991} ; 
(12) \cite{Huchtmeier2005} ; 
(13) \cite{Thuan1981} ; 
(14) \cite{Thuan1999b} ; 
(15) \cite{Davoust2004} ; 
(16) \cite{Walter2008} ; 
(17) \cite{Meurer1998} ; 
(18) \cite{Hunter2011} ; 
(19) \cite{VanEymeren2009} ; 
(20) \cite{Lopez-Sanchez2012} ; 
(21) \cite{Cannon2004} ; 
(22) \cite{DeBlok2006} ; 
(23) \cite{BegumChengalur2005} ; 
(24) \cite{Ekta2009} ; 
(25) \cite{Pustilnik2002} ; 
(26) \cite{VanZee1998} ; 
(27) \cite{Paturel2003} ; 
(28) \cite{EktaChengalur2010} ; 
(29) \cite{Moles1994} ; 
(30) \cite{Draine2007} ; 
(31) \cite{Galametz2011} \\
\noindent                                                                                           
{\bf References for H$_2$ masses}                                                                   
(1) Cormier et al., A\&A, in press ; 
(2) \cite{Israel2005} ; 
(3) \cite{Sage1992} ; 
(4) \cite{Kobulnicky1995} ; 
(5) \cite{Leroy2007} ; 
(6) \cite{Leroy2006} ; 
(7) \cite{Bettoni2003} ; 
(8) \cite{Young1995} ; 
(9) \cite{Albrecht2004} ; 
(10) \cite{Greve1996} ; 
(11) \cite{Schruba2012} ; 
(12) \cite{Walter2001} ; 
(13) \cite{Bottner2003} ; 
(14) \cite{Taylor1998} ; 
(15) \cite{Gratier2010} ; 
(16) \cite{Dale2001b} ; 
(17) \cite{Leroy2005} ; 
(18) \cite{Wilson2012} ; 
(19) \cite{Draine2007} ; 
(20) \cite{Galametz2011} ; 
(21) \cite{Young2011} ; 
(22) \cite{Young1989} ; 
(23) \cite{Leroy2009} \\
{\bf Note on mass references :} The references given here are where the original measurement was found. If a correction was applied, the quoted masses are the corrected masses (see $^b, ^c, ^d$) and not the original masses. If no correction was applied, the quoted masses are the original masses. Note that some galaxies have a molecular gas mass but no reference : this means that the molecular gas mass has been corrected (see $^d$) but that no measurements of the molecular gas mass is available in the literature. \\                                                                                                                                                                                                                                                                                                                                                                                                                                                                                                                                                                                                                                                                                        
}                                                                                                   
\end{table}                                                                                         
\end{landscape}                                                                                     

\addtocounter{table}{+1}     
\begin{table}[h!tbp]                                                                                
\begin{center}                                                                                      
\caption{Dust masses for the KINGFISH and G11 samples.}        
\label{t:Dust}                                                                                      
{\footnotesize                                                                                      
 \begin{tabular}{lcc}                                                                               
\hline                                                                                              
\hline                                                                                              
Name &                          M$_{dust}$ [\msun] &                     Uncertainty (\%) \\
\hline                                                                                              
{\bf KINGFISH} & & \\                                                                               
NGC0337         & 1.60$\times 10^{  7    }$ &                   8.14 \\
NGC0584         & - & - \\
NGC0628         & 2.57$\times 10^{  7    }$ &                  12.42 \\
NGC0855         & 1.84$\times 10^{  6    }$ &                  22.90 \\
NGC0925         & 2.52$\times 10^{  7    }$ &                  14.44 \\
NGC1097         & 8.29$\times 10^{  7    }$ &                   8.05 \\
NGC1266         & 8.43$\times 10^{  6    }$ &                   7.27 \\
NGC1291         & 1.64$\times 10^{  7    }$ &                  15.67 \\
NGC1316         & 9.35$\times 10^{  6    }$ &                  10.99 \\
NGC1377         & 1.47$\times 10^{  6    }$ &                  10.67 \\
NGC1404         & - & - \\
IC0342          & 3.95$\times 10^{  7    }$ &                   7.34 \\
NGC1482         & 2.54$\times 10^{  7    }$ &                   7.91 \\
NGC1512         & 3.09$\times 10^{  7    }$ &                  27.99 \\
NGC2146         & 4.75$\times 10^{  7    }$ &                   8.23 \\
HoII            & 7.58$\times 10^{  4    }$ &                   7.90 \\
DDO053          & 1.04$\times 10^{  4    }$ &                  13.38 \\
NGC2798         & 1.24$\times 10^{  7    }$ &                   6.86 \\
NGC2841         & 5.91$\times 10^{  7    }$ &                   9.11 \\
NGC2915         & 3.38$\times 10^{  5    }$ &                  21.07 \\
HoI             & 2.38$\times 10^{  5    }$ &                   8.27 \\
NGC2976         & 1.89$\times 10^{  6    }$ &                   9.07 \\
NGC3049         & 7.69$\times 10^{  6    }$ &                  20.60 \\
NGC3077         & 1.13$\times 10^{  6    }$ &                  15.35 \\
M81dwB          & 2.81$\times 10^{  4    }$ &                  63.98 \\
NGC3190         & 1.82$\times 10^{  7    }$ &                  11.09 \\
NGC3184         & 3.16$\times 10^{  7    }$ &                  12.82 \\
NGC3198         & 3.76$\times 10^{  7    }$ &                  13.32 \\
IC2574          & 2.79$\times 10^{  6    }$ &                  47.44 \\
NGC3265         & 1.84$\times 10^{  6    }$ &                  11.68 \\
NGC3351         & 1.93$\times 10^{  7    }$ &                   6.82 \\
NGC3521         & 7.08$\times 10^{  7    }$ &                  12.08 \\
NGC3621         & 1.89$\times 10^{  7    }$ &                  12.70 \\
NGC3627         & 4.04$\times 10^{  7    }$ &                   7.52 \\
NGC3773         & 7.56$\times 10^{  5    }$ &                  12.46 \\
NGC3938         & 4.65$\times 10^{  7    }$ &                  13.22 \\
NGC4236         & 2.67$\times 10^{  6    }$ &                  14.51 \\
NGC4254         & 5.83$\times 10^{  7    }$ &                   9.45 \\
NGC4321         & 7.23$\times 10^{  7    }$ &                  11.64 \\
NGC4536         & 3.11$\times 10^{  7    }$ &                  11.81 \\
NGC4559         & 9.56$\times 10^{  6    }$ &                  11.95 \\
NGC4569         & 1.58$\times 10^{  7    }$ &                   6.72 \\
NGC4579         & 3.30$\times 10^{  7    }$ &                   5.98 \\
NGC4594         & 2.06$\times 10^{  7    }$ &                  17.07 \\
NGC4625         & 1.99$\times 10^{  6    }$ &                  14.87 \\
NGC4631         & 3.48$\times 10^{  7    }$ &                   9.40 \\
NGC4725         & 4.45$\times 10^{  7    }$ &                   9.86 \\
NGC4736         & 5.64$\times 10^{  6    }$ &                  11.28 \\
DDO154          & - & - \\
NGC4826         & 4.31$\times 10^{  6    }$ &                   8.08 \\
DDO165          & - & - \\
NGC5055         & 5.39$\times 10^{  7    }$ &                   8.21 \\
NGC5398         & 3.06$\times 10^{  6    }$ &                  43.26 \\
NGC5408         & 8.50$\times 10^{  4    }$ &                  25.68 \\
NGC5457         & 6.62$\times 10^{  7    }$ &                  15.15 \\
NGC5474         & 3.73$\times 10^{  6    }$ &                  12.30 \\
NGC5713         & 2.78$\times 10^{  7    }$ &                   7.68 \\
NGC5866         & 6.05$\times 10^{  6    }$ &                   9.06 \\
NGC6946         & 7.27$\times 10^{  7    }$ &                   6.24 \\
NGC7331         & 1.05$\times 10^{  8    }$ &                   8.57 \\
NGC7793         & 6.41$\times 10^{  6    }$ &                  13.41 \\
& & \\                                                                                              
\hline                                                                                              
{\bf G11} & & \\                                                                                    
M83             & 1.10$\times 10^{  7    }$ &                   8.14 \\
NGC1808         & 2.54$\times 10^{  7    }$ &                  34.24 \\
NGC7552         & 7.24$\times 10^{  7    }$ &                  16.10 \\
\hline                                                                                              
\end{tabular}                                                                                       
}                                                                                                   
\end{center}                                                                                        
\end{table}                                                                                         
\addtocounter{table}{-1}                                                                            
\begin{table}[h!tbp]                                                                                
\begin{center}                                                                                      
\caption{(continued) Dust masses for the KINGFISH and G11 samples.}                                 
{\footnotesize                                                                                      
 \begin{tabular}{lcc}                                                                               
\hline                                                                                              
\hline                                                                                              
Name &                          M$_{dust}$ [\msun] &                     Uncertainty (\%) \\
\hline                                                                                              
M82             & 6.85$\times 10^{  6    }$ &                  33.02 \\
NGC1068         & 2.77$\times 10^{  8    }$ &                  36.05 \\
NGC0891         & 5.84$\times 10^{  7    }$ &                  11.84 \\
MGC+02-04-025   & 1.01$\times 10^{  8    }$ &                  47.08 \\
NGC7469         & 1.55$\times 10^{  8    }$ &                  19.16 \\
NGC5256         & 2.01$\times 10^{  8    }$ &                  62.93 \\
NGC5953         & 2.25$\times 10^{  7    }$ &                  18.90 \\
M51             & 3.74$\times 10^{  8    }$ &                  21.82 \\
NGC3995         & 5.75$\times 10^{  7    }$ &                  23.47 \\
NGC3994         & 5.27$\times 10^{  7    }$ &                  34.81 \\
NGC6052         & 5.84$\times 10^{  7    }$ &                  32.58 \\
NGC1222         & 1.33$\times 10^{  7    }$ &                  26.71 \\
NGC7674         & 9.83$\times 10^{  7    }$ &                  29.69 \\
NGC4670         & 1.13$\times 10^{  6    }$ &                  21.12 \\
\hline                                                                                              
\end{tabular}                                                                                       
}                                                                                                   
\end{center}                                                                                        
{\footnotesize                                                                                      
\noindent                                                                                           
The dust masses for the DGS sample are given in RemyRuyer et al., in prep. \\                                                                                                                                                                                                                                     
}                                                                                                   
\end{table}

\section{G/D with morphological type, stellar mass and star formation rate}\label{ap:galparam}

\begin{figure}[h!tbp]
\begin{center}
\includegraphics[width=8.8cm]{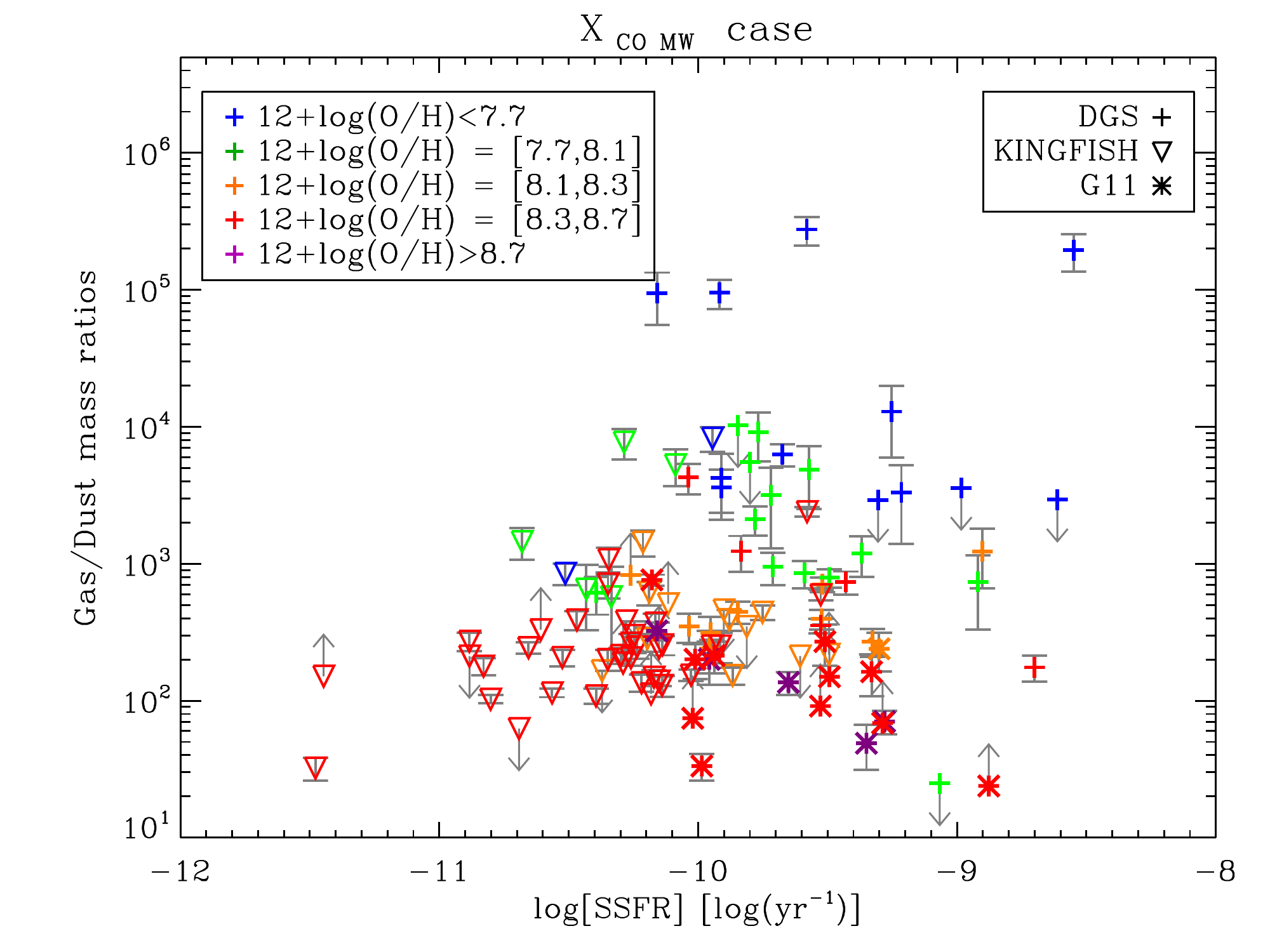}
\includegraphics[width=8.8cm]{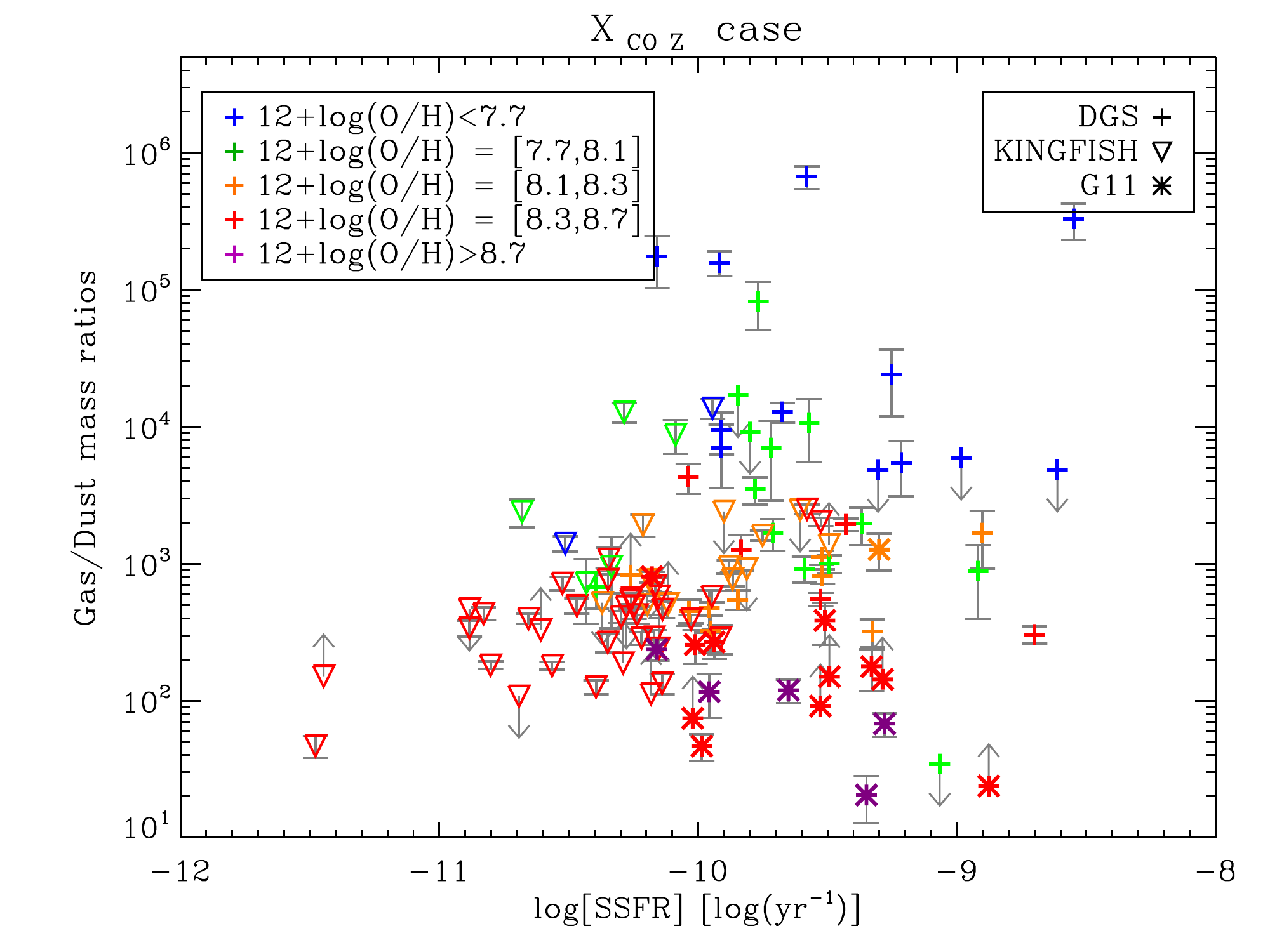}
\caption{{\revised G/D as a function of SSFR for the 2 values of \xco: \xcogal\ ({\it top}) and \xcoz\ ({\it bottom}). The colours code for the metallicity of the galaxies and the symbols differentiate between the three samples: DGS (crosses), KINGFISH (downward triangles) and G11 (stars).}}
\label{f:GDvsSSFR}
\end{center}
\end{figure}

\begin{figure}
\begin{center}
\includegraphics[width=8.8cm]{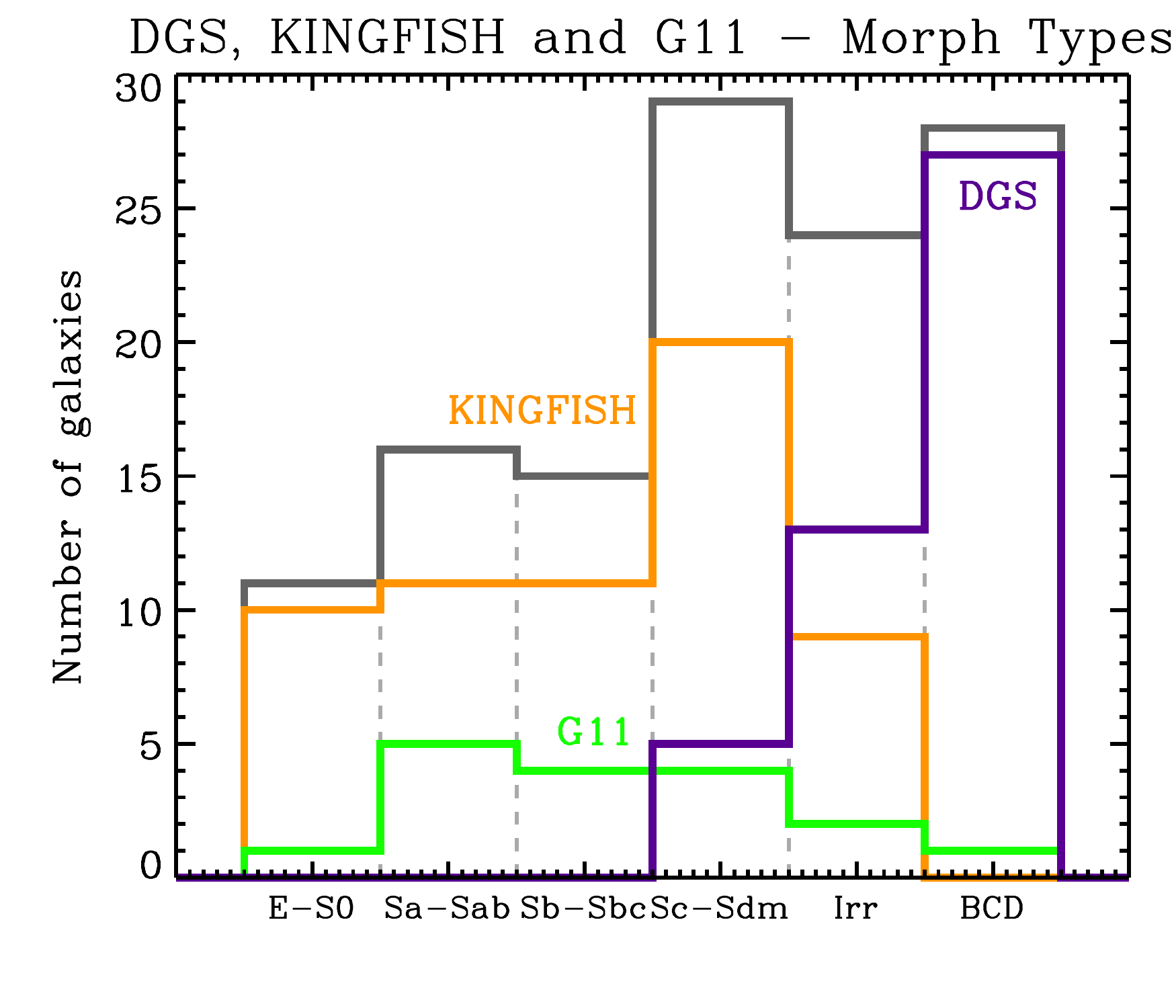}
\includegraphics[width=8.8cm]{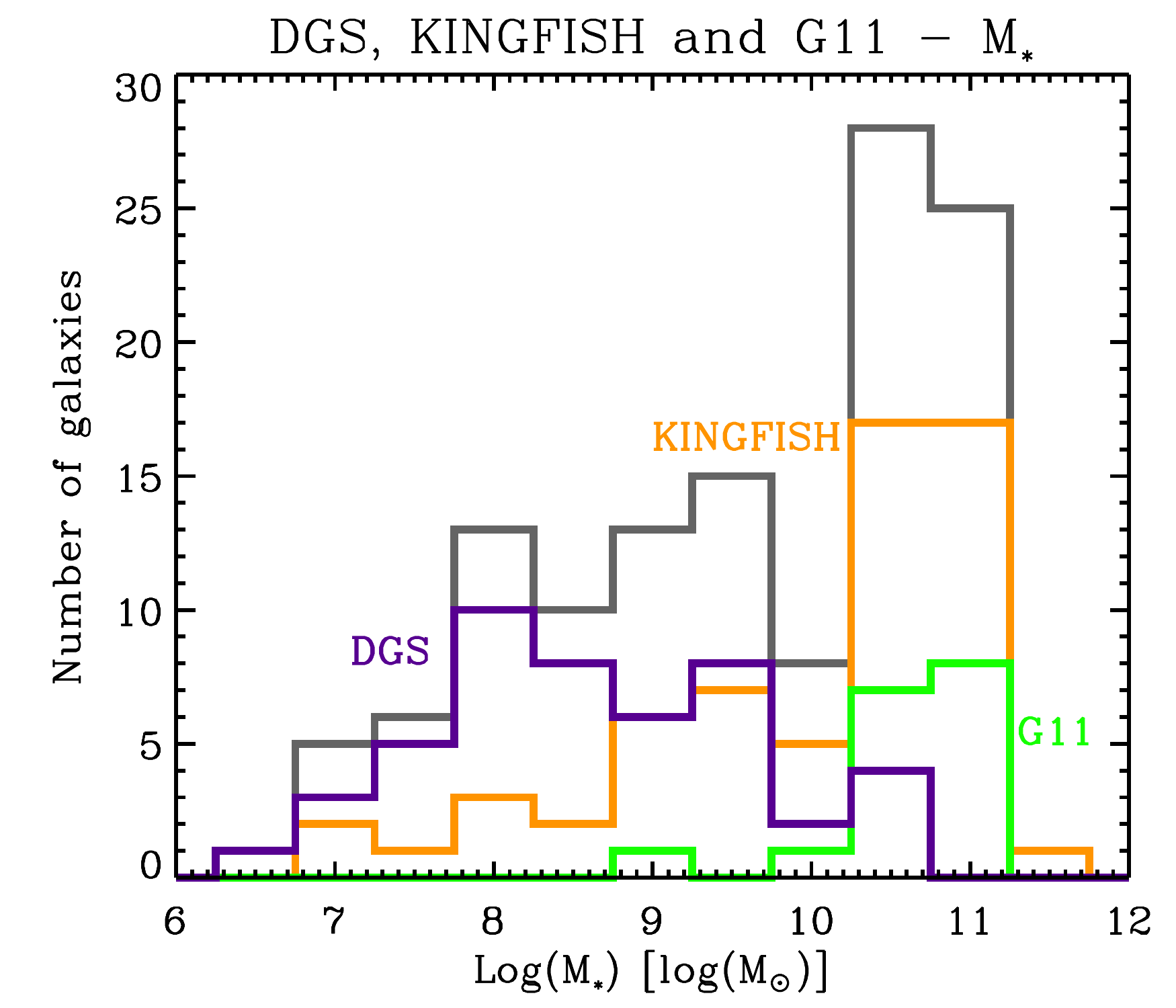}
\includegraphics[width=8.8cm]{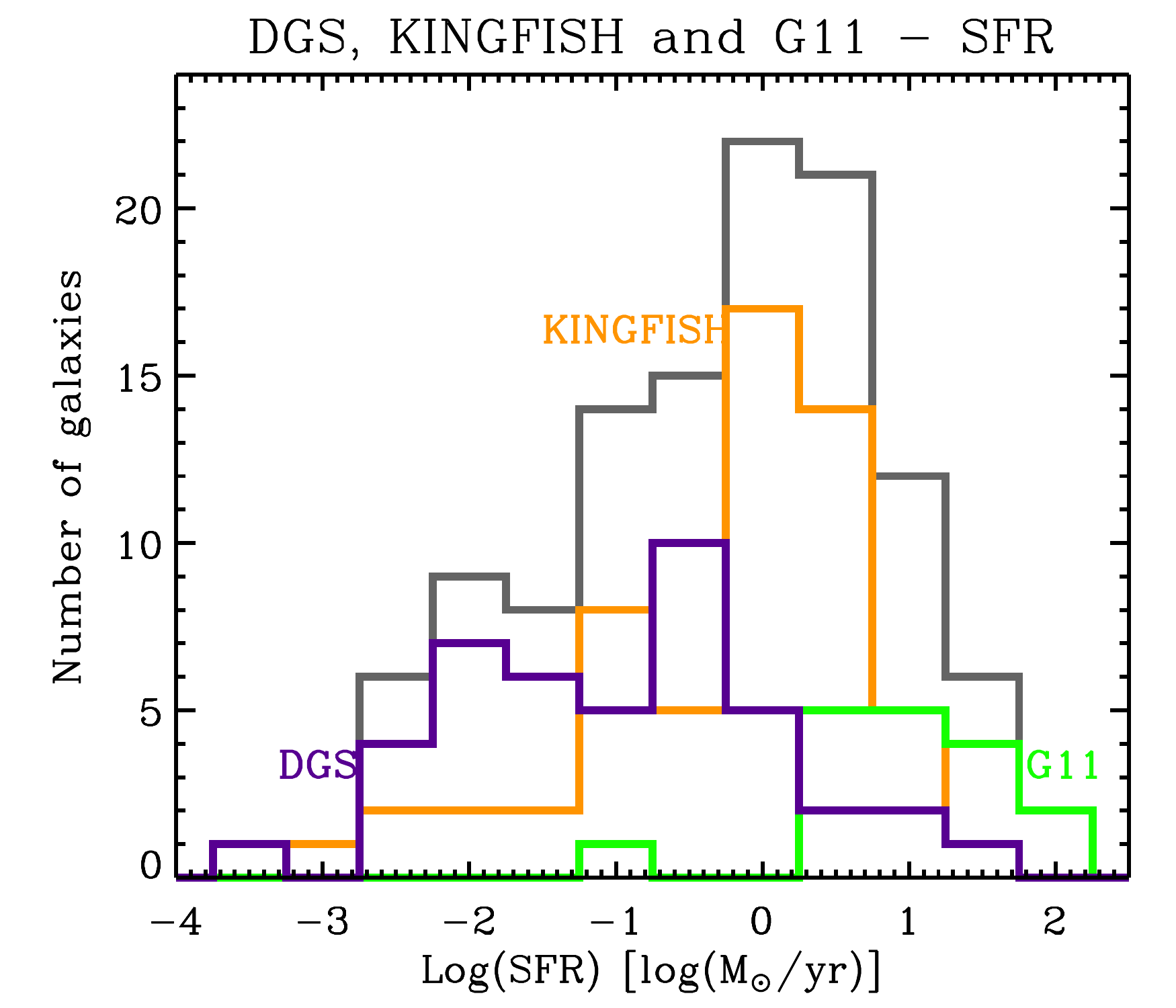}
\caption{{\revised Morphological type distributions {\it (top)}, stellar mass distributions {\it (center)}, and star formation rate distributions {\it (bottom)} of the DGS (purple), KINGFISH (orange) and G11 (green) samples. In each panel, the total distribution is indicated in grey.}}
\label{f:histosMorMsSFR}
\end{center}
\end{figure}

\begin{figure*}
\begin{center}
\includegraphics[width=8.8cm]{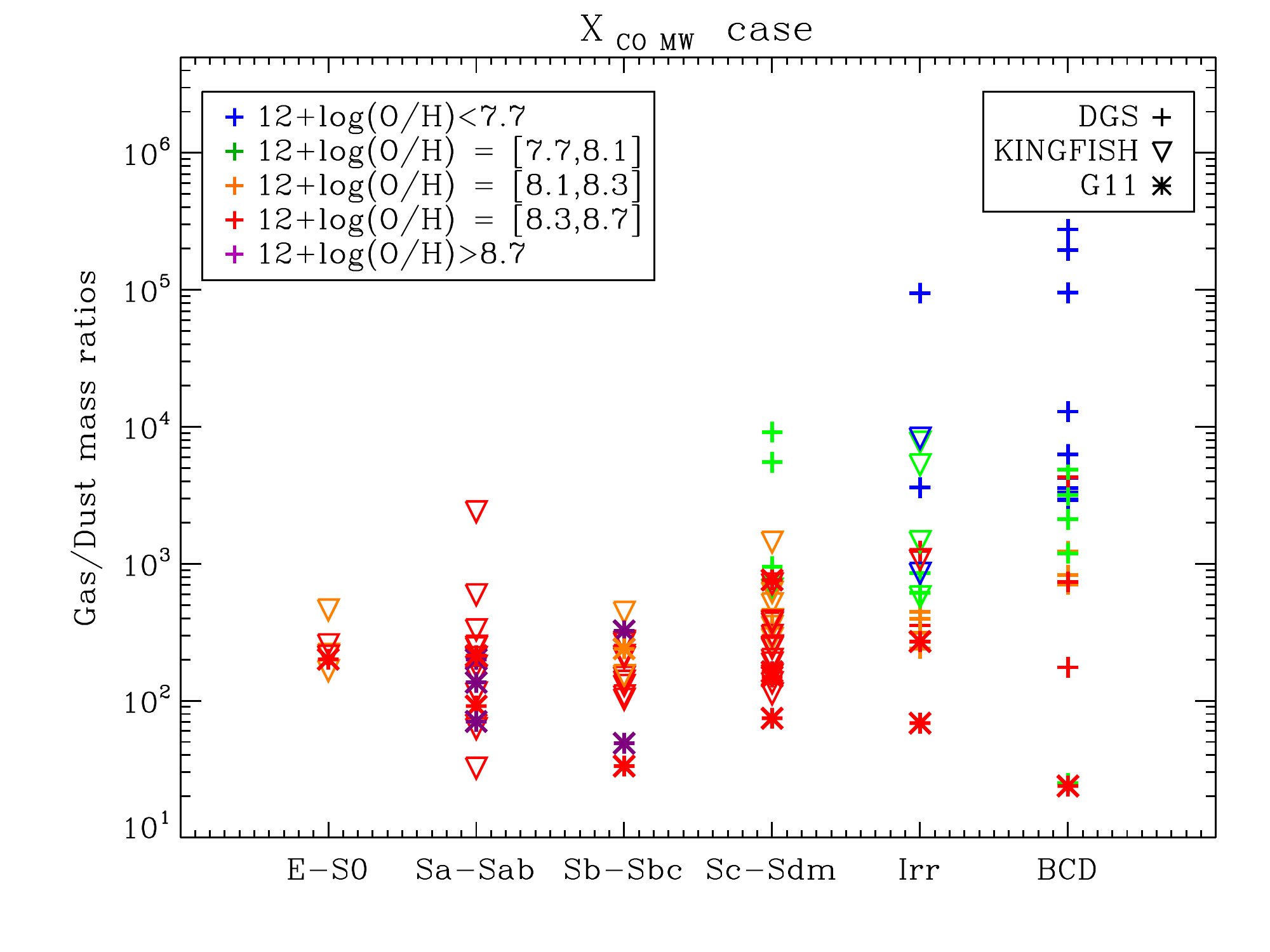}
\includegraphics[width=8.8cm]{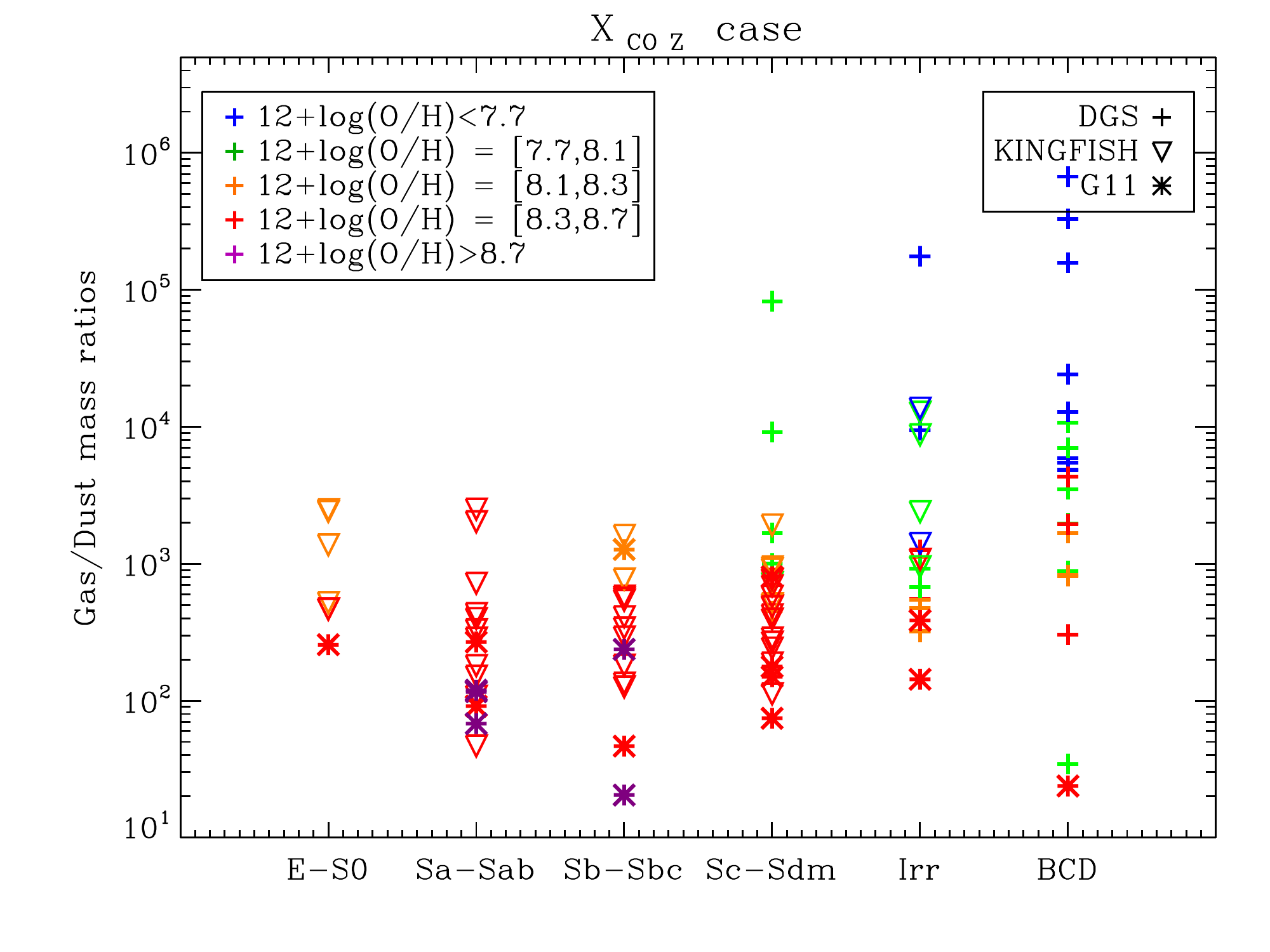}
\includegraphics[width=8.8cm]{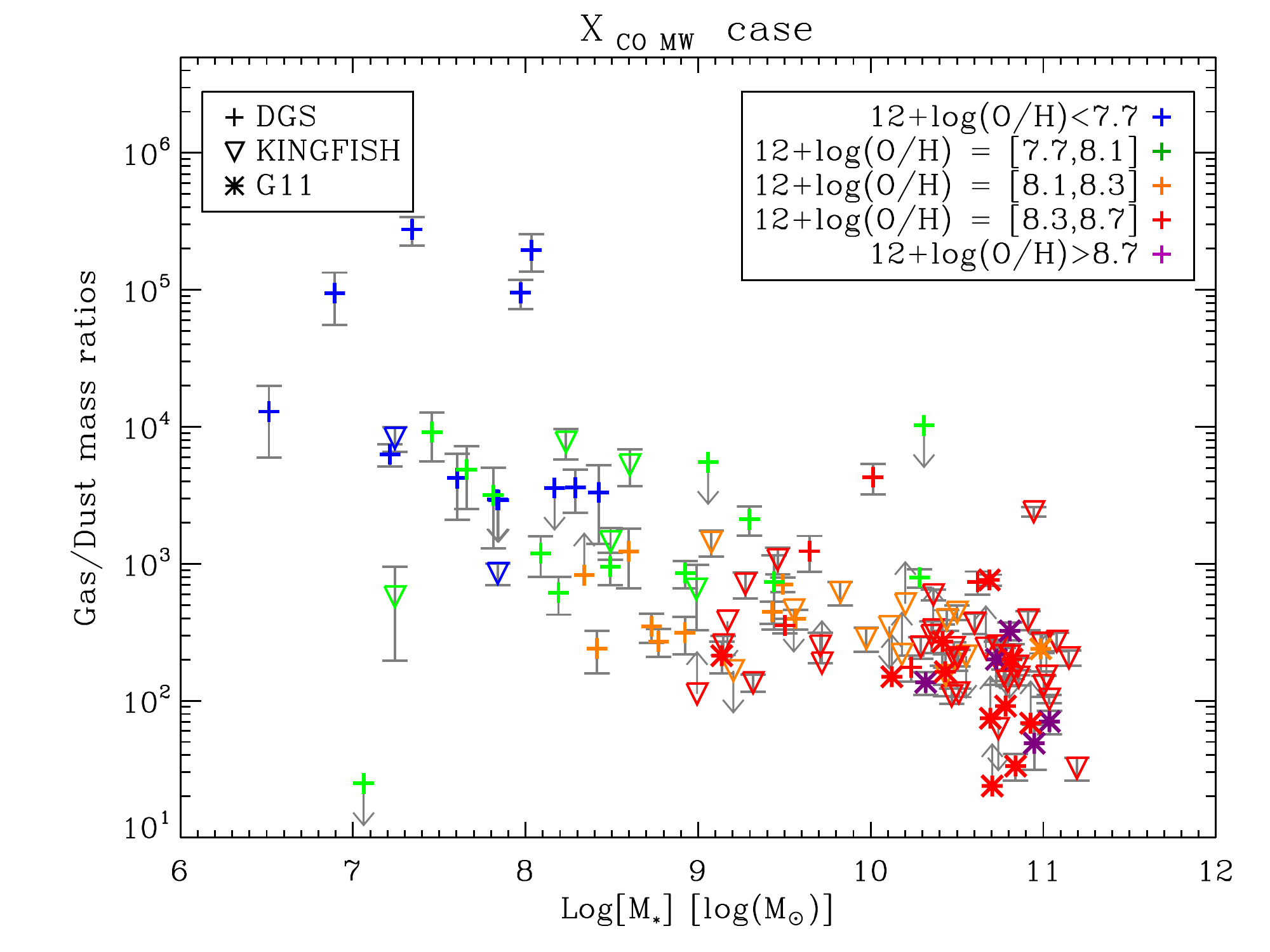}
\includegraphics[width=8.8cm]{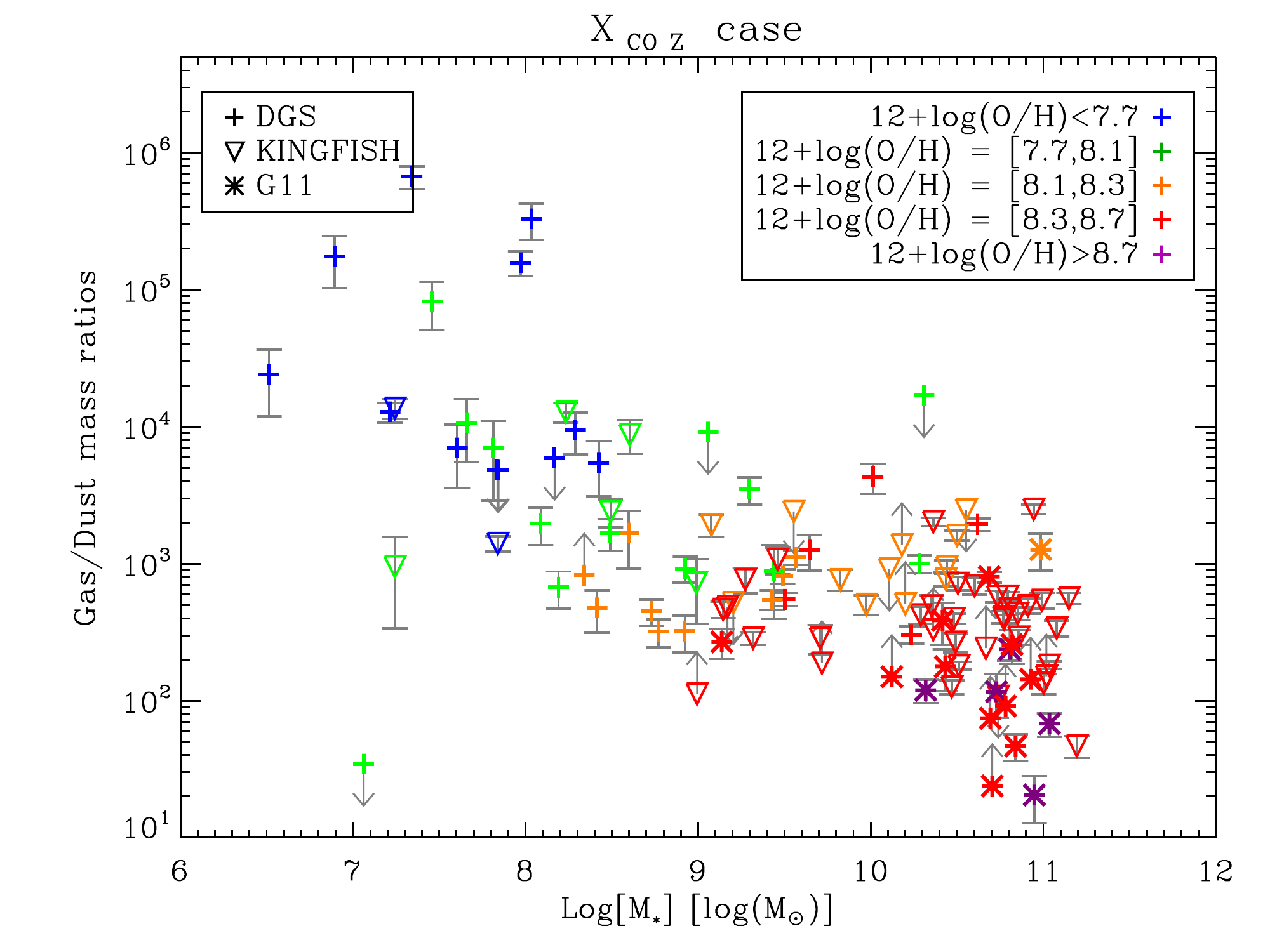}
\includegraphics[width=8.8cm]{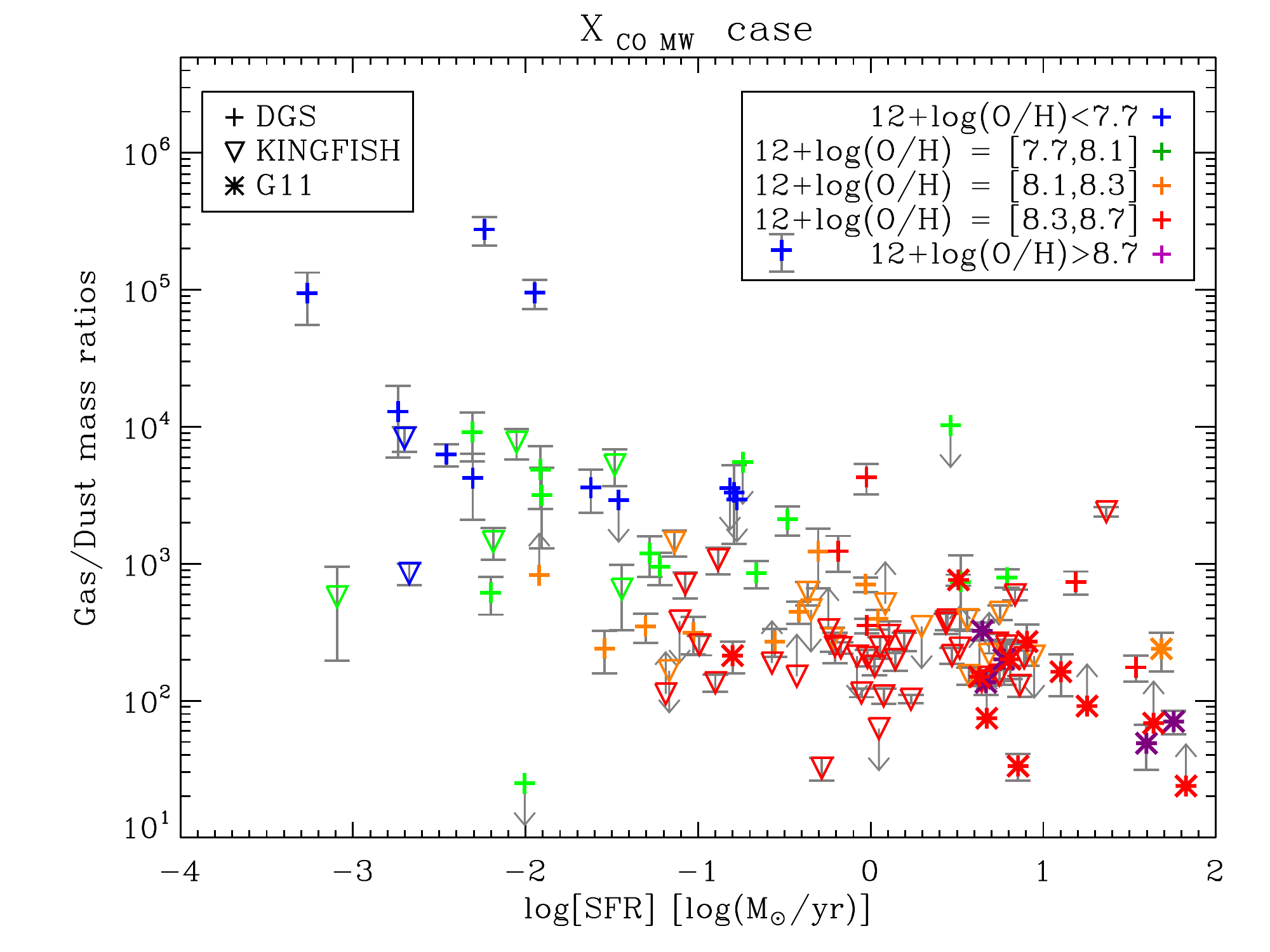}
\includegraphics[width=8.8cm]{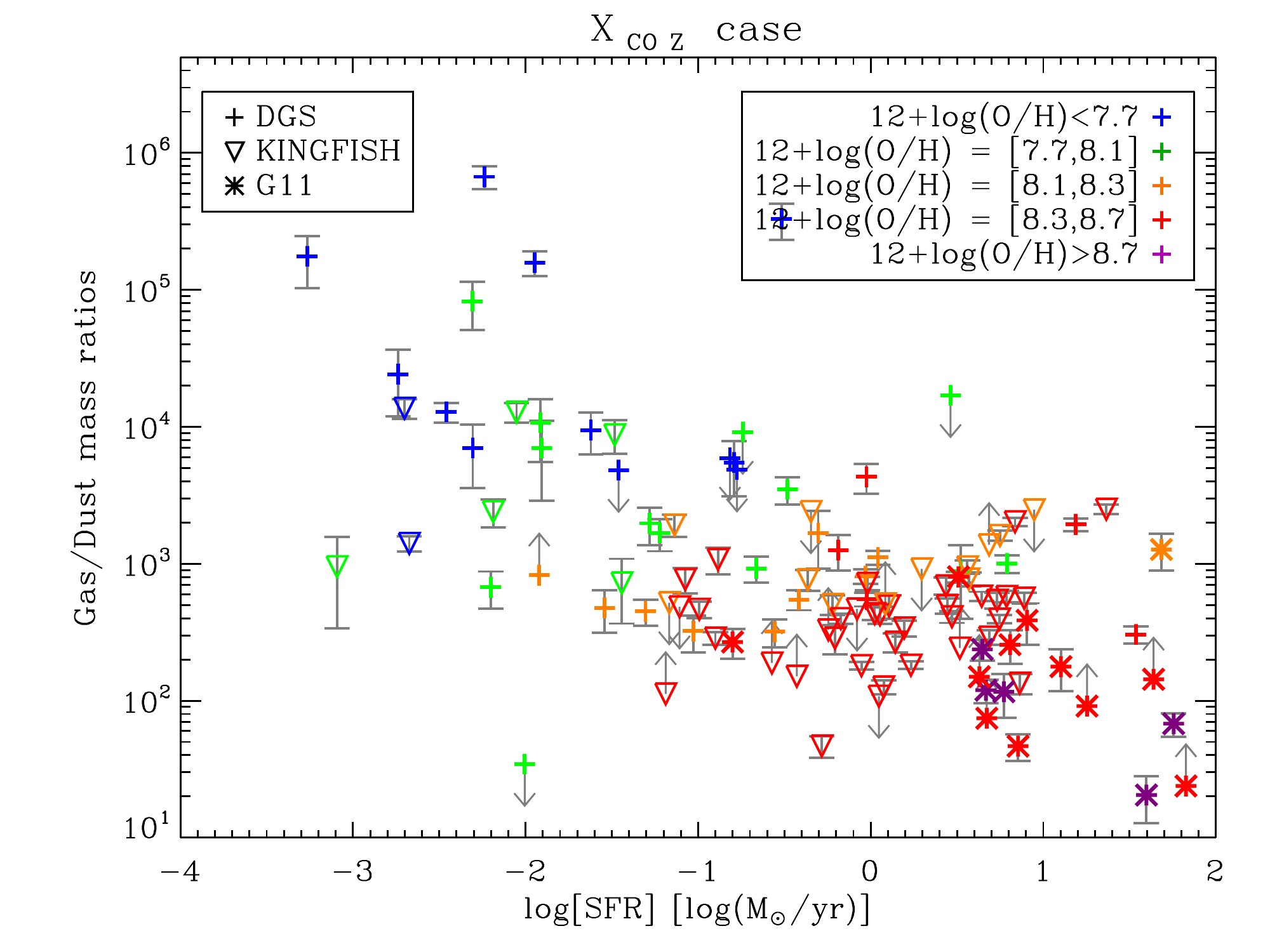}
\caption{{\revised G/D as a function of morphological type {\it (top row)}, stellar mass {\it (center row)} and star formation rate {\it (bottom row)} for the 2 values of \xco: \xcogal\ ({\it left column}) and \xcoz\ ({\it right column}). The colours code for the metallicity of the galaxies and the symbols differentiate between the three samples: DGS (crosses), KINGFISH (downward triangles) and G11 (stars).}}
\label{f:GDvsparam}
\end{center}
\end{figure*}

{\revised In this appendix we present how the sample distributes in morphological types, stellar masses and star formation rates (Fig. \ref{f:histosMorMsSFR}). The morphological types were taken from NED. The stellar masses were obtained from the formula of \cite{Eskew2012} using the 3.6 and 4.5 \mic\ IRAC flux densities. Whenever IRAC data was not available, we compute synthetic IRAC photometry from the SEDs. The SFR were converted from TIR luminosities using the \cite{Kennicutt1998} formula. 
We also present the G/D as a function of these three parameters (Fig. \ref{f:GDvsparam}) and as a function of the specific star formation rate (SSFR, Fig. \ref{f:GDvsSSFR}). SSFR is defined as the star formation rate divided by the stellar mass: SFR/$M_*$.}

\end{document}